%% file: manuscript.tex
\documentclass[preprint,12pt,5p,twocolumn]{elsarticle}

\def\review{preprint}
\def\rev{rev}

\def\figstyle{color}
\def\bw{bw}

\ifx\review\rev
 \usepackage{lineno,rotating}
\fi

\usepackage{amssymb}

\biboptions{comma,square,numbers}
\journal{Nuclear Instruments and Methods A}

\begin{document}

\begin{frontmatter}



\title{Time Projection Chambers for the T2K Near Detectors}

\input{authors}
\begin{abstract}
\small{
The T2K experiment is designed to study neutrino oscillation properties 
by directing a high intensity neutrino beam produced at 
J-PARC in Tokai, Japan, towards the large Super-Kamiokande detector 
located 295 km away, in Kamioka, Japan.
The experiment includes a sophisticated near detector complex, 
280 m downstream of the neutrino production target in order to measure
the properties of the neutrino beam and to better understand neutrino
interactions at the energy scale below a few GeV.
A key element of the near detectors is the ND280 tracker, consisting of 
two active scintillator-bar target systems surrounded by three large 
time projection chambers (TPCs) for charged particle tracking. 
The data collected with the tracker is used to study charged 
current neutrino interaction rates and kinematics prior to oscillation, 
in order to reduce uncertainties in the oscillation measurements
by the far detector. 
The tracker is surrounded by the former UA1/Nomad dipole magnet and the 
TPCs measure the charges, momenta, and particle types of
charged particles passing through them.
Novel features of the TPC design include its rectangular box layout
constructed from composite panels, the use of bulk micromegas 
detectors for gas amplification, electronics readout based on a 
new ASIC, and a photoelectron calibration system.
This paper describes the design and construction of the TPCs, 
the micromegas modules, the readout electronics, 
the gas handling system, and shows the 
performance of the TPCs as deduced from measurements 
with particle beams, cosmic rays, and the calibration system.}
\end{abstract}

\ifx\review\rev
\fi

\end{frontmatter}

\ifx\review\rev
 \linenumbers
\fi

\input{chap01}
\input{chap02}
\input{chap03}
\input{chap04}
\input{chap05}
\input{chap06}
\input{chap07}
\input{chap08}
\input{chap09}





\bibliographystyle{ieeetr}
\begin{flushleft}
\rm{\small{\bibliography{manuscript}}}
\end{flushleft}


\end{document}

%% file: authors.tex
\author[geneva]{N.\,Abgrall}
\author[lpnhe]{B.\,Andrieu}
\author[cea]{P.\,Baron}
\author[geneva]{P.\,Bene}
\author[infnbari]{V.\,Berardi}
\author[cea]{J.\,Beucher}
\author[triumf,uvic]{P.\,Birney}
\author[cea]{F.\,Blaszczyk}
\author[geneva]{A.\,Blondel}
\author[uvic]{C.\,Bojechko}
\author[cea]{M.\,Boyer}
\author[geneva]{F.\,Cadoux}
\author[cea]{D.\,Calvet}
\author[infnbari]{M.G.\,Catanesi}
\author[valencia]{A.\,Cervera}
\author[cea]{P.\,Colas}
\author[cea]{X.\,De\,La\,Broise}
\author[cea]{E.\,Delagnes}
\author[cea]{A.\,Delbart}
\author[geneva]{M.\,Di\,Marco} 
\author[cea]{F.\,Druillole}
\author[lpnhe]{J.\,Dumarchez}
\author[cea]{S.\,Emery}
\author[valencia]{L.\,Escudero}
\author[triumf]{W.\,Faszer}
\author[geneva]{D.\,Ferrere}
\author[geneva]{A.\,Ferrero}
\author[uvic]{K.\,Fransham}
\author[uvic]{A.\,Gaudin}
\author[cea]{C.\,Giganti}
\author[cea]{I.\,Giomataris}
\author[cea]{J.\,Giraud}
\author[triumf]{M.\,Goyette}
\author[triumf]{K.\,Hamano}
\author[ubc,ipp]{C.\,Hearty}
\author[triumf]{R.\,Henderson}
\author[cea]{S.\,Herlant}
\author[barcelona]{M.\,Ieva}
\author[ubc]{B.\,Jamieson}
\author[barcelona]{G.\,Jover-Ma\~nas}
\author[uvic,triumf]{{D.\,Karlen}\corref{cor}}\ead{karlen@uvic.ca}
\author[triumf]{I.\,Kato}
\author[triumf]{A.\,Konaka}
\author[aachen]{K.\,Laihem}
\author[triumf,uvic]{R.\,Langstaff}
\author[padova,infnpad]{M.\,Laveder}
\author[cea]{A.\,Le\,Coguie}
\author[lpnhe]{O.\,Le\,Dortz}
\author[triumf]{M.\,Le\,Ross}
\author[triumf,uvic]{M.\,Lenckowski}
\author[barcelona]{T.\,Lux}
\author[cea]{M.\,Macaire}
\author[triumf,ubc]{K.\,Mahn}
\author[geneva]{F.\,Masciocchi}
\author[cea]{E.\,Mazzucato}
\author[infnpad]{M.\,Mezzetto}
\author[triumf]{A.\,Miller}
\author[cea]{J.-Ph.\,Mols}
\author[valencia]{L.\,Monfregola}
\author[cea]{E.\,Monmarthe}
\author[uvic]{J.\,Myslik}
\author[cea]{F.\,Nizery}
\author[triumf]{R.\,Openshaw}
\author[geneva]{E.\,Perrin}
\author[cea]{F.\,Pierre}
\author[cea]{D.\,Pierrepont}
\author[uvic]{P.\,Poffenberger}
\author[lpnhe]{B.\,Popov}
\author[infnbari]{E.\,Radicioni}
\author[geneva]{M.\,Ravonel}
\author[cea]{J.-M.\,Reymond}
\author[cea]{J.-L.\,Ritou}
\author[uvic]{M.\,Roney}
\author[aachen]{S.\,Roth}
\author[barcelona]{F.\,S\'anchez}
\author[cea]{A.\,Sarrat}
\author[geneva]{R.\,Schroeter} 
\author[aachen]{A.\,Stahl}
\author[valencia]{P.\,Stamoulis}
\author[aachen]{J.\,Steinmann}
\author[aachen]{D.\,Terhorst}
\author[lpnhe]{D.\,Terront}
\author[uvic]{V.\,Tvaskis}
\author[cea]{M.\,Usseglio}
\author[lpnhe]{A.\,Vallereau}
\author[cea]{G.\,Vasseur}
\author[ubc]{J.\,Wendland}
\author[geneva]{G.\,Wikstr\"om}
\author[cea]{M.\,Zito}

\cortext[cor]{Corresponding authour}

\address[ipp]{Institute of Particle Physics, Canada}
\address[triumf]{TRIUMF, Vancouver, Canada}
\address[ubc]{Department of Physics and Astronomy, 
University of British Columbia, Vancouver, Canada}
\address[uvic]{Department of Physics and Astronomy, 
University of Victoria, Victoria, Canada}

\address[cea]{Irfu/DSM, CEA-Saclay, 91191 Gif/Yvette CEDEX France}
\address[lpnhe]{LPNHE, IN2P3-CNRS, 75252 Paris CEDEX 05, France}

\address[aachen]{III. Physikalisches Institut, RWTH Aachen University, Aachen, Germany}

\address[infnbari]{INFN, Sezione di Bari, Bari, Italy}
\address[infnpad]{INFN, Sezione di Padova, Padova, Italy}
\address[padova]{University of Padova, Padova, Italy}

\address[barcelona]{Institut de F\`{\i}sica d'Altes Energies, Barcelona, Spain}
\address[valencia]{IFIC, University of Valencia and CSIC, Valencia, Spain}

\address[geneva]{Physics Section, University of Geneva, Switzerland}

%% file: chap01.tex
\section{Introduction}
\label{sec-intro}

Over the past decade, the phenomenon of neutrino oscillation has been 
firmly established from observations of neutrinos produced by cosmic 
rays in the atmosphere\cite{Ashie:2005ik}, 
by the sun\cite{Ahmad:2002jz}, 
by nuclear reactors\cite{Abe:2008ee},
and by accelerators\cite{Ahn:2006zza,Adamson:2008zt}.
The goals of the T2K experiment\cite{Itow:2001ee} 
are to improve the measurements of 
the “atmospheric” (2-3) mixing parameters by an order of magnitude, and to 
increase the sensitivity to 1-3 mixing, possibly observing this for the first 
time. 
If the experiment finds evidence for 1-3 mixing, 
this will open the possibility of measuring leptonic CP-violation in 
the future.

\subsection{T2K and the off axis near detector}
\label{ss-intro-t2k}

The T2K experiment is designed with an off-axis neutrino beam 
configuration\cite{Mann:1993zk,Helmer:1994ac}, 
providing a relatively narrow band beam peaked at about 700 MeV, 
so that the far detector is located at the first oscillation maximum. 
Near detectors, located 280~m downstream of the production target,
are designed to ensure that the neutrino beam properties 
are well understood so that the experiment can reach its ultimate 
sensitivity.
On the neutrino beam axis, the INGRID detector monitors the neutrino
beam profile.
Along the off-axis direction towards the far detector, the ND280
detector measures the interaction rates, neutrino spectra, and
neutrino interaction kinematics.

The ND280 detector consists of several detector systems
contained within the former UA1/Nomad dipole magnet which provides
a magnetic field of approximately 0.2~T.
Innermost are the PiZero detector, specifically designed to study
neutral current interactions that produce $\pi^0$ particles and 
a tracker, consisting of two fine-grained scintillator detectors (FGDs)
that act as active neutrino targets interleaved
with three time projection chambers (TPCs).
Electromagnetic calorimeters surround these detectors within the magnet coil
and planes of scintillators are inserted within the magnet yoke
to act as a muon range detector.

\subsection{ND280 tracker}
\label{ss-intro-nd280}

The ND280 tracker is designed to study charged current neutrino interactions.
At 700\,MeV, a sizable fraction of neutrino interactions 
are charged current quasi-elastic (CCQE), 
in which the neutrino energy can be determined by measuring 
the momentum of the charged lepton. 
For 2-3 mixing studies, the spectrum of $\nu_\mu$ interactions observed 
in the near detector will be used to estimate the unoscillated 
spectrum at the far detector, and $\nu_\mu$ interaction kinematics 
will be studied to help model background from non-CCQE interactions
in the far detector. 
For 1-3 mixing studies, the near detector will measure 
the $\nu_e$ contamination in the beam, 
an important and irreducible background at the far detector.

\subsection{Tracking performance requirements}
\label{ss-intro-tpcperformance}

Neutrino energy estimation in CCQE events is limited at about the 10\% level 
due to the Fermi motion of the struck nucleons. 
For this reason, a relatively modest momentum resolution goal is 
set to be $\delta$(p$_\perp$)/p$_\perp < 0.1 $p$_\perp$ [GeV/c]
(perpendicular to the magnetic field direction).
The overall momentum scale, however, needs to be known at the level of 2\%, 
in order not to limit the  precise determination of $\Delta m^2_{23}$. 
The ionization energy loss of electrons in 1 atm Argon gas is roughly 
45\% larger than for muons over the momentum range of interest. 
To deduce the $\nu_e$ contamination of the beam, the 
resolution in ionization energy loss needs to be better than 10\%.

\subsection{TPC system design overview}
\label{ss-intro-designoverview}

The tracker performance goals can be reached with
time projection chambers\cite{Marx:1978zz}
operated in a 
magnetic field of 0.2 T with a sampling length of 700 mm and 
pad segmentation of 70 mm$^2$, providing space point resolution of
about 0.7~mm.
For gas-amplified readout of the ionization electrons, 
the collaboration decided to use bulk micromegas 
detectors\cite{Gio:2006nim}.
To fit the geometry of the UA1/Nomad magnet, 
a rectangular design for the TPCs was required. 

\begin{figure}[htp]
\centering
\includegraphics[width=0.45\textwidth]{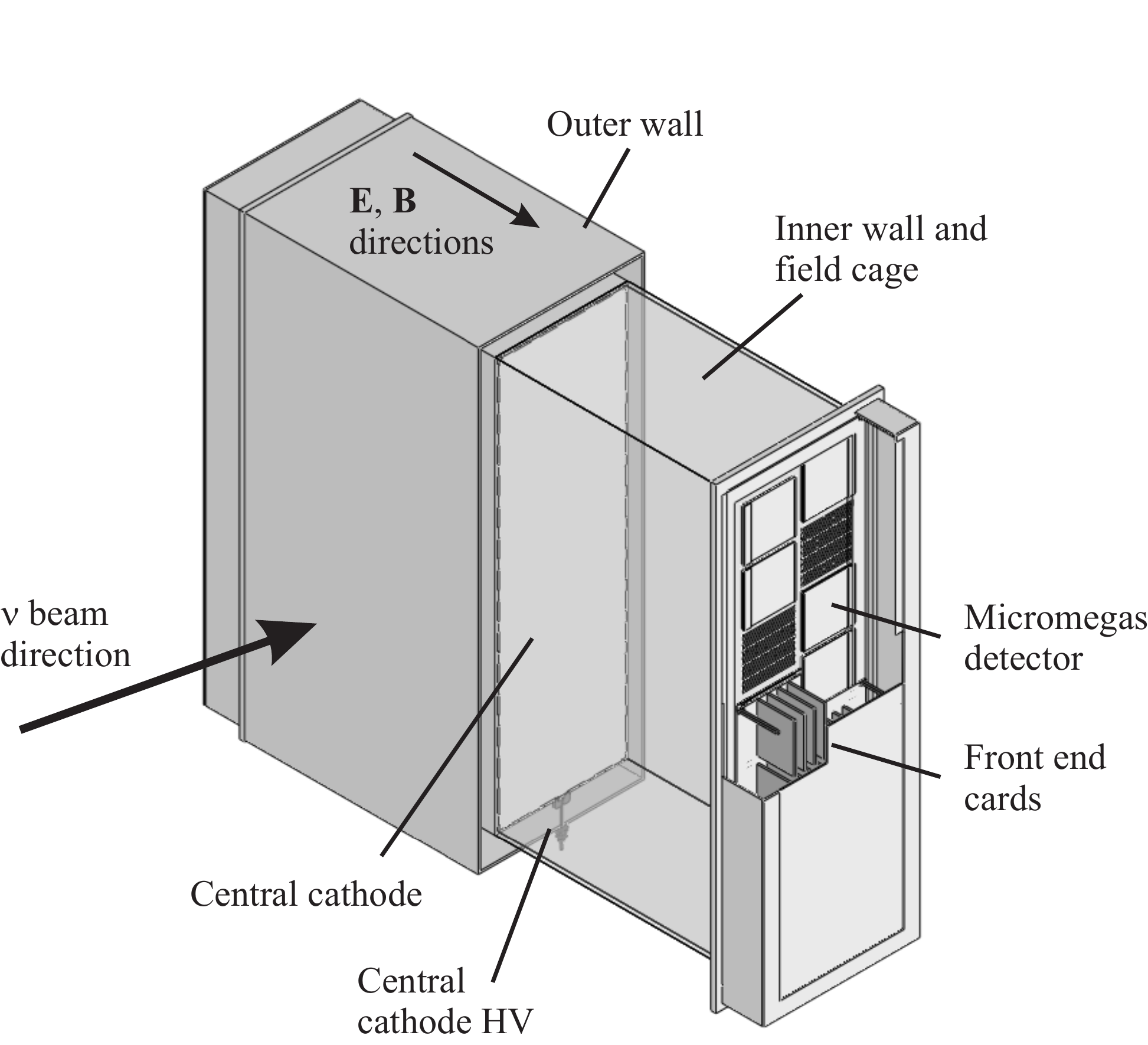}
\caption{Simplified cut-away drawing showing the main aspects of
the TPC design.
The ND280 off-axis detector uses a right handed coordinate system with
$z$ in the horizontal plane along the neutrino beam direction, and $y$ in
the vertical direction.}\label{fig:intro-design}
\end{figure}

A double box design was selected, 
in which the walls of the inner box form the field cage, 
and the walls of the outer box are at ground potential, 
with CO$_2$ acting as an insulator between. 
The walls are made from composite panels, and
the inner box panel surfaces are machined to form a copper strip pattern, 
in order to produce a uniform electric drift field. 
A simplified drawing of the TPC design is shown in 
Fig.~\ref{fig:intro-design}.

The gas system is designed to maintain a stable mixture
in the inner volume and a constant positive pressure with respect to the
outer volume. 
The inner gas mixture, Ar:CF$_4$:iC$_4$H$_{10}$ (95:3:2) and
referred to as ``T2K TPC gas'' in this document,
was chosen for its high speed, low diffusion, and good performance
with micromegas detectors.
There are twelve micromegas modules that tile each
readout plane in two offset columns, so that inactive regions
are not aligned.
Front end electronics cards that plug into the back of the micromegas modules
digitize buffered analog data and send zero suppressed data out of
the detector with optical links.
A photoelectron calibration system is incorporated into the design to
generate a control pattern of photoelectrons from the cathode.

The next six sections describe these TPC subsystems in detail,
followed by a report on the overall performance of the TPCs.

%% file: chap02.tex
\section{Mechanical structure}
\label{sec-mech}

\begin{figure}[htp]
\centering
\ifx\figstyle\bw
 \includegraphics[width=0.4\textwidth]{fig02_bw.jpg}
\else
 \includegraphics[width=0.4\textwidth]{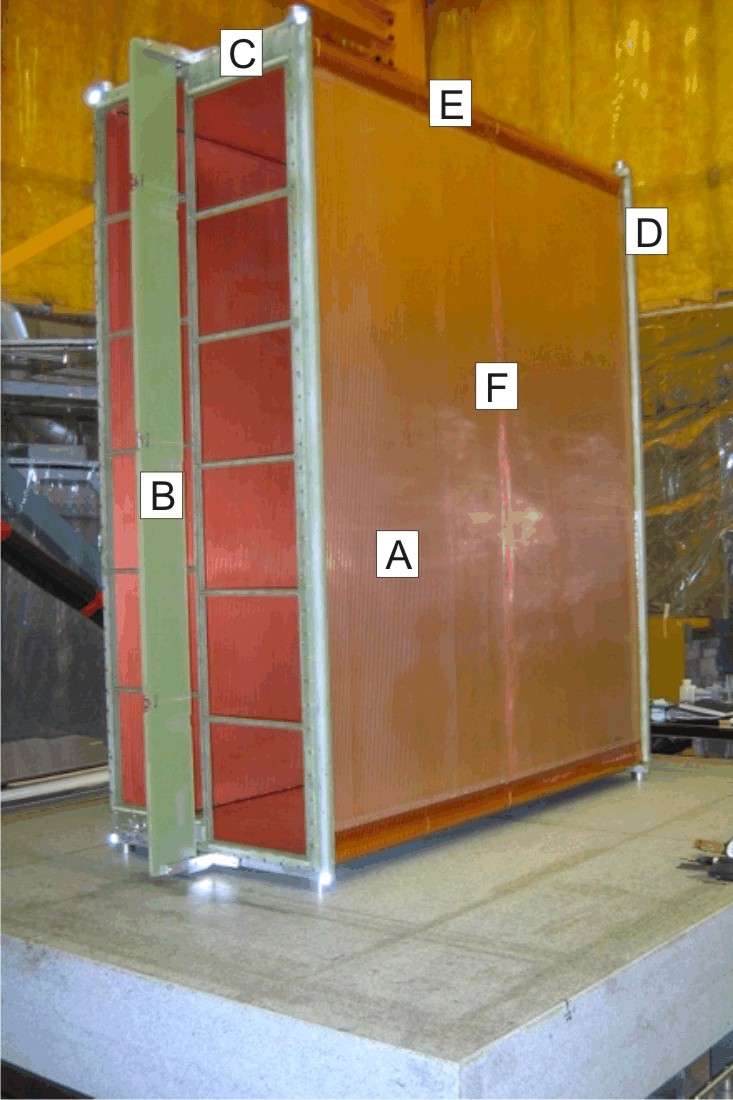}
\fi
\caption{Inner box on the granite table in the TRIUMF clean room.
A: one of inner box walls; B: module frame stiffening plate;
C: module frame; D: inner box endplate; E: field-reducing corners;
F: central cathode location.}
\label{fig:mech-inner-box}
\end{figure}

A TPC module consists of two gas-tight boxes, one inside the other.
The inner box (Fig.~\ref{fig:mech-inner-box})
is subdivided by the cathode located at its midpoint, 
and supports the twelve micromegas modules that are located in a plane
parallel to the cathode at each end. The walls joining the cathode 
and the micromegas are covered with a series of conducting strips 
joined by precision resistors, forming a voltage divider that creates
the uniform electric field along the drift direction. The inner box
is constructed from G10 and G10/rohacell laminates. 

\begin{figure}[htp]
\centering
\ifx\figstyle\bw
 \includegraphics[width=0.4\textwidth]{fig03_bw.jpg}
\else
 \includegraphics[width=0.4\textwidth]{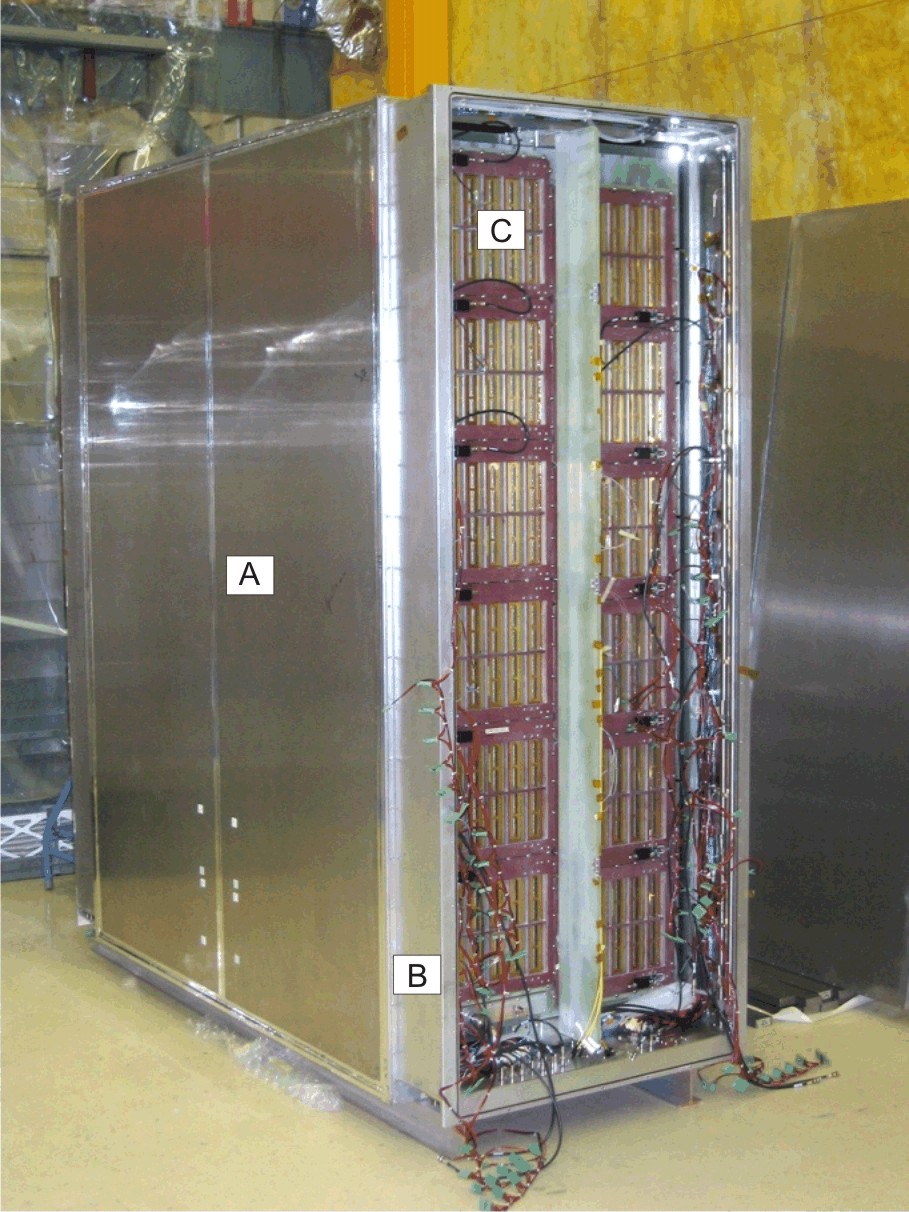}
\fi
\caption{Outer box with the different components labeled.
A: one of the outer box walls; B: service spacer; C: one of the
micromegas modules inserted into the module frame.}
\label{fig:mech-outer-box}
\end{figure}

The outer box (Fig.~\ref{fig:mech-outer-box}), 
which is made from aluminum and aluminum/rohacell 
laminates, contains a CO$_2$ atmosphere that provides the electrical
insulation between the inner box and ground, and excludes atmospheric
oxygen from entering the inner volume. 

The following sections outline the requirements on the design, 
the design itself, and details of the construction. 

\subsection{Requirements}
\label{ss-mech-req}

There are three principle requirements that drive the design of the TPC:
\begin{itemize}
\item The electric field distortions resulting from imperfections in the
construction of the module should produce distortions in the reconstructed
positions of primary electrons 
of approximately 0.2~mm or less, small compared
to the nominal space point resolution of approximately 0.7~mm,
and small enough to not affect the momentum scale by more than 2\%.

\item The module must be sufficiently gas tight to keep the oxygen level in 
the drift volume below about 10~ppm. It is also desirable
to keep the CO$_2$ concentration
in the inner box low enough that the purification filters that remove 
the CO$_2$ have a lifetime of at least one month. 

\item The maximum electric field in the region between the inner and 
outer boxes should be at least a factor of 3 below the nominal 
breakdown in CO$_2$ of 20~kV/cm. 
\end{itemize}

The design described below achieves these goals while maximizing the physics
performance:
\begin{itemize}
\item The design limits the amount of material. This is especially 
important for the walls between the tracking volume and the FGD, which
multiple scatter the leptons used to infer the neutrino 
energy spectrum.

\item The design maximizes the tracking volume given the overall 
envelope available within the ND280 detector. Particular effort was 
made to maximize the active tracking length along the neutrino beam 
direction.
\end{itemize}

Studies were performed with Comsol 
Multiphysics\footnote{http://www.comsol.com}
to optimize the field cage design and
to translate the tolerance on the electric field distortions
into tolerances on the design and 
construction. These studies were largely two-dimensional, consisting of a
horizontal slice through the center of the module. 
The resulting distortions in tracking
were obtained by drifting charges from various locations. 
Distortions were found to be minimized by having the strips on the
inside and outside surfaces aligned.
The key 
tolerances arising from these studies are:
\begin{itemize}
\item The resistor pairs that form the voltage divider between the
central cathode and the micromegas must be matched to within an 
rms of 0.1\%.

\item The central cathode should be flat to within 0.1~mm.

\item The micromegas plane should be flat to within 0.2~mm.

\item The central cathode and micromegas planes should be parallel to
within 0.2~mm.
\end{itemize}

The latter two points imply a corresponding stiffness of the inner box
to the operating over pressure of the drift gas, and to the 
weight of the front end electronics.   

The relative alignment of the cathode and the micromegas perpendicular to the
drift direction is not as critical, nor is the flatness of the top, bottom, 
front and back inner box walls. 

In addition, the modules were constructed entirely out of non-magnetic 
materials to not distort the magnetic field after it was mapped.
All materials that are exposed in the inner volume were tested
in smaller TPCs to ensure that they do not introduce electronegative
impurities to the drift gas.

\subsection{Design}
\label{ss-mech-design}

\subsubsection{Field cage, including central cathode}

The central cathode and the top, bottom, front, and back walls of the inner 
box are 13.2~mm thick copper-clad-G10/rohacell laminated panels. 
The panels have solid G10 frames, 
which allow for solid mechanical connections and
ensure that the rohacell is not exposed to the drift volume. 

The copper cladding on the inner and outer surfaces is divided into 
10~mm wide strips on an 11.5~mm spacing. A pair of selected 
20~M$\Omega$ resistors joins adjacent strips, forming a voltage 
divider between the cathode and the mounting point of the micromegas. 
The inner and outer surface strips are 
aligned and
jumpered together so that the
voltage differential between the central cathode and ground is 
supported by the CO$_2$ and not the inner box walls. 

The maximum electric field occurs at the edges where the 
walls meet. To reduce this field to acceptable levels, these
edges are rounded off using curved G10 parts covered with 
kapton sheets with copper strips on an 11.5~mm spacing. 
The resulting maximum electric field is 5~kV/cm. These
strips also provide the electrical connection between the copper strips
on adjacent walls. 

The central cathode has two mesh-covered cutouts to allow gas flow 
between the two drift volumes.
The cathode copper surface holds the aluminum targets used 
as part of the laser calibration system, as described in
section~\ref{sec-calib}.

The ends of the four walls are glued into endplates, solid G10
rectangular frames. These hold the O-rings that make the gas seal
between the module frames and the field cage. Thin G10 sheets 
with copper strips are mounted to the inner edges of the endplates
to continue the field cage to the micromegas mounting surface. 

Excluding the module frames, the inner box has exterior dimensions
of $1808 \times 2230 \times 854$~mm in $x \times y \times z$. 
Along $z$, the neutrino beam direction, the
interior dimension is 772~mm. The active tracking length is 
approximately 
720~mm
after excluding the 15~mm closest to each wall, 
in which the field is insufficiently uniform,
and 20~mm between the two columns of micromegas modules.
The maximum drift distance
from central cathode to micromegas is 897~mm.
Given the nominal cathode voltage of -25~kV and micromegas voltage
of -350~V, the drift field is approximately 275~V/cm, close to the
saturation point for T2K TPC gas.

\subsubsection{Module frame}

The twelve micromegas modules are mounted into individual cutouts in
the module frame at each end of the inner box. O-rings provide the gas
seal between the micromegas and the frame. 

The module frame---which by itself is not rigid---is held planar by 
its screwed connection to the endplate, and by a 158~mm thick stiffening
plate that is screwed to the module frame at three locations along
its axis. 

The module frame also supports the gas inlet and outlet manifolds, 
and the optical packages of the laser calibration system. 

\subsubsection{Outer box}

The front, back, top, and bottom walls of the outer box are formed 
from 14.3~mm thick aluminum-rohacell laminated panels. The panels have solid 
aluminum frames that allow for solid mechanical connections 
and which ensure that the rohacell is not exposed to the gas 
volume. 

An endplate---a solid aluminum rectangular frame---is glued to each
end of the box. The service spacer, described in the next section, 
is screwed to each endplate, with an O-ring making the gas seal. The 
outer box endplates are the support points for mounting the TPC
into the ND280 detector, and are the points where the inner box
is connected to the outer. 

Because the inner and outer boxes have quite different thermal expansion
coefficients, the inner box supports allow for limited relative motion
between the boxes. 

The gap between the inner and outer boxes is 68~mm on three surfaces, and
118~mm on the bottom, where extra space is required for the cathode HV
connection. The overall size of an outer box (including the service 
spacers) is 
$2302 \times 2400 \times 974$~mm. The endplates are 12~mm wider in $z$ than 
the front/back walls when the box is at atmospheric pressure. This ensures
that even at the maximum CO$_2$ over-pressure of 5~mbar, the resulting 
5~mm bowing of each of the front and back walls does not impinge upon the
stay-clear region between a TPC module and the neighboring components in 
ND280.

The outer and inner box walls combined have a thickness of 3.3\% of
a radiation length for particles entering the TPC in the beam direction
and away from the cathode and edges of the active volume.

\subsubsection{Service spacers}

A service spacer is screwed with an O-ring to each outer box endplate. 
The service spacer, together with a removable cover, forms the
volume which encloses the front end electronics. The distance from 
the micromegas outer surface to the inner surface of the 
service spacer cover is 200~mm. This volume is in the CO$_2$ atmosphere.

All services, other than the central cathode HV, enter the 
detector through feedthroughs in the bottom of the service spacer. 
These include gas, cooling water, electronics power, readout and 
calibration, micromegas voltage, and temperature and other monitoring
information. 

\subsection{Machining and assembly}
\label{ss-mech-machining}

Parts for the three TPC modules were machined primarily at TRIUMF, with
some components produced at university and commercial machine shops. 
There were 160 SolidWorks\footnote{http://www.solidworks.com}
drawings used to machine and assemble the
parts for a TPC module. Some modifications were made to the design
after the first module to simplify construction. There were no changes
with respect to function or performance. 

The project duration from first machining of parts to the shipment of the 
final module was 33 months. Approximately half of this was spent on the first
module, and half on the second and third. Although a conceptual design was
completed before the start of machining, detailed drawings were produced
in parallel with the construction project. 

The modules were assembled in the TRIUMF clean room, in many cases using 
a large, flat granite table. 

The follow sections detail a few aspects of the fabrication of components
and their assembly into completed modules.  

\subsubsection{Router}

Many of the components that went into a module are large and relatively flat, 
including the inner and outer box walls and endplates, the central cathode, 
module frames, and stiffeners. 
These components were machined at TRIUMF on a 
Multicam 5-504-R Moving-Gantry router.
It has a 10 foot square bed, with movable gantry head, 
with vacuum hold-down used for almost all parts. 

This router was relatively new to TRIUMF, and its commissioning coincided
with the start of production of parts for this project. The learning curve
was quite extensive, but tolerances of approximately 0.1~mm were 
eventually achieved over distances of 1--2~m.  

The parts for the 
second and third modules required approximately 900 hours each on the router,
plus comparable time on smaller CNC and manual machines. 

\subsubsection{Panel lamination}

A significant fraction of all components were fabricated from laminated
panels, including the inner box and outer box walls, the central cathode,
and the service spacer covers. 

Inner box panels were assembled from five 
FR5 or G10 frame pieces, 1/32~inch
G10, copper-clad on one surface, and rohacell. The
FR5 bars were water-cut from plates ground to $11.60 \pm 0.05$~mm thickness.
The G10 frame pieces were used only on the first module and were 
milled to thickness on the router. This process was time consuming
and did not achieve the tolerance of the ground FR5. 

Outer box panels used four or five aluminum frame pieces machined from 
$1/2$~inch tooling plate, 1/32~inch aluminum sheets, and rohacell. 

\begin{figure}[ht]
\centering
\ifx\figstyle\bw
 \includegraphics[width=0.4\textwidth]{fig04a_bw.jpg}
\else
 \includegraphics[width=0.4\textwidth]{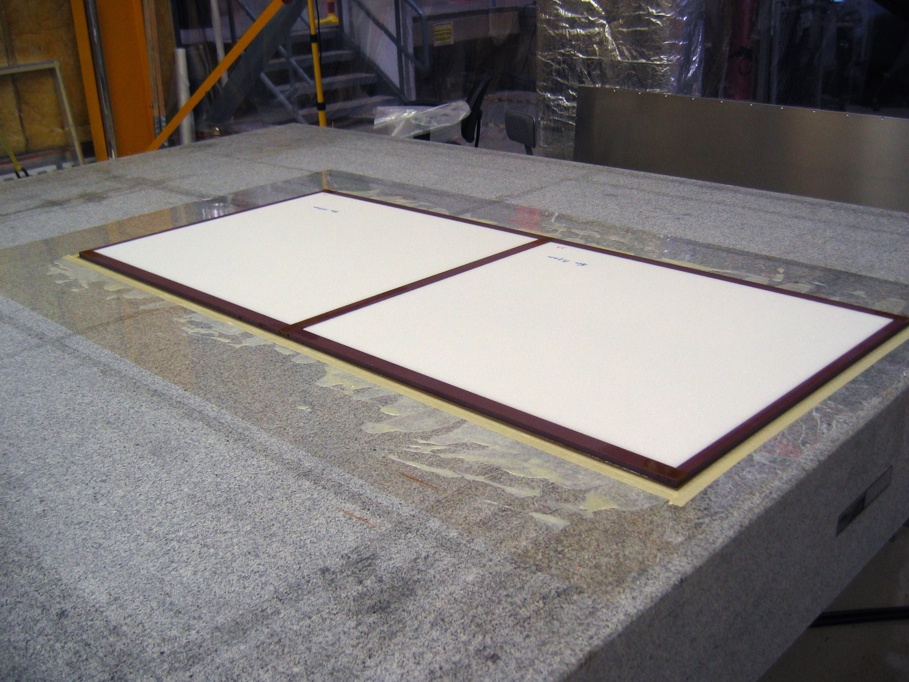}
\fi
\vskip 0.05in
\ifx\figstyle\bw
 \includegraphics[width=0.4\textwidth]{fig04b_bw.jpg}
\else
 \includegraphics[width=0.4\textwidth]{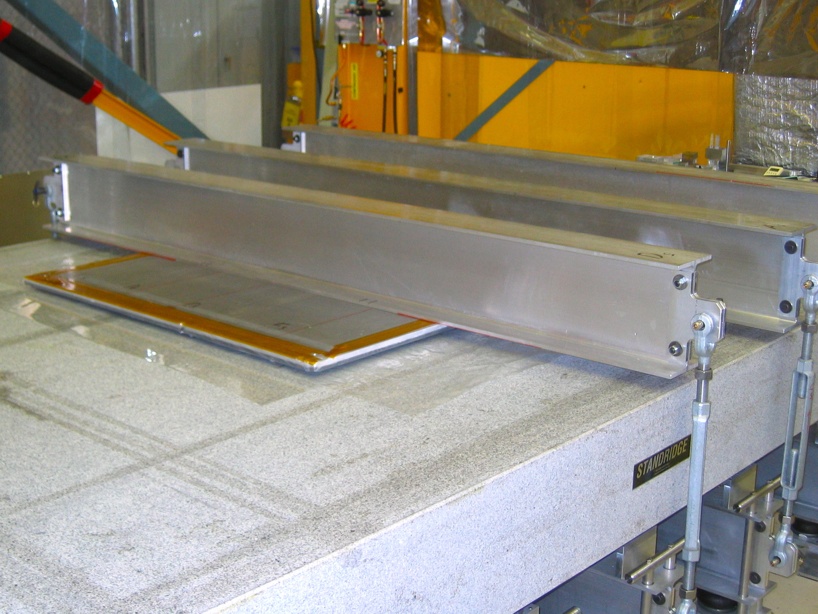}
\fi
\caption{An inner box panel being laminated. The central FR5 bar
provides a solid mounting point for the cathode.}
\label{fig:mech-panel}
\end{figure}

There were four major steps to create a panel from these components.
\begin{enumerate}
\item A skin was placed on the granite assembly table and uniformly
covered with a thin layer of epoxy. The four or five frame pieces
were then screwed together on the panel, together with one or two
slightly over-thickness
pieces of rohacell (Fig.~\ref{fig:mech-panel}). 
The parts were pressed over night using 
one-inch thick 
aluminum press plates covered with a poron layer. 

\item The rohacell was skimmed on the router to match the thickness of 
the frame pieces, and the excess skin extending beyond the frame
pieces was trimmed.

\item The second skin was glued onto the panel using the granite table
and press.

\item The router was used to do final machining, including overall 
transverse dimension, grooves for central cathode mounting, and 
the cutting of the copper surfaces of the inner box panels into 
the field-forming strips. 
\end{enumerate}

\subsubsection{O-ring seals in G10 and other joints}

Permanent joints in both the inner and outer boxes were 
screwed and glued with epoxy for strength and to form a gas seal.
Joints that required disassembly used O-rings. These joints included 
micromegas to module frame; module frame to endplate; service 
spacer to endplate; and cover to service spacer. 

The screws were non-magnetic stainless 316 in almost all cases.
Stainless 18-8 was used if 316 was not available. 

The inner box O-ring joints required an O-ring groove machined into 
G10. Due to the fibrous nature of the material, a machined G10 
surface did not give an adequately smooth surface. Instead, the groove
was milled over-depth, filled with a low-viscosity optical epoxy, then machined
to final depth and hand-polished. This technique gave good results, 
even over the 17~m of 1/16~inch O-ring grooves needed for micromegas
mounting per TPC readout plane. A typical inner box leak is
$40 \pm 10$~cc/min at an overpressure of 4.5~mbar.

Due to concern about tapped holes in G10 stripping with repeated use, 
brass inserts were used for the holes corresponding to the O-ring
grooves in G10. 

\subsubsection{Inner box assembly}

A number of alignment steps were performed
prior to the gluing of the inner box.  The endplate, module frame, 
and module frame stiffener for each end were assembled together on the granite
table and the screws adjusted to flatten the module frame to 
within $\pm$ 25~$\mu$m.

The targets for the laser calibration system were manually placed on the 
central cathode in locations indicated by scribe marks engraved by the
router as part of the cathode machining. The targets were die-cut circles
and strips made from aluminum tape with conducting adhesive. 
The actual locations
were then surveyed using a camera mounted on the router head. 

The four outer walls and central cathode of the inner box 
were screwed together without
glue, then the walls were 
shimmed with kapton tape (typically one or two layers) 
to ensure that all four were the same length (in $x$) 
to within $\pm$ 25~$\mu$m. The endplates were then screwed in place
and checked for flatness against the granite table. Additional 
kapton tape shims were added as necessary between the inner box 
walls and the endplates to ensure the endplates were 
flat to within $\pm$ 25~$\mu$m. The walls were drilled and doweled
before being disassembled. The central cathode dowels constrained 
it in $y$ and $z$ but not $x$. 

The final step prior to gluing was the soldering of the voltage-divider
resistors onto the outer surface of the top panel. 

The box was reassembled with a thin film of glue on all wall edges, 
except that of the central cathode, which was left free to move within
the screw clearance. After adjustment, it was centered between the 
two endplates to within 50~$\mu$m at all locations around its 
circumference. 

A total of 14 gluing steps (one per day) were required to glue the
inner box. The gas tightness of the box was ensured by two separate
glue seals of each edge. All gluing was done with the inner box on 
the granite table to keep it square and flat. 

After the completion of gluing, the field-reducing corners were mounted
and the kapton strips soldered between adjacent walls. 

The module frame and stiffener were bolted in place, and the micromegas
modules mounted. The inner box was on the granite table for these 
steps as well: even after gluing, a box was not rigid until the 
module frame and micromegas were mounted. 

After gas tightness and high-voltage tests of the micromegas and the 
central cathode, the inner box was ready for final assembly.

\subsubsection{Resistor chain}

The resistor chain between the central cathode and the micromegas in 
each drift volume is formed from 79 pairs of 20~M$\Omega$ resistors.
The resistors were standard 1\% tolerance surface-mount pieces, 
measured and sorted to give pairs with combined resistances matched
to an RMS of 0.17~per mil.

The resistors were soldered between strips
on the outer surface of the bottom panel 
prior to the inner box assembly. The kapton corner strips were 
soldered with jumpers to the panels to complete the field cage. 

The panels required extensive cleaning to remove
solder flux between adjacent strips that would otherwise pull the
resistance out of tolerance. 

The inner box was designed so that all gluing could be done without
epoxy connecting adjacent copper strips. Although the glue itself 
is not conducting, moisture adhering to its surface can 
produce conductances high enough to distort the field. In one module, 
excess glue was used in one of the gluing stages, producing significant
distortions during initial testing. The effect disappeared after the
box spent two weeks in dry gas. 

The measured RMS of the strip-to-strip resistance, after cleaning
(and drying), was 0.5~per mil or better. 

\subsubsection{Inner box mechanical tolerances}

The thicknesses of the second and third central cathodes, 
constructed using ground FR5, were uniform to within 
$\pm$ 40~$\mu$m. The first, which used milled G10, was uniform to
$\pm$ 110~$\mu$m. In the absence of distortions induced during the
installation and alignment of the central cathode, 
these values would be the corresponding values for the 
achieved flatness. Although the cathode was optically surveyed after
installation, the targets used on it did not allow for high precision
in that dimension, so the planarity was not directly measured. 

Three target locations were measured on the outer surface of each
micromegas during the final optical survey of a TPC module. 
Fitting the 36 locations to a plane gave an RMS deviation 
from planarity that ranged from 85~$\mu$m to 120~$\mu$m for 
five of 
the 
six readout planes, and 180~$\mu$m for the other.  
Contributions include the flatness of the inner box endplates, to 
which the module frames are mounted, $\pm$ 25~$\mu$m maximum
deviation, and variations in the module frame thickness, 
$\pm$ 35~$\mu$m. Although the screws connecting the stiffening rib 
to the module frame were initially adjusted to flatten it to 
within $\pm$ 25~$\mu$m, some of these screws were subsequently
inadvertently changed, contributing to the deviation, particularly 
in one case. 

The maximum deflection of the micromegas surface was measured to be
75~$\mu$m for an inner box overpressure of 4~mbar. The 
deflection of the surface under the normal operating pressure of
$0.40 \pm 0.03$~mbar is therefore small compared to the intrinsic flatness. 
The motion of the module frame under the weight of the front end 
electronics was less than 40~$\mu$m. 

The process of centering the cathode
between the endplates ensured that it was parallel to the endplates
to within $\pm$ 50~$\mu$m. The cathode and micromegas plane were
therefore parallel to better than 100~$\mu$m.

\subsubsection{Service spacer fabrication}

The fabrication method of the service spacers
was changed after the experience with the first of the six that 
were made.
For the first one, the flanges that mate with the outer box and 
the service spacer covers were each machined from a plate of 
aluminum, then welded together with four plates to complete the 
walls. The holes for service feedthroughs were machined prior
to welding; the final machining after welding included 
the O-ring 
grooves on both flanges and final dimensions. 
The substantial amount of required welding warped the part to the extent that
the flange that mated with the outer box endplate needed to be shimmed
with epoxy to ensure a good O-ring seal. 

For the subsequent service spacers, each of the four sides were machined from
1.25--3~inch thick
plates of aluminum, then welded 
prior to final machining. These parts did not require
shimming. 

\subsubsection{Outer box assembly}

The assembly of an outer box, excluding service spacers, was similar
to the assembly of the field cage. The four walls were first shimmed
using kapton tape to within $\pm$ 50~$\mu$m of the same length. The
four walls and two 
endplates were then screwed together without glue and the endplates 
were tested for flatness against the granite table. Kapton tape shims
were added between the walls and endplates until the maximum 
deviation from flatness of an endplate was no more than 75~$\mu$m. 
This tolerance was more than adequate to ensure a uniform compression 
of the O-ring between the outer box and service spacers. 

The service spacers and covers were temporarily mounted
to test the gas tightness of the outer box, then removed for
the final assembly. After the edges were verified to be gas tight, 
they were covered with conducting aluminum tape to ensure good
electrical contact between walls and with the endplates. 

\subsubsection{Final assembly}

The inner box was rolled into the outer box on a pair of long 
rails temporarily mounted on the outer box endplates. After 
attaching the inner box supports, the rails were withdrawn. 

The service spacers were then mounted on each end, and the 
front end electronics installed and tested. 

\subsection{Materials and suppliers}

A partial list of the materials used in the TPC construction
is shown in table~\ref{tab:mech-suppliers}.

\ifx\review\rev
 \begin{sidewaystable}
 \centering
\else 
 \begin{table*}
\fi
\caption{A partial list of materials and suppliers used in the 
mechanical construction.}
\footnotesize{
\begin{tabular}{|p{3.25in}|p{3.25in}|} \hline \hline
Material & Supplier \\ \hline
Epoxy: Epon 826 resin, 	Versamid 140 resin	&				E. V. Roberts, 18027 Bishop Ave, Carson CA \\ \hline
Type 316 stainless steel screws	& McMaster-Carr, Fabory Metrican, Pacific Fastener \\ \hline
Buna-n o-rings with vulcanized Joint & Wriason Seals \\ \hline
51IG Rohacell Sheet	& Rohm Industries available from Scion Industries LLC, 3693 East County Road 30, Fort Collins CO \\ \hline
20 Mohm surface mount, resistors 2512 package &			Queale Electronics 485 Burnside Rd East, Victoria, B.C. \\ \hline
Aluminum; bars were Mic-6 cast Al tooling plate $0.500 \pm 0.005$ inch & Copper and Brass Sales (Thyssenkrupp Materials), 19044 Ð95A Ave, Surrey, B.C. \\ \hline
G10 bare and copper clad one side; Skins were 1/32~inch thick G10 clad with .0014Ó thick Cu	 & Current Inc 30 Tyler St., East Haven CT \\ \hline
Flexible circuit boards, 1 oz copper on 2 mil polyimide film & Tech Etch, 45 Aldrin Rd, Plymouth, MA \\ \hline
Threaded brass inserts & Yardley Products, 10 West College Ave, Yardley PA \\ \hline
Nitto AT-5105E Aluminum tape with conductive adhesive &		Supplied, slit and punched by Carolina Tape Supply Corp, 502 19 St Pl S.E., Hickory NC\\ \hline
FR5 ground to thickness	$11.6 \pm 0.05$ mm & VRC (vonroll) Passwangstrasse 20, CH4226 Breitenbach, Switzerland \\ \hline
BC600 Optical Epoxy & Saint-Gobain, 12345 Kingsman Road, Newbury, OH \\ \hline
G10 Fiberglass tubing & Sabic Polymershapes, 104-11 Burbridge St, Coquitlam, B.C.\\ \hline

\hline \hline
\end{tabular}
}
\label{tab:mech-suppliers}
\ifx\review\rev
 \end{sidewaystable}
\else 
 \end{table*}
\fi

%% file: chap03.tex
\section{Gas system}
\label{sec-gas}

As described in section~\ref{sec-mech}, 
each TPC module consists of an inner volume (TPC) 
containing the drift space for the primary electrons, 
and an outer volume (Gap) to insulate the grounded outer box from the 
high-voltage field-cage, and to reduce diffusion of atmospheric contaminants 
into the drift gas.
The gas system is required to supply the appropriate gas mixtures to the 
modules, while maintaining required flow rates, differential pressure 
constraints, gas purity constraints, and gas mixture composition stability.
Systems also have to be in place to protect the gas system, detector chambers 
and personnel from dangerous conditions caused by component failures or 
operator errors.

\subsection{Requirements}
\label{ss-gas-req}

Each of the three TPC volumes contains 3000~litres (L), and each of the three 
Gap volumes contains 3300~L.
The Gap CO$_2$ flow rate needs to be high enough to prevent significant 
build-up of atmospheric contaminants due to diffusion into the outer volume.
This is not a very strenuous constraint.
HV breakdown and leakage currents in the field cage can easily be avoided
by maintaining H$_2$O concentrations less than a few tenths of a percent.
Similarly, diffusion of N$_2$ and O$_2$ from the Gaps into the TPCs can be kept
at insignificant levels by maintaining less than a few percent of air
contamination in the Gaps.
To allow a complete 5~volume-change purge of the Gaps in a reasonable
1.5~days, we chose a maximum total flow capacity of 22.2~L/min
(7.4~L/min per Gap).   

The TPC flow rates have to be large enough to prevent significant 
build up of contaminants diffusing into the TPC volumes and associated 
gas system components.
Since the TPC volumes are surrounded by the CO$_2$ 
filled Gap volumes, the major contaminant will be CO$_2$.
Based on estimates of likely diffusion rates of external contaminants, 
the TPC gas system was designed for an operating flow of 10~L/min/TPC 
(30~L/min total flow), corresponding to 5~TPC-volume flushes per day.
To reduce gas operating costs, the system was designed to purify and 
recycle a major portion of the TPC exhaust gas.
An upper limit on the recycling ratio is determined by the requirement
to maintain TPC pressure during rapid atmospheric pressure increase.
The maximum expected atmospheric pressure increase rate of 
10~mbar/hour requires a minimum fresh input gas flow rate of 
1.5~L/min, corresponding to a 95\% recycle ratio at our nominal 
10~L/min/TPC flow rate.

Differential pressure between the inner TPC volume and the outer 
Gap volume causes the walls of the TPC volume to deflect.
These deflections cause changes to the electric field in the drift
region, which could cause unacceptable distortions to the electron drift.
To avoid this,
the TPC gas system was designed to maintain TPC to Gap 
differential pressure stability to less than $\pm$0.10~mbar during normal 
operations.  
To reduce the ingress of CO$_2$ and other contaminants from the 
surrounding Gap volume, the TPC volume is operated at a slight 
overpressure of 0.4~mbar with respect to the Gap volume.
 
The outer Gap volume walls are 
also sensitive to deflections.
Tests indicated we could operate with up to 5~mbar differential 
pressure between the Gap volume and atmosphere.
However, the gas system has pressure relief bubblers to protect
both the inner TPC and the outer Gap volumes from damaging
pressure excursions due to failure of other gas system components.
These bubblers are referenced to atmospheric pressure.
The TPC pressure with respect to atmosphere is the sum of the
TPC-Gap differential pressure plus the Gap-atmosphere differential 
pressure.
In order for the bubblers to protect the TPC volume at the TPC-Gap 
differential $\pm$5~mbar level, the Gap-atmosphere operating 
differential pressure is kept to less than 1~mbar.

The drift velocity and gas gain of T2K TPC gas 
are sensitive to the stability of the gas mixture composition and to 
CO$_2$, N$_2$ and H$_2$O contamination.  
In addition O$_2$ impurities have to be minimised to avoid
electron attachment.  
To meet the required performance of the TPCs, conservative 
limits for mixture stability and these contaminants were established. 
These limits were based on Magboltz\cite{Biagi1999234} 
simulations combined 
with some prototype tests.
The gas mixture stability requirements are Ar = (95.00 $\pm$ 0.03)\% , 
CF$_4$ = (3.00 $\pm$ 0.009)\%, and  iC$_4$H$_{10}$ = (2.00 $\pm$ 0.02)\%.
Gas purity requirements are O$_2$ $<$ 10~ppm, H$_2$O $<$ 100~ppm and 
CO$_2$ $<$ 100~ppm.

\subsection{Design}
\label{ss-gas-design}

\begin{figure*}[htp]
\centering
\ifx\figstyle\bw
 \includegraphics[width=1.0\textwidth]{fig05_bw.pdf}
\else
 \includegraphics[width=1.0\textwidth]{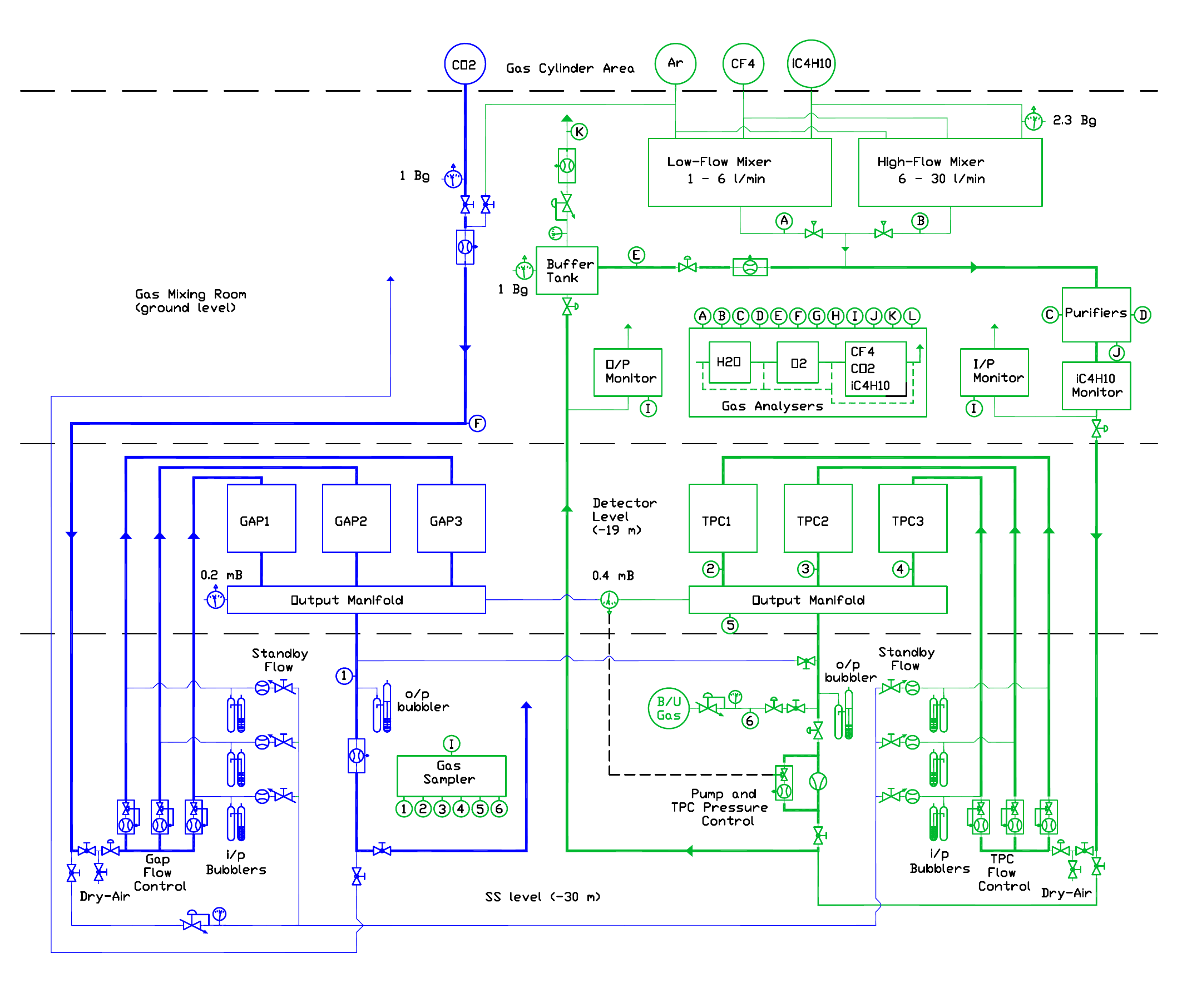}
\fi
\caption{Simplified schematic of the gas system.
The horizontal lines show the division of the equipment between the
gas supply area and gas mixing rooms on the surface, the detector
level, and the service stage 
level below the detector. The left (right) part of the
figure shows the components for the Gaps (TPCs).
}\label{fig:gas-schematic}
\end{figure*}

A simplified schematic of the gas system is shown 
in Fig.~\ref{fig:gas-schematic}.  
There are two gas systems; a simple one-pass flow-controlled CO$_2$ 
system for the Gaps, and a more complicated pressure-controlled, 
gas recycling system with three-component gas-mixers, 
to supply gas to the TPCs.
The major gas system components are at 4~different locations; 
an external gas cylinder enclosure and a temperature controlled gas
mixing room at ground level, the TPC modules and differential pressure 
transducers 19~meters below the surface in the detector hall, 
and the gas input distribution 
and exhaust system on the service stage level (SS level) 
below, 30~meters below the surface.

Each raw gas is supplied from two banks of cylinders, 
equipped with a pressure-actuated automatic switchover to switch to 
the reserve bank when the on-line bank is depleted.  
Supply pressure is set to 1~bar above atmosphere (barg) for the 
Gap CO$_2$ supply and to 2.3~barg~for the Ar, CF$_4$ and iC$_4$H$_{10}$
supplies to the TPC gas mixers.
Since isobutane at 3.3~bar liquefies at 23$^\circ$C, the isobutane cylinders 
and all gas tubes and gas system components containing unmixed isobutane 
are heated to maintain the temperature greater than 30$^\circ$C. 

CO$_2$ is delivered to the Gap input mass flow 
controllers\footnote{All mass flow controllers and mass flow meters are 
digital devices using MODBUS protocol, manufactured by Bronkhorst, 
www.bronkhorst.com}
(MFCs) at 1~barg.
The MFCs are typically set to deliver a flow of 2~L/min to each Gap.
In normal operation, to eliminate the 1.3~mbar pressure head 
that would arise from
the weight of the CO$_2$ in a 22~m vertical tube for a surface exhaust, 
the CO$_2$ gas is exhausted into a building air-exhaust grille located 
7~metres below the centre of the TPCs.
This results in a Gap-atmosphere differential pressure of about 0.2~mbar 
at the operating flow of 2~L/min/Gap, due to the combined effect of 
the negative pressure head and the flow-related back-pressure necessary 
to push the gas through the impedance of the exhaust tubing and gas 
system components.  A mass flow meter (MFM) near the supply input and another
MFM near the exhaust output measure the flows to check for possible leaks.

For the TPC system, the fresh Ar, CF$_4$ and iC$_4$H$_{10}$ components 
are delivered to the mixer, where they are mixed to the desired 
concentrations, and then join with the recirculating gas returning 
from the TPCs.  
The gas then flows through the purifiers and isobutane monitor to 
the 60~m long, 21~mm ID,  SS tube to the
gas system equipment on the SS level, 10~m below the detector.
MFCs on the SS level control the flow rate at 10~L/min to each of the TPCs.
The exhaust from the TPCs flows to a common output manifold where 
the pressure with respect to the Gap output manifold is measured.  
From there, the gas flows through the pump and back through a 
60~m long, 21~mm ID, tube to the buffer tank in the ground-level mixing 
room.  
A back-pressure regulator maintains a constant 1~barg pressure 
in the 110 L buffer tank, by exhausting to atmosphere the amount of 
gas that is required to maintain the set pressure.  
The balance of the gas is recirculated back through the system.  
The capacity of the buffer tank reduces pressure perturbations 
due to sudden flow changes, and also serves as a short-term reservoir 
to maintain TPC pressure if flow from the mixer is interrupted during 
periods of increasing atmospheric pressure.
                                             
To maintain mixture accuracy over a wide range of flow rates, 
the TPC gas system uses two separate mixers.  
The mixer MFCs are sized to deliver a maximum of 6~L/min of mixed 
gas for the low-flow mixer, and a maximum of 30~L/min for the 
high-flow mixer.
The high-flow mixer is typically used for purging the TPCs at 30~L/min, 
while the low-flow mixer typically delivers 3 L/min for our standard 
90\% recycling ratio.

Each purifier consists of 5.4~L of BASF R3-11 activated copper to 
remove O$_2$, plus 9.6~L of 5A molecular sieve to remove
CO$_2$ and H$_2$O.
The system has two purifiers, with valves to allow switching 
from a depleted online purifier to a regenerated reserve purifier.  
The 5A sieves are regenerated by heating to 180$^\circ$C 
while passing a flow of dry argon through the purifier, 
to expel the trapped CO$_2$ and H$_2$O.
The gas mixture is then changed to Ar/H$_2$ 95:05 to react with the
CuO in the activated copper, forming H$_2$O in the exhaust.
The purifier containers are thermally insulated from surrounding devices, 
and are regenerated in-situ.
The purifier materials also absorb CF$_4$ and iC$_4$H$_{10}$.
To prevent perturbation of the composition of the recirculating gas 
mixture when the purifiers are inserted into the gas stream, 
a freshly regenerated purifier is first pre-saturated by flowing 
exhaust gas from the buffer tank through the purifier until the 
gas composition at the input and output are the same.

To prevent flammable levels of isobutane entering the detector hall
and to ensure proper operation of the  micromegas, 
the isobutane monitor, an infrared 
analyser\footnote{Model IR2100, manufactured by General Monitors, 
www.generalmonitors.com}, 
continuously monitors the isobutane concentration in the gas 
flowing to the TPCs in the detector hall.   
If the isobutane concentration falls below 1\% or rises above 3\%, 
an interlock closes valves at the output of the isobutane monitor 
and the input of the buffer tank, shutting down gas flow 
to and from the TPCs in the detector hall.

The differential pressure between the TPC output manifold and the 
Gap output manifold is controlled by adjusting the flow through a 
pump-bypass MFC which controls flow from the pump output to the pump input.
This effectively controls the net pumping speed.  The signal from the
differential pressure transducer is fed to a 
proportional-integral-differential (PID) control device,  
which then outputs a signal to control the MFC valve.
The pump\footnote{Model MB-602, manufactured by Senior Aerospace 
Metal Bellows, www.metalbellows.com}
is a double metal-bellows type.
The two bellows are connected in series to reduce the slope of the 
pump's flow versus pressure curve, thus reducing the flow-range 
requirements of the MFC.

Oil-filled (dibutyl phthalate) pressure-relief bubblers on the 
input line to each TPC and each Gap provide protection against 
excessive high chamber pressures.
Similar pressure-relief bubblers on the output lines of each 
of the Gap output manifold and TPC output manifold provide 
protection against excessive low chamber pressures.
As these bubblers are the last line of defence,
the oil levels are set such that they will only bubble if all
other interlock systems, controls and devices have failed to act 
(e.g. component failures or power outages).  
To prevent air being sucked into the TPCs through the output bubbler 
when the TPCs are isolated from the gas system, a cylinder of mixed gas 
feeds gas through a precision low-pressure 
regulator\footnote{Type ZM-R/15S-GD-V020, 
manufactured by Zimmerli Messtechnik AG, www.zimmerliag.com}
to prevent the TPC 
pressures from falling below 0.1 mbarg.  A valve connecting this backup 
system to the TPC output line automatically opens whenever the TPCs are 
isolated and whenever electrical power is lost.

The gas analyser module consists of an oxygen 
analyser\footnote{Series 511, manufactured by MecSens S.A., 
now owned by Nirva Industries, www.nirva.ch},
an H$_2$O analyser\footnote{Picoview, manufactured by 
Manalytical Ltd., www.manalytical.com}
and a multi-gas infrared 
analyser\footnote{MGA3000, manufactured by ADC Gas Analysis Ltd., 
www.adc-analysers.com}
for analyzing 
CF$_4$, iC$_4$H$_{10}$ and CO$_2$ content.
A set of valves allows analysis of the gas from any one of 12~different 
ports (A through L in Fig.~\ref{fig:gas-schematic}) on the gas system.
The analyser flow rate is controlled by an MFC in the analyser module.
A pump is used to create sufficient pressure for gas sourced from ports 
J through L, which are all low (near atmospheric) pressure sources.
Any combination of the 3 analysers can be connected to the selected gas source.
On the SS level, another set of valves 
(ports 1 through 6 in Fig.~\ref{fig:gas-schematic}), pump, 
and MFC permit sampling gas directly from the TPCs, 
Gap exhaust and mixed-gas backup cylinders.
This system directs the source gas through a 60~m long, 4.6~mm ID SS 
tube to port ``I'' on the analyser module or either of the two gas 
monitor chambers in the mixing room at ground level,
described in section~\ref{ss-gas-monitor}.

During long shutdowns, a ``standby'' system supplies a 0.6~L/min flow of 
argon to each TPC and each Gap, to prevent absorption of atmospheric 
water and oxygen on the copper-clad internal surfaces of the chambers.
This system uses only manual valves and manual flow controllers and 
pressure regulators, to ensure that the system continues to operate 
during power outages.  The output gas is exhausted to atmosphere at 
ground level, to prevent possibly dangerous argon concentrations in 
the detector hall if the hall ventilation is off.

A dry-air purge system is provided to allow purging of the chambers 
before opening the chamber covers to repair or replace components inside.
A dry-air supply is connected to manual valves at the input to the Gap 
and TPC flow controllers.  These MFCs are then used to control the flow 
of air to the chambers during the purge.

\subsection{Control system}
\label{ss-gas-control}

The gas system contains more than 250~active devices such as valves, 
flow meters, flow controllers, pressure transducers, pressure 
regulators, gas analysers, pumps, switches, thermocouples and heaters.  
A programmable logic controller\footnote{Modicon Quantum 43412A, 
manufactured by Telemecanique/Schneider Electric, www.schneider-electric.us}
(PLC) controls and/or monitors 
approximately 170~of these devices that can accept and/or supply 
electronic signals.  
The graphical user interface is provided using EPICS\footnote{Experimental 
Physics and Industrial Control System, www.aps.anl.gov/epics/}
software.
A MIDAS\footnote{Data acquisition system, https://midas.psi.ch/}-based 
program transfers information from the PLC database
to the ND280~slow control system.  
This allows the ND280~slow control system to record and display the 
status of gas system devices and display gas system alarm warnings.

To prevent operator error or component failures from causing 
dangerous or undesirable states of the gas system, a comprehensive 
set of interlocks is programmed in the PLC code.
Virtually every device which can be controlled by the PLC has
multiple interlock conditions which can force the device into
a safe state unless the conditions are satisfied.
The general philosophy of the interlock logic is to
prevent undesirable states of the gas system, while
allowing maximum operator freedom, and minimizing interruptions
to TPC data-taking.

To ensure a safe state that cannot be overridden by remote operators,
there are several manual valves distributed at critical places in 
the gas system.  These valves are equipped with micro-switch 
read-backs so that their status can be displayed and included 
in the interlock conditions of devices controlled by the PLC.
Normally open or normally closed types of electronically 
controlled valves are chosen such that all valves revert to a 
safe state on loss of power.

\subsection{Monitor chambers}
\label{ss-gas-monitor}

For monitoring the supply and return gas of the TPCs, two independent
mini TPCs were constructed with a design similar to the large TPCs. 
The smaller micromegas modules used in the chambers 
where produced in the same way as the full size modules.
During normal operation, the gas monitor chambers use the 
same drift field and mesh voltage as the large TPCs. 

Each of these two chambers measures both the drift velocity 
and the gas amplification.
To measure the drift velocity there are two $^{90}$Sr sources 
above each chamber.
They produce two lines of tracks with a well defined separation distance 
perpendicular to the drift field. 
By measuring the time difference between the drift times 
of two lines, the drift velocity can be calculated. 
Each drift time measurement is triggered by signals from
scintillating fibres located directly below each chamber.
For the gain measurement there is one $^{55}$Fe source for each chamber. 

All measured values of gain, drift velocity and 
slow monitoring values (temperatures and pressures) 
are sent to the ND280 slow control system. 
For every measurement there are enough related ambient values 
available for applying corrections, so that they can be used as reference 
for the large TPCs.
Additionally, for special measurements the main parameters of the 
gas monitor chambers, drift field and mesh voltage can be changed 
via a MIDAS interface.

\subsection{Performance}
\label{ss-gas-performance}

The gas system has been operated with all three TPCs 
since beginning in January 2010.
After correcting initial problems with the isobutane temperature
control,
the gas system has operated smoothly.

The two main requirements for the gas system were stability of the 
TPC to Gap differential pressure, and gas quality.  
During normal operations, the TPC to Gap differential pressure 
has been maintained at $0.4\pm0.03$~mbar with occasional spikes of less 
than $\pm0.1$~mbar, typically caused by external winds 
perturbing the pressure at the Gap exhaust.  
Sudden changes of flow rate or buffer tank pressure can also 
cause small spikes in TPC-Gap differential pressure, but the system quickly recovers 
and stabilizes at the set-point pressure.  
The differential pressures are measured in the output manifolds, 
which have a volume several hundred times less than the chambers
they are connected to.
The combination of the much higher capacity of the chambers 
and the impedance of the connecting tubes implies that the 
pressure fluctuations in the chambers would be smaller than 
the numbers quoted above.

One aspect of gas quality is the presence of atmospheric contaminants 
such as H$_2$O and O$_2$ in the TPC gas mixture.
The O$_2$ and H$_2$O concentrations were each less than 2~ppm in 
the gas at the output of the purifier.  
Higher levels of H$_2$O or O$_2$ in the TPC exhaust gas are 
due to diffusion into, or outgassing from, the chambers and 
gas system components.
During the initial purge, after the TPCs had been exposed to 
atmospheric air for several months, liquid water was 
observed condensing at the output of the pump.  
The water levels in the exhaust gas fell to less than 
100~ppm after a week, and continued to slowly fall, reaching 10~ppm after 
2~months and less than 5~ppm after 5~months of gas flow.  
We suspect the internal copper surfaces of the TPCs absorb large 
amounts of water, which is slowly released over time.
The O$_2$ concentration in the exhaust gas decreased more quickly, 
reaching and stabilizing at $<2$~ppm after 2~weeks.
The CO$_2$ concentration varied from about 20~ppm immediately after 
insertion of a regenerated purifier, to about 120~ppm before 
switching to a fresh purifier 60 days later.

We experienced difficulty monitoring the CO$_2$ content of the gas.
The multi-gas IR analyser demonstrated strong interference between 
its component CO$_2$, CF$_4$ and iC$_4$H$_{10}$ analysers, as well as 
strong temperature dependence.
Changing the concentration of any one of the three gas components would 
result in large changes in the measured concentrations of the other two 
components.
Through extensive calibrations, we measured these cross-terms and 
developed a correction algorithm.
However, even with the corrections we can only measure CO$_2$ 
concentration to approximately $\pm$20 ppm and CF$_4$ concentrations 
to $\pm$0.15\%.
The isobutane component analyser drifted so badly that it was 
essentially unusable.
Fortunately, the isobutane monitor, installed to monitor isobutane 
levels for safety reasons, was capable of measuring 
isobutane concentration to about $\pm$0.1\%

The stability of the gas mixture circulating through the chambers 
depends on the stability of the mixture provided by the mixer, 
modified by any absorption or desorption of gas components by the 
TPCs or gas system components.   
The long term mixture ratios calculated from the flow rates reported by 
the mixer MFCs were $95.001\pm0.001$\% Ar, $2.9999\pm0.0008$\% CF$_4$ and 
$2.0000\pm0.0007$\% iC$_4$H$_{10}$.
The actual error in mixing is probably dominated by the MFCs' response 
to temperature changes.
For our typical long term MFC temperature stability of $\pm1.5$~$^\circ$C, 
the manufacturer's specifications for the MFCs used in the mixers \
indicate a mixture stability of $95.00\pm0.01$\% Ar, 
$3.000\pm0.007$\% CF$_4$, and $2.000\pm0.005$\% 
iC$_4$H$_{10}$, satisfying our original goals for mixture stability.

Materials in the gas system and chambers exposed to the gas
can absorb gas components which
will be further absorbed or desorbed with
changes in temperature or pressure.  
In particular, the materials in the purifiers absorb 19~L of 
CF$_4$ and 32~L of iC$_4$H$_{10}$ at 20~$^\circ$C, 
at atmospheric pressure.
Tests of the purifiers indicate that a 0.1~bar increase in pressure 
causes the purifiers to absorb 0.5~L of CF$_4$ and 0.5~L of 
iC$_4$H$_{10}$ from the gas stream.
During operation, the purifier pressure typically varies by $\pm$30~mbar,
which would cause the CF$_4$ and iC$_4$H$_{10}$ concentrations to vary
by 0.002\%, well within the required mixture stability.
Attempts to measure the effects of temperature changes on the 
purifier were difficult due to the low
thermal conduction and high heat capacity of the purifier materials.
During normal operation,
the bulk of the purifier is temperature controlled to better 
than $\pm0.5^\circ$C, and the temperature of the 
bottom third of the purifier varies by           
$\pm1^\circ$C due to cooling or heating by 
the returning recirculating gas.
This level of temperature variation is not expected to produce gas mixture
variations beyond the stability goals.

%% file: chap04.tex
\section{Micromegas modules}
\label{sec-mm}

\subsection{Performance requirements}
\label{ss-mm-requirements}
The total detection surface to be covered for the three TPCs is about 9\,m$^2$
with a pad segmentation of 70\,mm$^2$. For practical reasons, a modularity of
twelve bulk micromegas detectors per readout plane was chosen. This defines
a surface per module of about 0.12\,m$^2$ with the requirement of small
dead areas in order to obtain maximum sampling length for tracks. In addition,
a very good planarity of the detection surface, better than 150\,$\mu$m,
is demanded to minimize drift electric field distortions near the 
anode planes.

Very good gas gain 
uniformity at the level of a few \% is desirable in order to meet
the TPC performance requirements. The modules on a same readout plane are
operated at the same voltage to minimize drift 
electric field distortions near the detection plane. This requires
good reproducibility from one module to another
of the small, $\cal O$(100\,$\mu$m), amplification gap. Therefore, 
well controlled manufacturing processes and detector robustness
was necessary.

With low-noise readout electronics, the bulk micromegas detectors
can be operated at a moderate gain of 1000
with detection efficiencies close to 100\%
for minimum ionizing particles. This 
has the advantage of reducing the probability for sparks in the amplification 
region thereby providing detection reliability, operation stability and 
small dead time.

The design and construction of the T2K TPC bulk micromegas modules followed 
extensive tests in 2006 and 2007 which validated the physics 
performance that could be reached which such detectors \cite{Bou:2007nim,Anv:2009nim}.  

\subsection{Design}
\label{ss-mm-design}
The detector modules of the T2K TPC were built using the bulk micromegas
technology invented in 2004 by a CERN-Saclay collaboration \cite{Gio:2006nim}. 
This technique provides an excellent solution to minimize the unavoidable
dead areas on the edges of a module and allows large detection areas with 
excellent gas gain uniformity to be built. Moreover, such detectors can be 
manufactured in a single process, reducing the production time and cost.   

The bulk micromegas technique consists in laminating a woven mesh on a 
Printed Circuit Board (PCB) covered by a photoimageable film. At the end of 
the process, the micromesh is sandwiched between two layers of the same 
insulating material. The detector then undergoes UV exposure with an 
appropriate mask, followed by chemical development. A thin, 
few millimeter wide border at the edge can thus be obtained avoiding the 
need of an external additional frame to support the stretched micromesh.
  
\begin{figure}[htp]
\centering
\ifx\figstyle\bw
 \includegraphics[width=0.45\textwidth]{fig06_bw.jpg}
\else
 \includegraphics[width=0.45\textwidth]{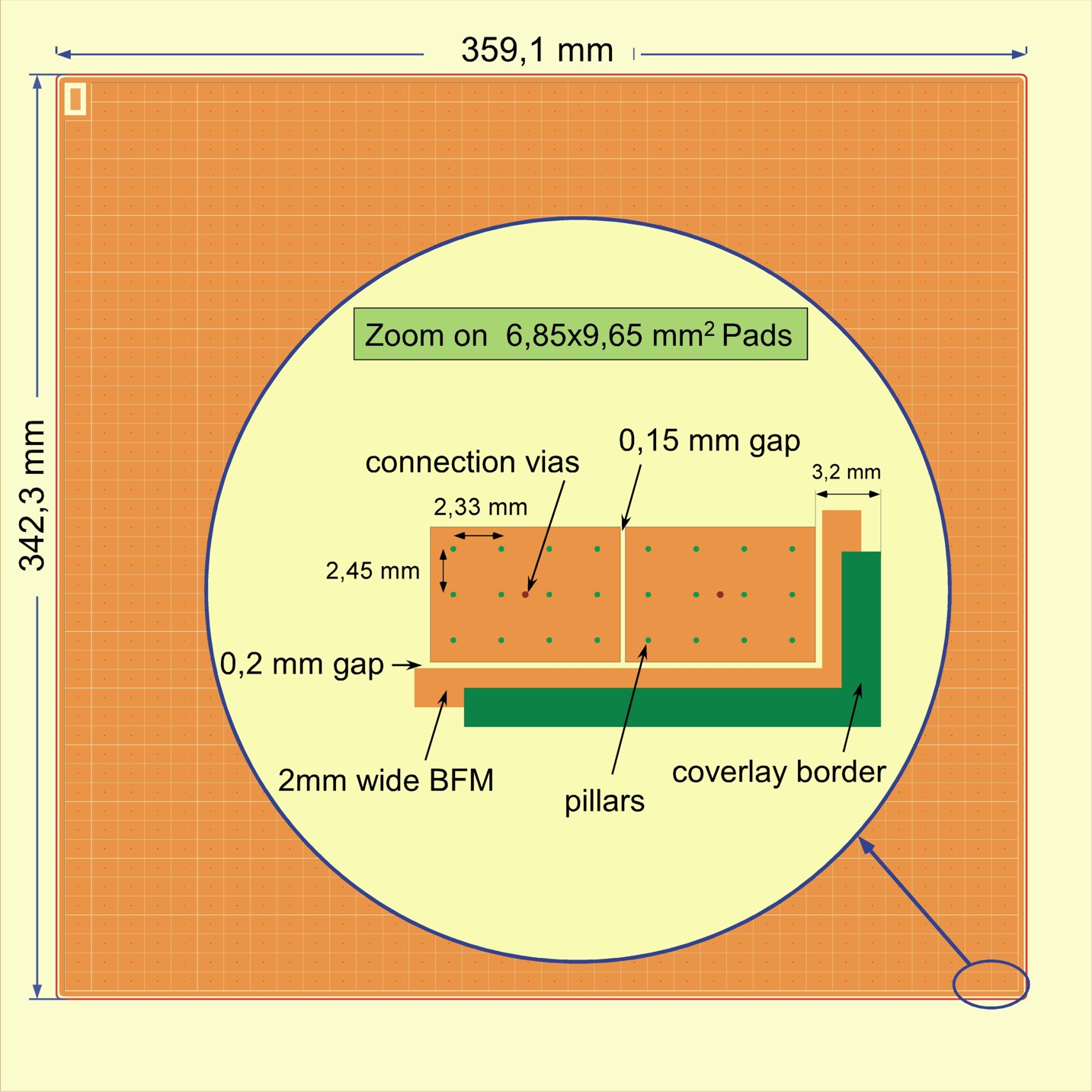}
\fi
\caption{View of the PCB from the anode pads side.
}
\label{fig:mm_pcb}
\end{figure}

Each bulk micromegas module of a TPC contains 1728 pads arranged in 48 
rows of 36 pads and covers a sensitive area of $36 \times 34$\,cm$^{2}$. 
The PCB is segmented into 
1726 rectangular pads on a pitch of $9.8 \times 7.0$\, mm$^{2}$, including
150\,$\mu m$ insulation between them, as shown in Fig.~\ref{fig:mm_pcb}.
In one corner, a two-pad equivalent 
area is reserved for the micromesh voltage supply connection from the 
backplane of the PCB. The thickness of the PCB 
is 2.2\,mm and comprises 
three layers of FR4 with blind vias in the inner layer. This solution avoids 
the gas tightness problems arising from the conventional two-layer structure 
with vias sealed with epoxide resins. The top conductive layer forming the 
anode pad plane is made of 25\,$\mu m$ thick copper deposited on FR4. The 
other three conductive layers are used for the routing network, 
grounding and pad-readout connectors. Finally, to minimize electric field 
line distortions near the edges of the TPC and between adjacent modules, a 2\,mm wide copper strip, 
called Border Frame Mesh (BFM), located on the micromesh plane, surrounds the active area of the 
detector. The BFM is at the same potential as the micromesh.  

\subsection{Production techniques}
\label{ss-mm-production}
The 72 bulk micromegas modules which instrument the three TPCs were produced 
between May 2008 and August 2009 by CERN/EN-ICE-DEM. A sandwich of two 
layers of 64\,$\mu$m Pyralux PC1025 photoimageable polyimide 
by DuPont\footnote{DuPont Electronic Polymers LP, 14 Tw Alexander Dr, Durham, 
NC 27709, USA}, 
a woven micromesh and finally a layer of Pyralux
were laminated on the PCB. The micromesh, manufactured by 
BOPP\footnote{BOPP, Bachmannweg 21, CH-8046, Zurich, Switzerland}, 
was made of 18\,$\mu$m diameter 304L stainless 
steel wires. After weaving, its thickness was reduced by 20-30\% by lamination.
The wires are spaced with a pitch of 63\,$\mu$m (400\,LPI). 
During the manufacturing process, the micromesh was held
on an external frame with a tension of about 12\,N. This procedure guaranteed
sufficient flatness of the micromesh during lamination 
and thereby a uniform amplification gap over the entire 
sensitive area of the detector module. At the end of the photoimaging process,
the micromesh is held in place by a 2\,mm coverlay border
and by 20736 regularly distributed pillars, maintaining the amplification 
gap of 128\,$\mu$m. The pillars, 12 per pad, are cylindrical with a 
diameter of about 0.5\,mm. The active area represents about 93\% of the 
module surface. After development, the bulk micromegas detector underwent
cleaning and baking processes to achieve complete polymerization of the Pyralux
material.  

\begin{figure}[htp]
\centering
\ifx\figstyle\bw
 \includegraphics[width=0.45\textwidth]{fig07_bw.jpg}
\else
 \includegraphics[width=0.45\textwidth]{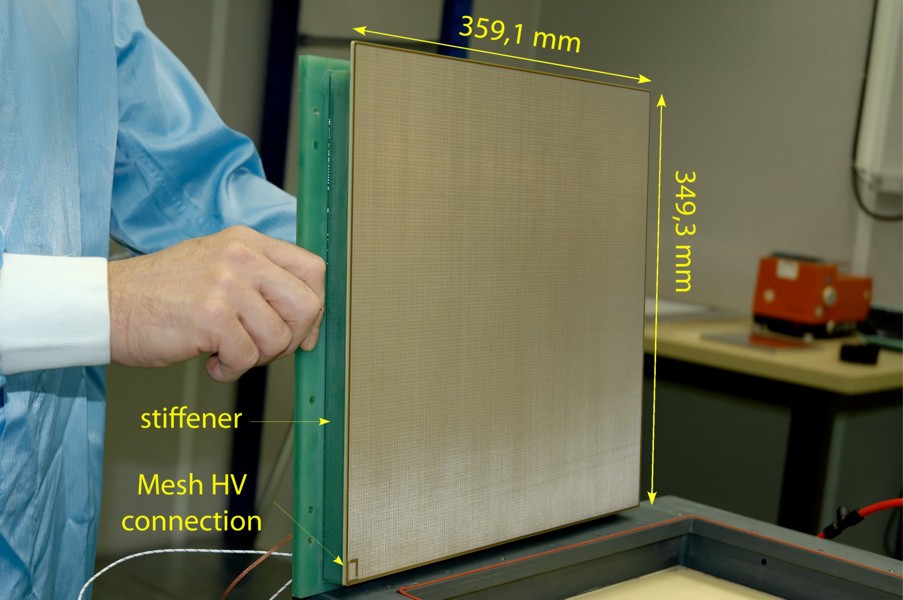}
\fi
\caption{A bulk micromegas detector module for the T2K TPC.}
\label{fig:mm_module}
\end{figure}

After protecting the sensitive surface with a 
melamine plate, the outer coverlay and the PCB frame were cut so as to 
leave a 3.2\,mm inactive external border. Twenty four pad-readout 
connectors
were then soldered on the back plane of the PCB.
Finally, to ensure the mechanical rigidity of the module and a planarity 
better than 150\,$\mu$m, the PCB was reinforced by a stiff frame made of 
FR5 
and bonded on the connector face. This structure hosts the seal for gas 
tightness and provides also anchorage for a guiding, fixation and 
extraction system for the readout electronics. A completed bulk micromegas 
detector module ready to be mounted on the T2K TPC is shown in 
Fig.~\ref{fig:mm_module}.
        
\subsection{Test bench}
\label{ss-mm-testbench}
A dedicated test bench was used at the T2K micromegas
production laboratory at CERN 
to characterize each 
bulk micromegas detector and to validate its performance. It 
consisted of an 
automated X-Y scanning system which allowed to measure the response of a 
single module pad when illuminated by a collimated 185\,MBq $^{55}$Fe 
source. The detector module was inserted in a gas-tight box  
providing a 4\,cm long drift region. The 
volume of the test box was about 8 litres, filled with a 10\,L/h flow
of T2K gas mixture. The pressure
of the gas within the test box
was maintained at a pressure 1\,mbar over the atmospheric pressure. 

The test box was made of G10 material and was mounted on an aluminum support. 
An aluminized mylar cathode was used to obtain a uniform drift field.
The cathode was supported by a G10 grid 
to ensure flatness of the mylar foil. In addition, the electric field 
near the edges of the module was corrected by a 25\,mm
wide copper strip surrounding the 
active volume, 5\,mm away and centered along the electric field direction
between the cathode and anode planes. 
    
The 5.9\,keV 
X-ray source was mounted on an external motorized head, located behind the 
drift region. Between two consecutive measurements, the source was 
moved from one pad to the next one by precisely positioning  
the X-ray beam in front of the target pad centre to better than 
0.1\,mm in each direction.
Two weaker 3\,MBq $^{55}$Fe sources, placed in 
opposite corners of the test box and   
permanently irradiating a few pads, were used  
to monitor the detector gain stability during a scan. Gas 
temperature and pressure probes as well as the 
current drawn in the micromegas detector were also recorded continuously. In 
addition, a small bulk micromegas chamber was installed upstream of the 
test box to monitor the quality of the gas mixture. 

The movable X-ray source
had different collimations which allowed to choose the irradiation
intensity of the pads as well as the size of the illumination spot. This 
feature was used, for instance, to measure, for each module, the  
gain variation as a function of the voltage applied on the micromesh, 
averaged over several pads in the central part of the detector. 
T2K prototype readout electronic cards were used to digitize the
detector signals (see section~\ref{sec-fee}).
A dedicated data acquisition program, synchronized with the motorized 
head, allowed detector scans to be performed in an automated way, thus 
improving the scan efficiency and measurement reliability.    
The duration of a complete scan of the 1726 active pads 
of a module was typically 6 hours for about 1000 recorded events per 
pad. The test bench facility allowed to calibrate and test up to 5 modules 
per week. 
        
\begin{figure}[htp]
\centering
\ifx\figstyle\bw
 \includegraphics[width=0.45\textwidth]{fig08_bw.jpg}
\else
 \includegraphics[width=0.45\textwidth]{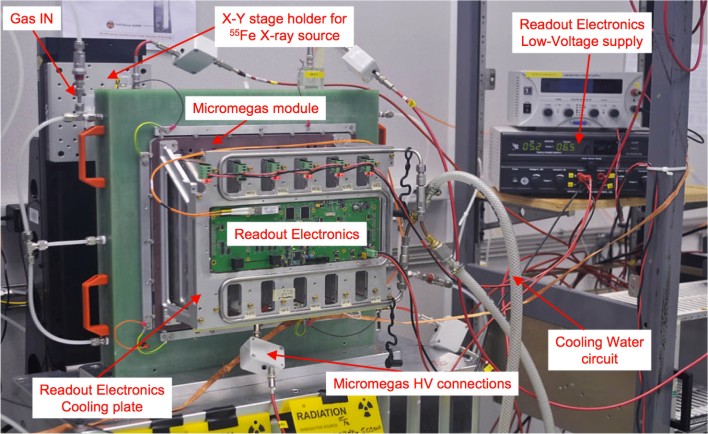}
\fi
\caption{Test bench setup for the micromegas modules. }
\label{fig:test_bench}
\end{figure}

\subsection{Quality control and test bench results}
\label{ss-mm-qc-tb}
We describe in this section the various quality controls that were performed
at the different production stages of a bulk micromegas detector 
and present 
test bench results obtained for the characterization of the modules 
as well as relevant performance reached with the bulk micromegas 
technology.    

\subsubsection{Quality controls}
\label{sss-mm-qc}
Quality checks done during production included visual, optical controls 
and measurements of mechanical tolerances of the different components used as
well as electrical and gas tightness tests.
  
Each PCB was controlled and
required to have a thickness within 0.1\,mm of the nominal value 
and a sag smaller than 2\,mm. In addition, the four copper layers 
of the PCB were optically inspected. 

At the end of the lamination stage, before the PCB was cut to its 
final dimensions 
and the twenty four pad-readout connectors soldered on the backplane,
the detectors underwent an electrical test in air. Modules
were qualified if the global micromesh current did not exceed 2\,nA 
at -600\,V with the anode pads connected to ground. 

The gluing operation of the PCB on its stiffener was done in a class 1000
clean room located in the T2K micromegas production laboratory. 
The position of
the stiffener gasket with respect to the micromesh was controlled 
and required to be within 50\,$\mu$m. This was important for proper 
flatness and positioning of the micromesh along the drift 
direction in the TPC to ensure a uniform drift electrical field. 
   
The assembled detector module was inserted in a gas-tight box filled with 
dry air and forced to spark by gradually increasing the micromesh voltage.
In this way, residual dusts were burned out and most of the 
tiny surface asperities of the micromesh or of the copper pads smoothened.
This procedure proved to significantly reduce the probability of sparking 
when operating detectors with gas mixtures, leading to safer 
functioning conditions of the detectors in the T2K experiment.  
The identification of defective pads was obtained by requiring
the module to stand a voltage of about -900\,V in dry air with less than
a few sparks per hour. Over the entire bulk micromegas production, 
only 12 pads out of 124416 (0.01\%) were found faulty, showing the 
high-level quality of the manufacturing process. The detected defective 
pads were 
disconnected from the readout chain.

The final step of the production was the detector calibration and its
characterization using the automated test bench described in 
section~\ref{ss-mm-testbench}.

\subsubsection{Test bench results}
\label{sss-mm-tbr}
The X-Y scans of the produced modules were performed with -350\,V applied on 
the micromesh and with a drift electric field of 200\,V/cm, that is, 
in similar operating conditions as in the T2K experiment.

\begin{figure}[htp]
\centering
\includegraphics[width=0.45\textwidth]{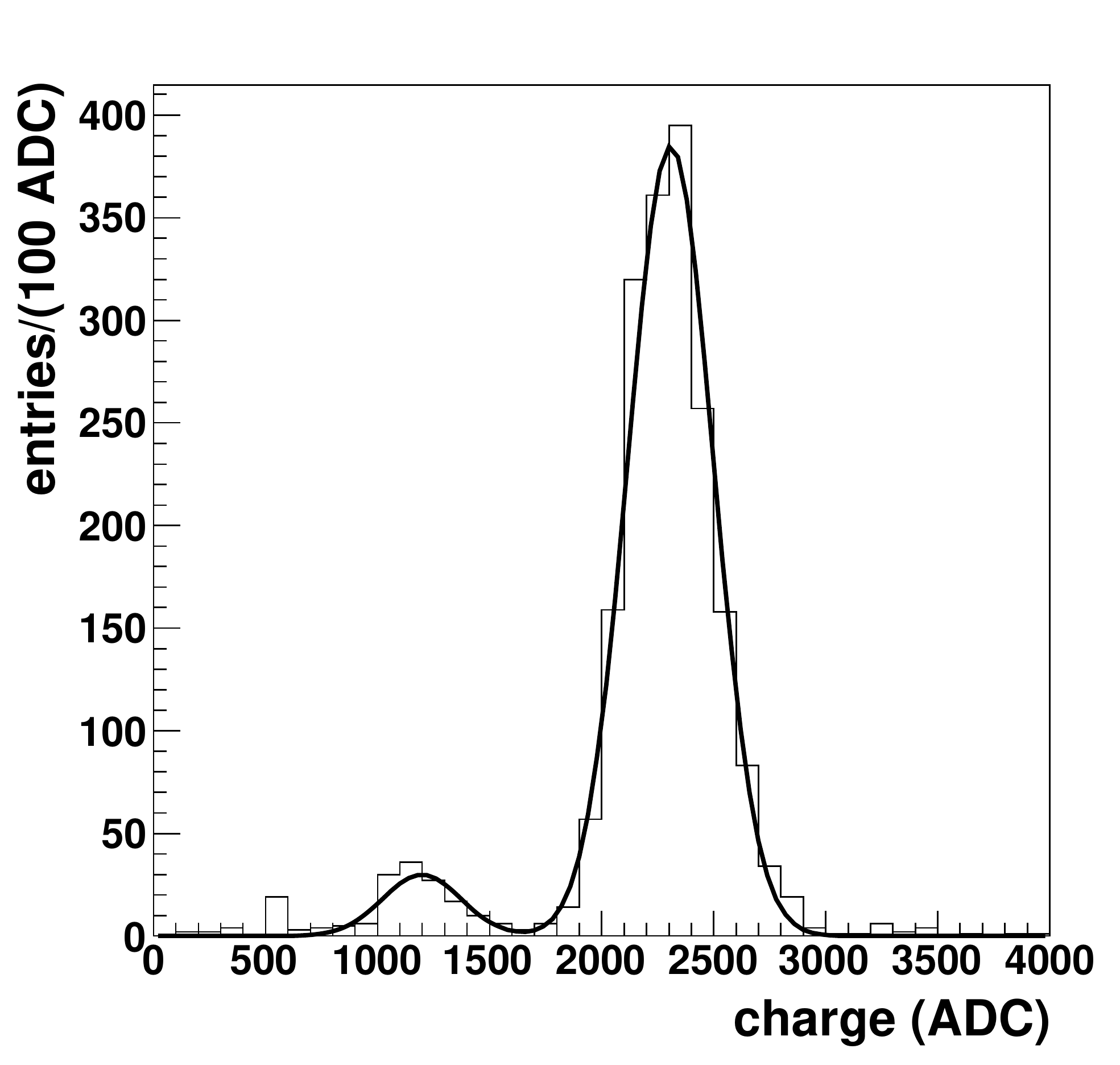}
\caption{$^{55}$Fe X-ray energy spectrum (in ADC counts) measured with 
a single pad. The energy resolution obtained at 5.9\,keV is 8.2\%}
\label{fig:mm_55fe}
\end{figure}

A typical energy
spectrum measured with the $^{55}$Fe source illuminating a single pad
is shown in Fig.~\ref{fig:mm_55fe}. A very good resolution of about 8\% at 
5.9\,keV 
was obtained allowing the 2.9\,keV escape line in argon to be clearly observed.

\begin{figure}[ht]
  \centering
  \begin{tabular}{c}
    \hskip -0.25 cm
    \includegraphics[width=0.45\textwidth]{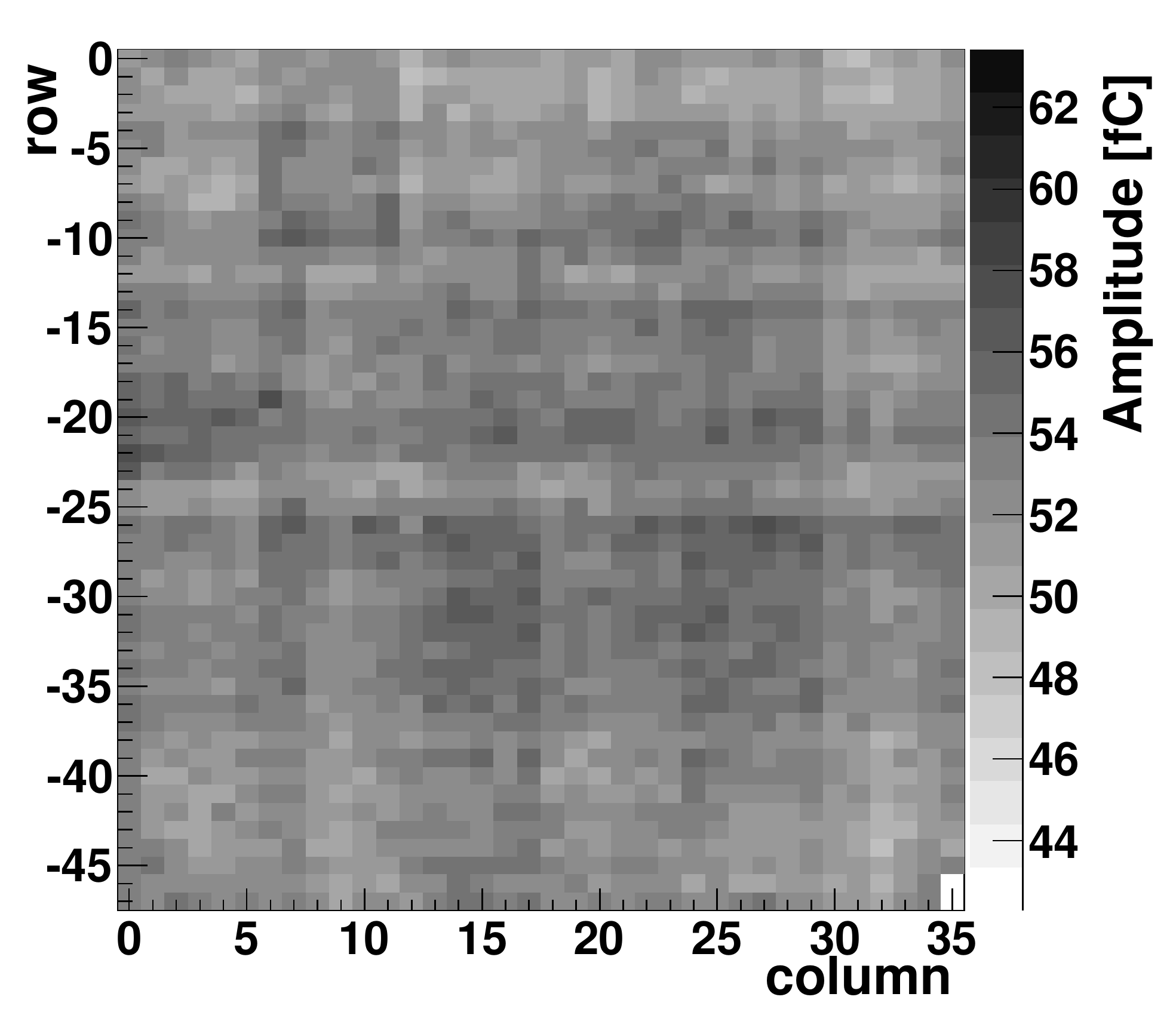}\\
    \hskip -0.25 cm
    \includegraphics[width=0.45\textwidth]{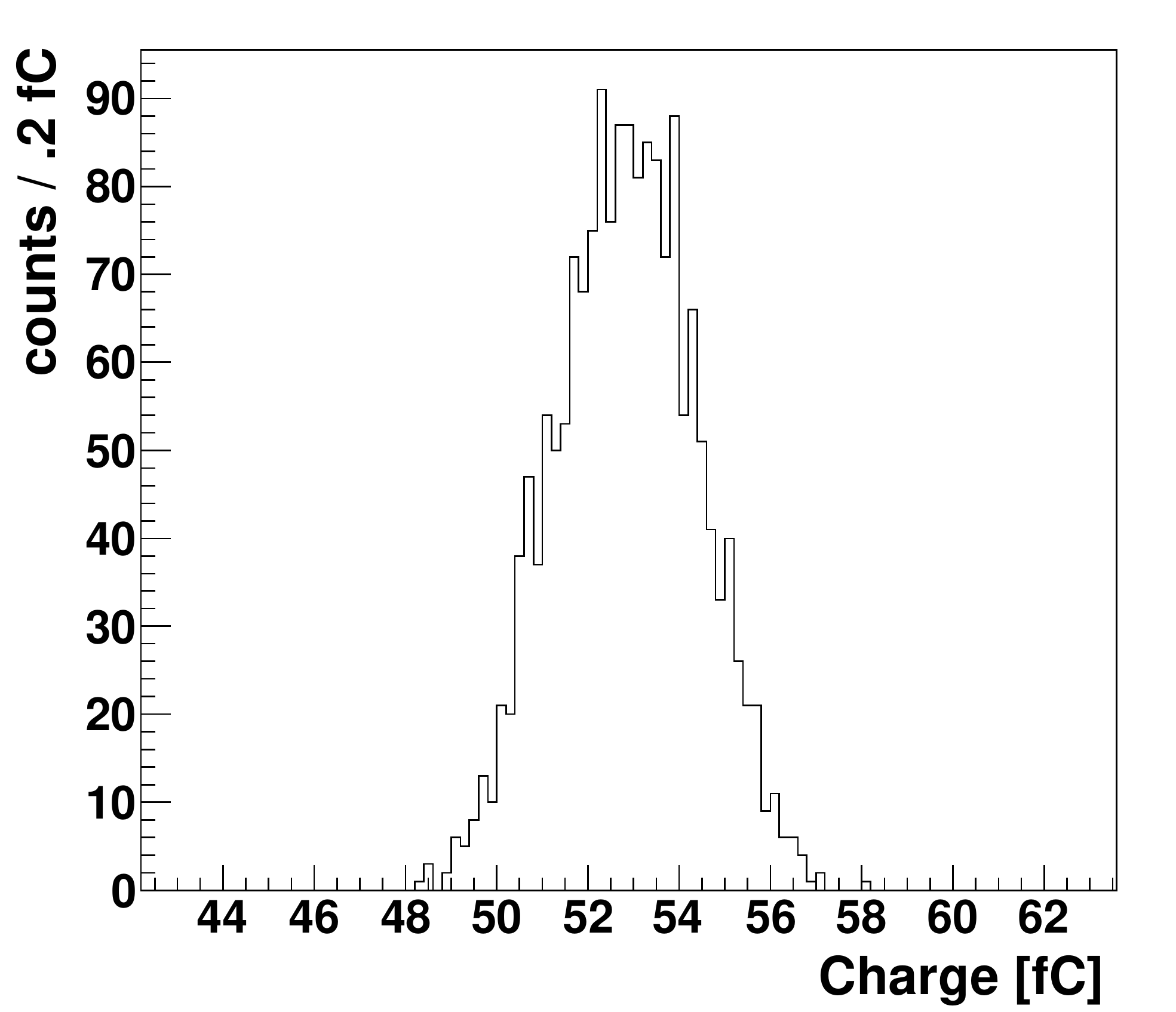}
  \end{tabular}
  \caption{Typical 2D map of collected charge (top) with the
           corresponding distribution for the 1726 pads (bottom) of a module. 
           The two 
           pads in the bottom right corner are used for the micromesh voltage 
           connection.}
  \label{fig:mm_scan_c}
\end{figure}

\begin{figure}[ht]
  \centering
  \begin{tabular}{c}
    \hskip -0.25 cm
    \includegraphics[width=0.45\textwidth]{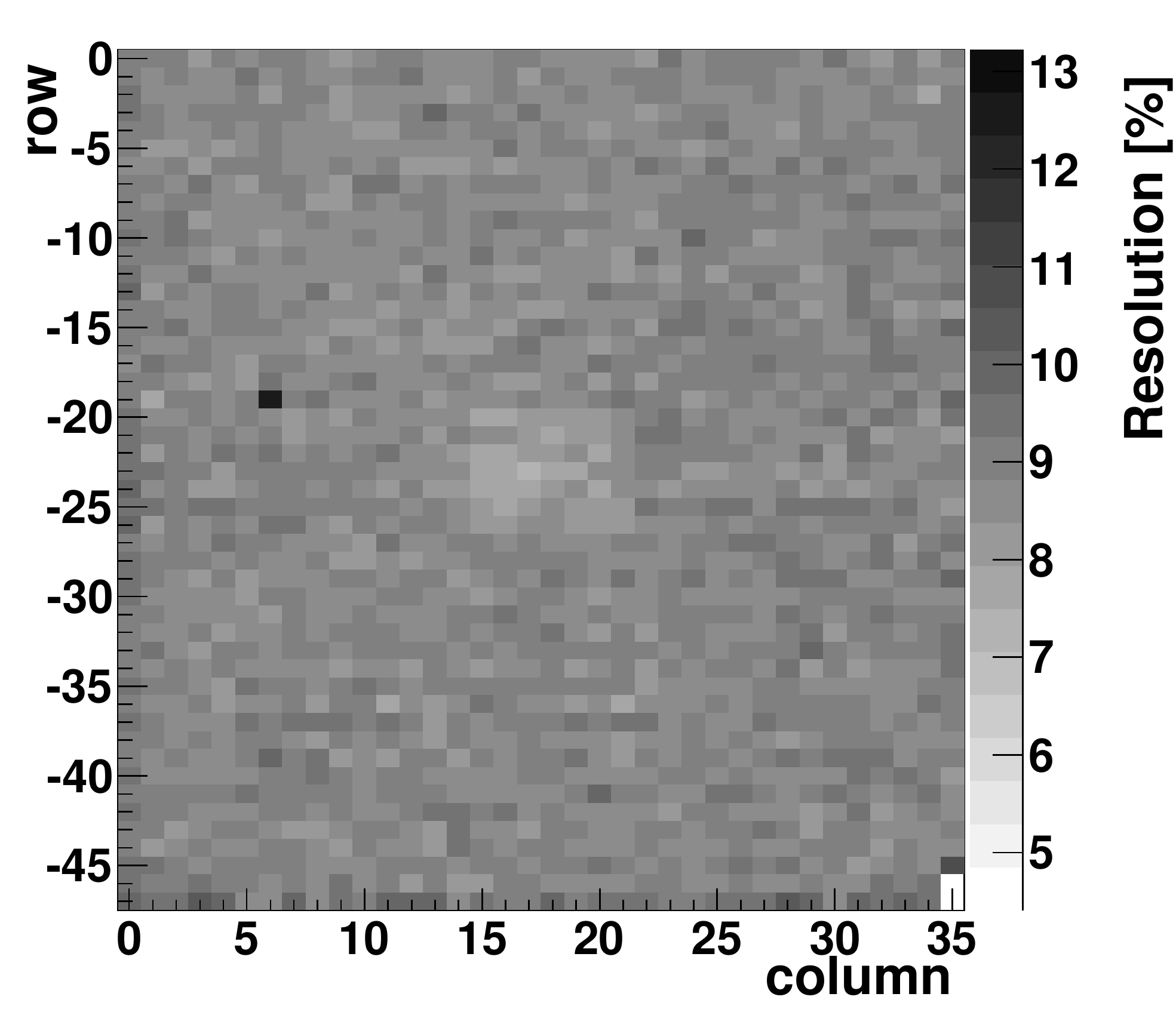}\\
    \hskip -0.25 cm
    \includegraphics[width=0.45\textwidth]{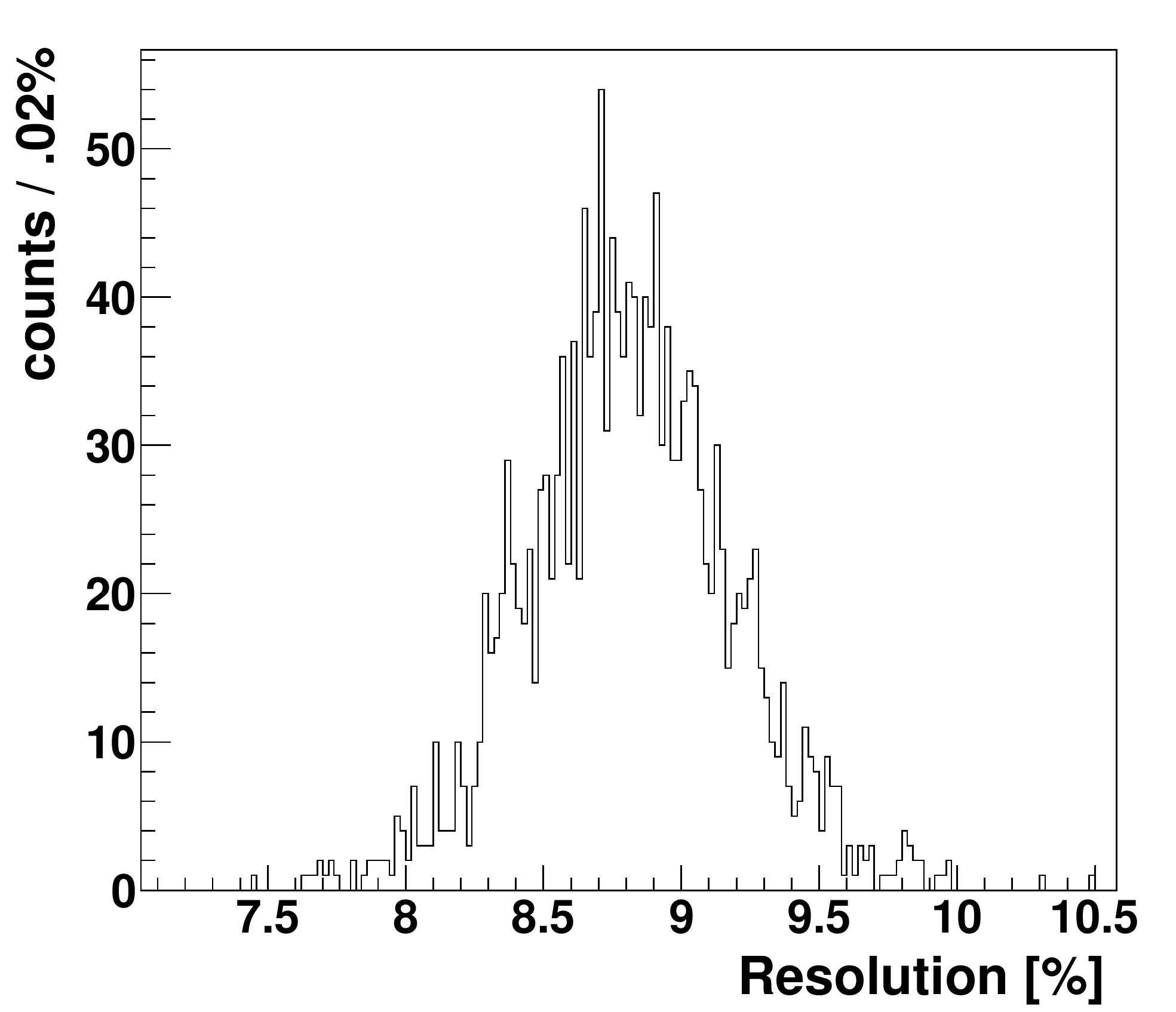}
  \end{tabular}
  \caption{Typical 2D resolution map at 5.9\,keV (top) with the corresponding
           distribution for the 1726 pads (bottom) of a module. The two 
           pads in the bottom right corner are used for the micromesh voltage 
           connection.}
  \label{fig:mm_scan_r}
\end{figure}

Fig.~\ref{fig:mm_scan_c} and Fig.~\ref{fig:mm_scan_r} show the main results of 
a complete scan of the 1726 
active pads of a module. The two-dimensional maps of the mean gain value 
and of the resolution measured at 5.9\,keV indicate the very good
response uniformity obtained with a bulk micromegas detector. 
The typical r.m.s.
dispersion of collected charge is better than 3\% over 
the entire surface of a detector while the r.m.s. dispersion of the energy
resolution at 5.9\,keV is about 6\%. These results were 
obtained after taking into account the small differences between the  
different channels in the 
routing path length from a pad to the readout connector.
These length differences resulted in slightly different
input capacitances for the channels.

Dispersion values in the average gain and resolution for the 72 modules, after 
correcting for atmospheric pressure variations 
between scans, were found to be better than 8\% and 3\%, respectively.  

\begin{figure}[ht]
\centering
\includegraphics[width=0.45\textwidth]{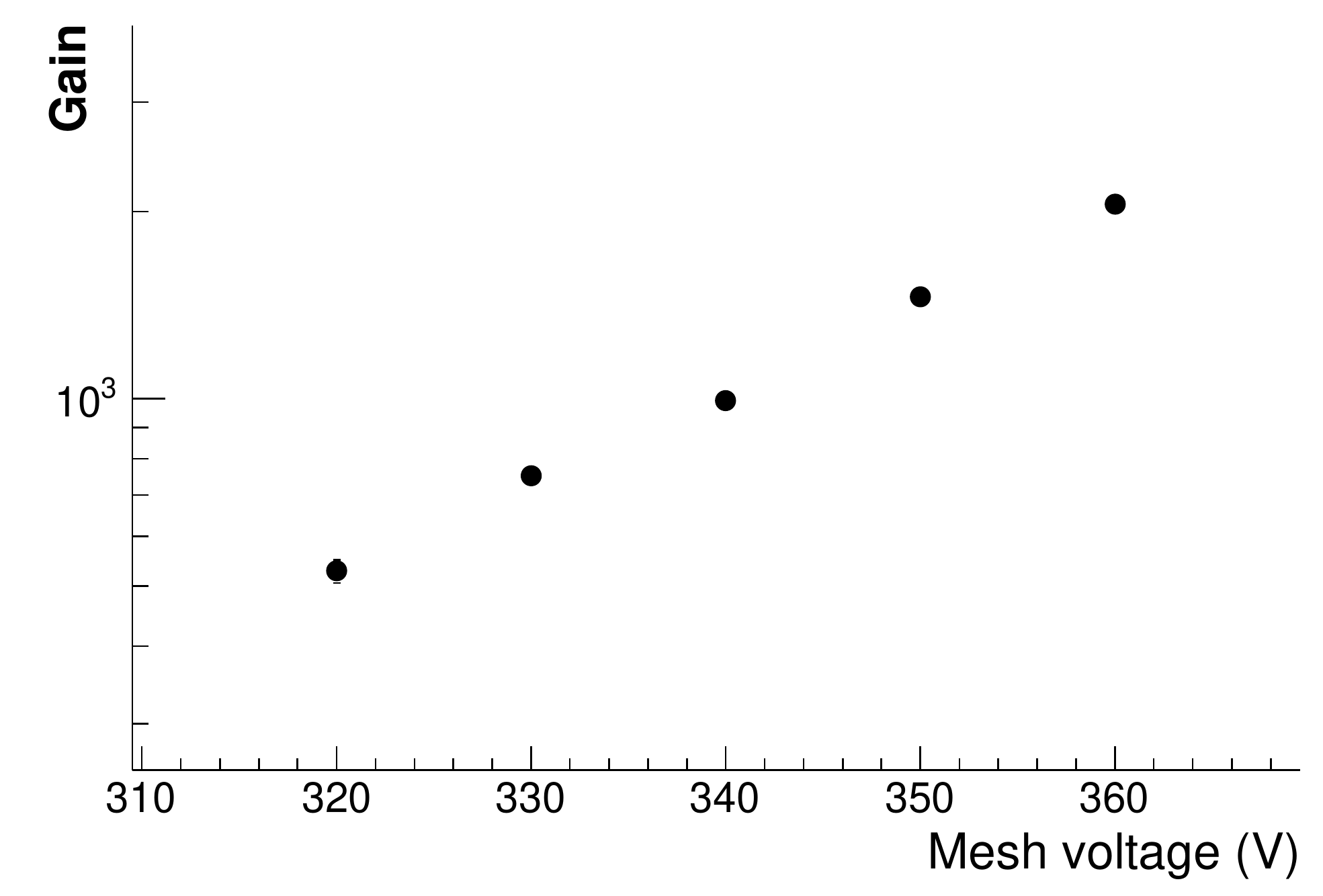}
\caption{Gas gain as a function of the micromesh voltage
for a T2K bulk micromegas detector.
}
\label{fig:mm_gain}
\end{figure}

Finally, the gain of the bulk micromegas detectors were measured
for voltages on the micromesh between -320\,V and -360\,V. The
electric field intensity in the amplification region ranged between
25.0\,kV/cm and 28.1\,kV/cm. For each gain measurement, the drift electric 
field was set to 200\,V/cm and the $^{55}$Fe source 
illuminated a well defined region in the central part of the detector. 
Fig.~\ref{fig:mm_gain} shows a typical gain curve obtained with the
T2K gas mixture.
Gain values were found to vary in the 500 to 2000 range. At the nominal 
operation voltage of -350\,V in the T2K experiment, the gas gain 
was measured to be about 1500 and the spark rate lower than 0.1/h. 
Such sparks were found to last typically a few milliseconds with 
a few volts drop ($<$ 5V) on the micromesh high voltage supply corresponding to
a current drawn of a few hundreds of nA ($<$ 500nA).   
 
Additional measurements like the micromesh transparency, cross-talk or
ageing effects were performed on prototype
detectors with characteristics identical to the ones of the T2K modules.
The main results are presented below.

\subsubsection{Micromesh transparency}
\label{sss-mm-transparency}

The electron transmission from the drift region to the amplification
region of a bulk micromegas detector is well known to depend on the ratio
of the corresponding electric fields as well as on the micromesh geometry. 
A measurement of the transparency was obtained by varying the 
drift field E$_{d}$
with respect to the field E$_{a}$ in the amplification 
region (Fig.~\ref{sss-mm-transparency}). The 
voltage on the micromesh was set to the nominal value of 
-350\,V corresponding to E$_{a}$\,=\,27.4\,kV/cm.
For values of E$_{a}$/E$_{d}$ below 100, the effective gain of the 
detector was found to drop rapidly due to the smaller electron 
transmission. In the nominal T2K operation region (E$_{a}$/E$_{d} = 100)$,
the transmission is maximum. This effect was well accounted for by a field 
calculation that took into account the 
pitch size and the wire dimensions of the micromesh.     

\begin{figure}[ht]
\centering
\includegraphics[width=0.45\textwidth]{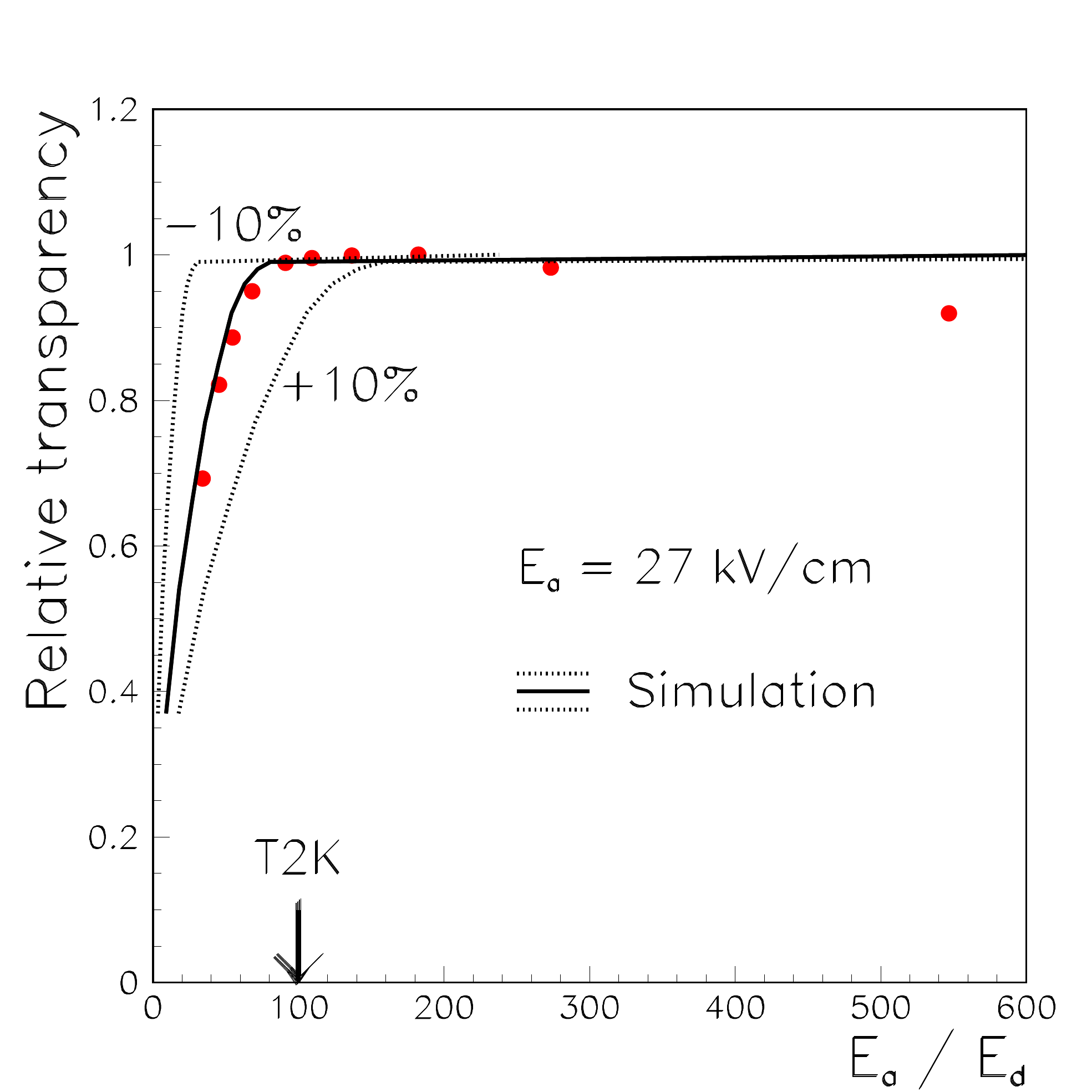}
\caption{Transparency is shown relative to its maximum
as a function of the E$_{a}$/E$_{d}$
ratio. The solid line represents the result of a simulation while the dotted 
lines show the transparency variation due
to a $\pm$10\% change of the micromesh wire size.
}
\label{trans_ieee}
\end{figure}

\subsubsection{Cross-talk}
\label{sss-mm-xtalk}

Small cross-talk effects in the detector PCB were observed by correlating 
the charges measured in adjacent pads. In the case of irradiation by a 
collimated $^{55}$Fe source, almost all the energy deposited in the 
detector is contained in a single pad. A cross-talk contribution at the level
of 1\% above the noise level was determined for lateral pads 
surrounding the irradiated pad. For this measurement, a prototype detector was 
operated at a higher gain of about 4000. The observed cross-talk contribution 
corresponds to a parasitic capacitance of a few pF, mainly due to the routing 
strips inside the detector PCB.

\subsubsection{Ageing}
\label{sss-mm-ageing}

Although ionization rates in the T2K TPCs are very low, dominated 
by cosmic rays and calibration triggers, it is 
important 
to understand whether ageing effects in a bulk micromegas detector may develop
with accumulated charge. Measurements were carried out on a small, 11\,cm 
diameter detector, using a 20\,mA X-ray gun. The spectrum of the 
incident photons peaked at a maximum energy of 8\,keV. The 
detector was operated at a gain of about 4000 in an  
Ar($95\%$)/$i$C$_{4}$H$_{10}(5\%)$ gas mixture. Typical micromesh currents 
of 5$\mu$A were measured on a 10\,cm$^{2}$ area 
during the X-ray exposure which lasted several days. The total accumulated 
charge density on the anode was 0.17\,C/cm$^{2}$, orders of magnitude more
charge than will be collected in life of the T2K experiment. 
To better control possible 
deterioration of the detector response, comparison measurements were 
performed regularly by irradiating
a 4\,cm$^{2}$ surface of the detector, covered by a 200\,$\mu$m thin Al foil.
This absorber allowed to reduce by a factor 50
the X-ray beam intensity on the detector. No significant ageing effect 
at the level of a few percent was observed during these tests. Similar 
conclusions were obtained when
the prototype was operated with the 
T2K gas mixture.

\subsection{High voltage systems}
\label{ss-mm-hv}
Each micromesh of the 72 bulk micromegas detectors is individually  
polarized through a RCR filter (500\,k$\Omega$-22n\,F-1.2\,$\Omega$) mounted 
on the backplane of the module.
In the T2K experiment, the high voltage is supplied by six 16 channel 
ISEG EHS F010n\_104\_SHV modules\footnote{iseg Spezialelektronik GmbH, Bautzner Landstr. 23, D-01454 Raderberg / OT Rossendorf, Germany} (one per TPC readout plane) that were 
modified to provide independent current trip settings for each channel, a 2nA
current setting and monitoring resolution, a 20mV precision voltage setting 
as well as the possibility to tag sparks by current or voltage over-threshold 
detection.

The three TPC cathodes are powered by separate rack mounted HV 
supplies\footnote{Bertan model 225-30R, Spellman High Voltage 
Electronics Corporation} at -25~kV.
The last strips in the field cages are individually powered by six channels of
an ISEG EHS 80 10n\_805 module at the same potential as the micromegas
modules.
Each strip is connected to ground through a 6.6\,M$\Omega$ 
resistance provided by resistors mounted within the TPCs, so that
the cathode and micromegas operating potentials can be changed independently.
These permanently mounted resistors also safely ensure that the cathode voltage
will not be present at the ends of the TPCs, under any circumstance.

%% file: chap05.tex
\section{Front-end electronics}
\label{sec-fee}

\subsection{Requirements}
\label{ss-fee-req}
The desired precision on track reconstruction calls for highly segmented detectors. Each of the 72 Micromegas modules is segmented into $36 \times 48$ pads leading to 124,416 electronic channels for the three TPCs. The desired dynamic range is 10 Minimum Ionizing Particle (MIP) with a signal to noise ratio of 100. Taking into account the gain of the detector and the capacitance of a pad, the smallest required charge measurement range is 0-120 fC and the equivalent noise contribution must not exceed 1000 electrons for 1 MIP. The desired integral non linearity is 1\% below 3 MIP and 5\% from 3 MIP to 10 MIP. The readout electronics has to acquire 500 time samples during the maximum drift time of electrons in the gas of the TPC (10 $\mu$s to 100 $\mu$s depending on gas mixture and drift field). An external trigger is provided to the TPCs and the maximum sustained event taking rate is 20 Hz. The raw event size is $\sim$120 MB and data must be reduced within a 50 ms time budget by a factor $\sim$1000 using zero-suppression and lossless compression. The front-end electronics operates in a closed environment without access during operation. High channel density and low power consumption are important goals to reduce demands on the water cooling system and minimize the mass of material placed inside the magnet.

\subsection{Design}
\label{ss-fee-design}
\begin{figure}[htp]
\centering
\includegraphics[width=0.45\textwidth]{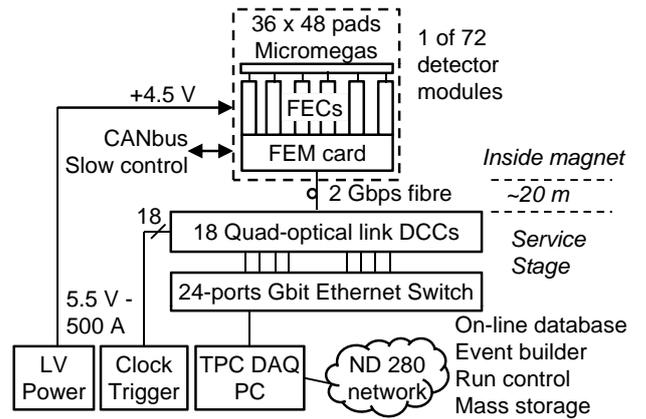}
\caption{Architecture of the TPC readout electronics.}\label{fig:fee-archi}
\end{figure}

The specific performance requirements and the particular detector layout led to the design of a new readout electronic chain optimized for this application. The front-end electronics has a modular structure and the same building block is duplicated 72 times. Each block of front-end electronics is composed of six Front-End Cards (FECs) and one Front-End Mezzanine (FEM) card directly mounted at the back of a Micromegas detector. The cornerstone of the readout system is a 72-channel application specific integrated circuit (ASIC) called ``AFTER'' which processes detector pad signals and buffer them in a 511-time bin switched capacitor array (SCA). Four AFTER chips and the required external ADC are mounted on a 288-channel FEC. The FEM board is used to aggregate the data of the 6 FECs of a detector module, perform zero-suppression, and send the remaining data over a 2 Gbps optical link to a back-end Data Concentrator Card (DCC). The 100 MHz experiment global clock, trigger information and configuration data are sent through the return path of this optical link. The FEM board is also connected to a CANbus network which is used for slow control and monitoring. The readout electronics requires a single 4.5 V power input. Specific mechanics is needed to hold the different cards and provides shielding and cooling. A schematic view of the TPC readout electronics is shown in  Fig.~\ref{fig:fee-archi}.

A set of 18 quad-optical link DCCs connected via a private gigabit Ethernet switch is used to collect the data received from the 72 front-end modules into a standard PC linked to the global data acquisition system of the experiment. The back-end electronics uses commercial Field Programmable Gate Array (FPGA) evaluation boards from Xilinx with custom add-ons and off-the-shelf networking products and computers. More detailed descriptions of the TPC front-end and back-end electronics are given in \cite{Cal:2009tns} and \cite{Cal:2010rtc} respectively.

\subsubsection{The AFTER chip}
\label{sss-after-chip}
\begin{figure}[htp]
\centering
\includegraphics[width=0.45\textwidth]{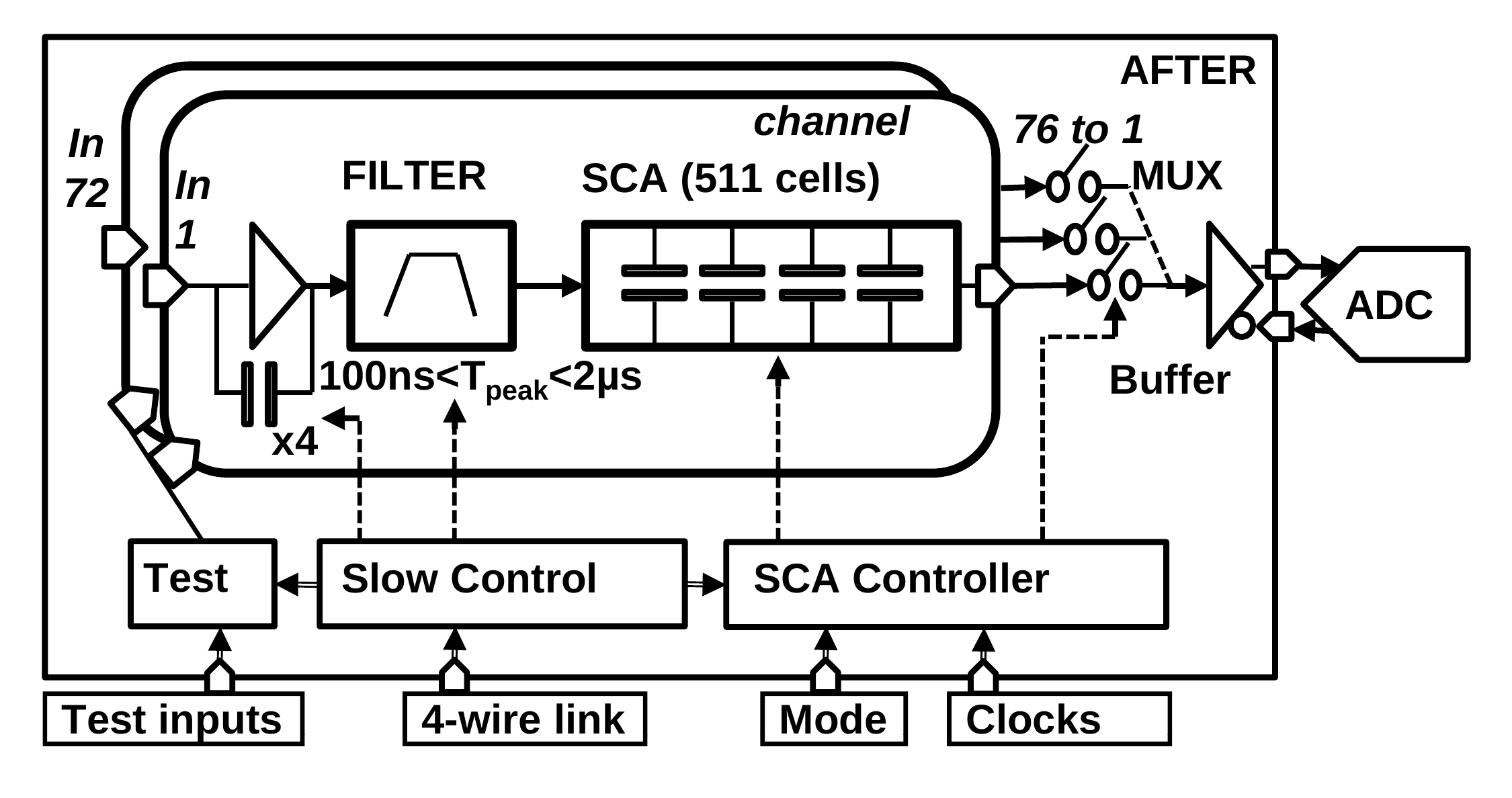}
\caption{Architecture of the AFTER chip.}\label{fig:after-archi}
\end{figure}

The AFTER chip performs the first concentration of the data from the inputs of its 72 channels to only one analog output, connected to an external ADC. Its architecture is made to optimally support various kinds of detectors and gas mixtures, using adjustable gain and shaping time parameters. These parameters and the control of the chip are managed by slow control, using a custom serial 4-wire link. Two inputs are available to allow electrical calibration and the test of one, several or all channels. 
Each channel (Fig.~\ref{fig:after-archi}.) integrates a front-end part dedicated to the collection of charges and the shaping of the detector signal as well as a switched capacitor array to sample and store the analog signal.

The front-end part is composed of:
\begin{itemize}
\item a Charge Sensitive Amplifier (CSA), based on single-ended folded cascode architecture, optimized for detector capacitances in the 20-30 pF range. The full charge range (120 fC, 240 fC, 360 fC or 600 fC) is determined by selecting one CSA feedback capacitor among four.
\item a Pole-Zero Cancellation (PZC) used to cancel the long duration undershoots at the shaped output. It introduces a zero to cancel the low frequency pole of the CSA and replaces it by a higher frequency pole.
\item a R.C2 filter associated with the PZC pole. This 2-complex pole Sallen-Key low pass filter provides a semi-Gaussian shaping of the analog channel with a very small undershoot (0.8\%). The peaking time of the global filter can be set among 16 values (from 100 ns to 2 $\mu s$) by using various combinations of resistors in the feedback network.
\item a Gain-2 amplifier which adjusts the voltage dynamic range of the chain (Gain of 2) and drives the channel SCA.
\end{itemize}
The analog memory is based on a switched capacitor array structure. It includes 72 effective channels plus 4 dummy channels (Gain 2 stage + SCA) which can be useful for common mode or fixed pattern noise rejection. The SCA is used as a 511-cell deep circular buffer in which the signal coming out from each analog channel is continuously sampled and stored at the sampling rate, which can be up to 50 MHz. This frequency is set to adjust the duration of the capture window for the given time structure of the beam and the drift velocity of electrons in the TPCs. Sampling is stopped by the FEM upon reception of an external trigger signal. Then, the 511 analog samples of each channel are sequentially read and multiplexed column by column at 20 MHz towards an external commercial 12-bit ADC. This reading operation is performed cell by cell, starting from the oldest sample. At the end of the readout phase, additional data corresponding to the address of the last read column is also multiplexed for control purpose. The readout of the full memory takes 2 ms, but it is possible to abort digitization after any number of columns have been acquired.
An on-chip buffer drives differentially the ADC inputs. It is designed to settle to 0.1\% of the final output voltage in less than half of the ADC clock period (i.e. 10 ns).

\subsubsection{The FEC}
\label{sss-fec}
The FEC comprises four AFTER chips and the necessary passive components to protect the electronics against high voltage surges caused by sparks that can occur in the detector. Two PhotoMos relays are used to optionally leave floating the 10 M$\Omega$ polarization resistor of each detector pad. In case of permanent short between the mesh of the detector and one or several pads, a group of 144 pads can be disconnected to avoid excessive current being drawn from the high voltage source. Each FEC is inserted directly at the back of the Micromegas detector and connections are established via four dual-row 80-pins 1.27 mm pitch connectors. This coupling method is very compact, cost effective (no flexible cables are needed) and causes minimal degradations to sensitive analog signals. The output of the four AFTER chips is digitized at 20 MHz by a quad-channel 12-bit ADC (AD 9229 from Analog Devices). The time required for digitizing SCAs is proportional to the number of cells being readout. Digitization takes about 2 ms when the full depth of the SCAs is used ($79 \times 512 \times 50$ ns cycles). The maximum possible event acquisition rate of the FEC is therefore $\sim$500 Hz. The FEC also comprises an on-board pulser for calibration and a silicon identification chip (DS 2438 from Maxim) that also performs temperature, supply current and voltage measurements.

\subsubsection{The FEM card}
\label{sss-fem}
The six FECs of each detector module are driven by the FEM board. This card provides the FECs with all the necessary clock and synchronization signals to drive the AFTER chips and it receives the data collected in the SCAs after digitization. The SCA of all 24 AFTER chips of a module are digitized in parallel and the peak bandwidth at the input of the FEM is 5.76 Gbit/s. Since this level was found too high to be sent off-detector directly, the FEM stores the raw data received from the FECs into a digital buffer composed of two 9-Mbit Zero-Bus Turnround (ZBT) static memory chips. Upon request from the back-end electronics, the FEM performs an optional pedestal equalization and returns the data via its 2 Gbps optical link in zero-suppressed or uncompressed format. The algorithm used for zero-suppression simply consists of applying a per-channel programmable threshold. Ten samples before threshold crossing and four samples after the last threshold crossing are kept to preserve the tails of the waveforms. Zero-suppression is applied on-the-fly when data are requested by the back-end electronics and is implemented in the FPGA of the FEM board (Xilinx Virtex 2VP4). It takes $\sim$10 $\mu s$ to retrieve the 511 time-bins of a channel from the memory of the FEM and apply zero-suppression. The data that remains are sent to the DCC. If no sample passed the threshold, an empty response packet is returned in less than 100 ns, while if all samples are above threshold, sending the 511 time bins of that channel takes $\sim$2.5 $\mu$s. The amount of time needed to readout the 1824 channels of a FEM varies from 18.5 ms to 22.8 ms depending on channel occupancy. The limitation imposed by the FEM on the event acquisition rate is thus of the order of 45 Hz. In addition to a fast memory and FPGA logic with multi-gigabit per second I/O capacity, the FEM comprises an 8-bit micro-controller attached via opto-isolators to a CANbus segment shared by the 24 FEM boards of a TPC. The micro-controller monitors the operating temperature, voltage and current of the FECs and FEM board. It enables power and controls the PhotoMos relays of each FEC individually. The slow control bus is also used to load FPGA and micro-controller firmware revisions in the on-board flash memory.         

\subsection{Test results}
\label{ss-fee-test}

The AFTER chip has been manufactured using the 0.35 $\mu$m CMOS technology from AMS. The chip contains 400,000 transistors on a 58 mm$^2$ die and is packaged in a 160-pin LQFP. All functionalities of the chip and its various modes of operation have been fully validated. The volume produced was 5300 chips and about 89\% of them passed the validation tests. Detailed results are given in \cite{Bar:2008nss}. 

The power consumption of the chip varies between 6.2 mW and 7.5 mW per channel, depending on the bias current (400 $\mu$A or 800 $\mu$A) of the preamplifier which can be set by an external resistor. The peaking time and the shape of the signal well reproduce the expected values as well as the dynamic range and the integral non-linearity (better than 1.2\% over all four charge ranges). The chip even operates perfectly at 100 MHz write frequency although it has been designed for 50 MHz. A complete noise characterization was performed by varying the input capacitor and shaping time. For an input capacitance smaller than 30 pF and a shaping time shorter than 200 ns (relevant values for T2K TPCs), the noise is smaller than 1000 electrons rms which meets our requirements. The on-chip crosstalk has been measured. It is derivative and its amplitude is less than $\pm$0.4\% decreasing with the distance between channels. The voltage drop in the SCA after 2 ms is less than 1 ADC count (164 electrons for the 120 fC range or 1/4096 of the whole dynamic range). The mean value is 0.29 ADC count and the effect of charge leak in the SCA remains negligible compared to noise.

\subsection{Mechanical support and cooling}
\label{ss-fee-mech}

\begin{figure}[htp]
\centering
\ifx\figstyle\bw
 \includegraphics[width=0.45\textwidth]{fig16_bw.jpg}
\else
 \includegraphics[width=0.45\textwidth]{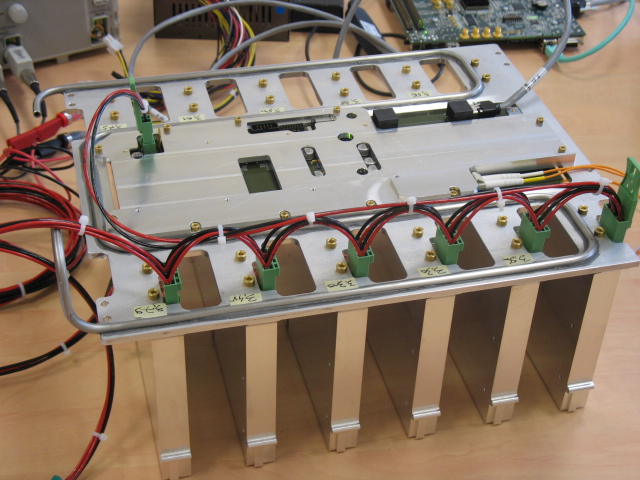}
\fi
\caption{Readout electronics and mechanics of one detector module.}\label{fig:fee-module}
\end{figure}

The overall dimensions of the block of electronics used to readout each detector module are 28 cm $\times$ 34 cm $\times$ 18 cm in depth. A picture of the readout electronics and mechanics of one module is shown in Fig.~\ref{fig:fee-module}.

Each FEC is enclosed in a shell-shaped aluminum carapace. The main component side of the FEM is covered with an aluminum heat sink and only a thin shield and dust protection flexible sheet is placed at the back of this card. The heat generated by the various active components is transferred by passive conduction to the carapaces via a thin layer of thermally conductive material. The front edge of the carapace of the FECs and the FEM are screwed on an aluminum plate which is cooled by circulating water. The weight of the electronics, mechanical support and cooling plates is $\sim9$ kg per module. The electronic cards represent 20\% of this mass.

%% file: chap06.tex
\section{Photoelectron calibration system}
\label{sec-calib}

In order to measure and monitor important aspects of the electron transport
in the TPCs, a calibration system that produces a control pattern of electrons
on the central cathode was incorporated into the TPC design.
Data from this system are used to precisely determine the electron 
drift velocity and to measure distortions in the electron drift due
to inhomogeneous and misaligned electric and magnetic fields.
The system can also be used to measure the absolute gain of the readout system.

\subsection{Requirements}
\label{ss-calib-req}

The requirements for understanding distortions in the T2K TPCs
differ from TPCs in colliding beam experiments which
are designed to accurately measure the curvature
of the radial trajectories of 
particles produced at the interaction point.
For those systems, lines of electrons produced by focused laser
beams ionizing the gas are commonly used to give information about the 
magnitude of distortions transverse
to the typical track directions.

Particles traverse the T2K TPCs in all possible orientations and
therefore it is important to measure distortions in all directions.
This is only possible if the control pattern of electrons are
produced as points, rather than lines.
The displacement of a point is an unambiguous measure of the distortion
direction and magnitude, whereas when a line image does not match the
original, there is not a unique set of displacements to explain the
distortion.

With the goal for the TPC momentum scale uncertainty of $<2$\%,
relative displacements as small as 0.1~mm can be important.
The displacement resolution for a single measurement can be larger than
this, however, since several measurements can be averaged.
To estimate the centre of a cloud of photoelectrons, using the charge sharing 
in neighbouring readout pads, requires knowledge of the transverse
distribution of the electrons which depends on the size of the 
source of photoelectrons and on the transverse diffusion.
Therefore a scheme to measure the diffusion should be included in the design.
Finally, the density of photoelectrons from the control pattern should be
similar to that produced by a minimum ionizing particle, so that the
micromegas gain and electronics readout parameters do not need to be
adjusted for calibration events.

\subsection{Design}
\label{ss-calib-design}

\begin{figure}[htp]
\centering
\includegraphics[width=0.45\textwidth]{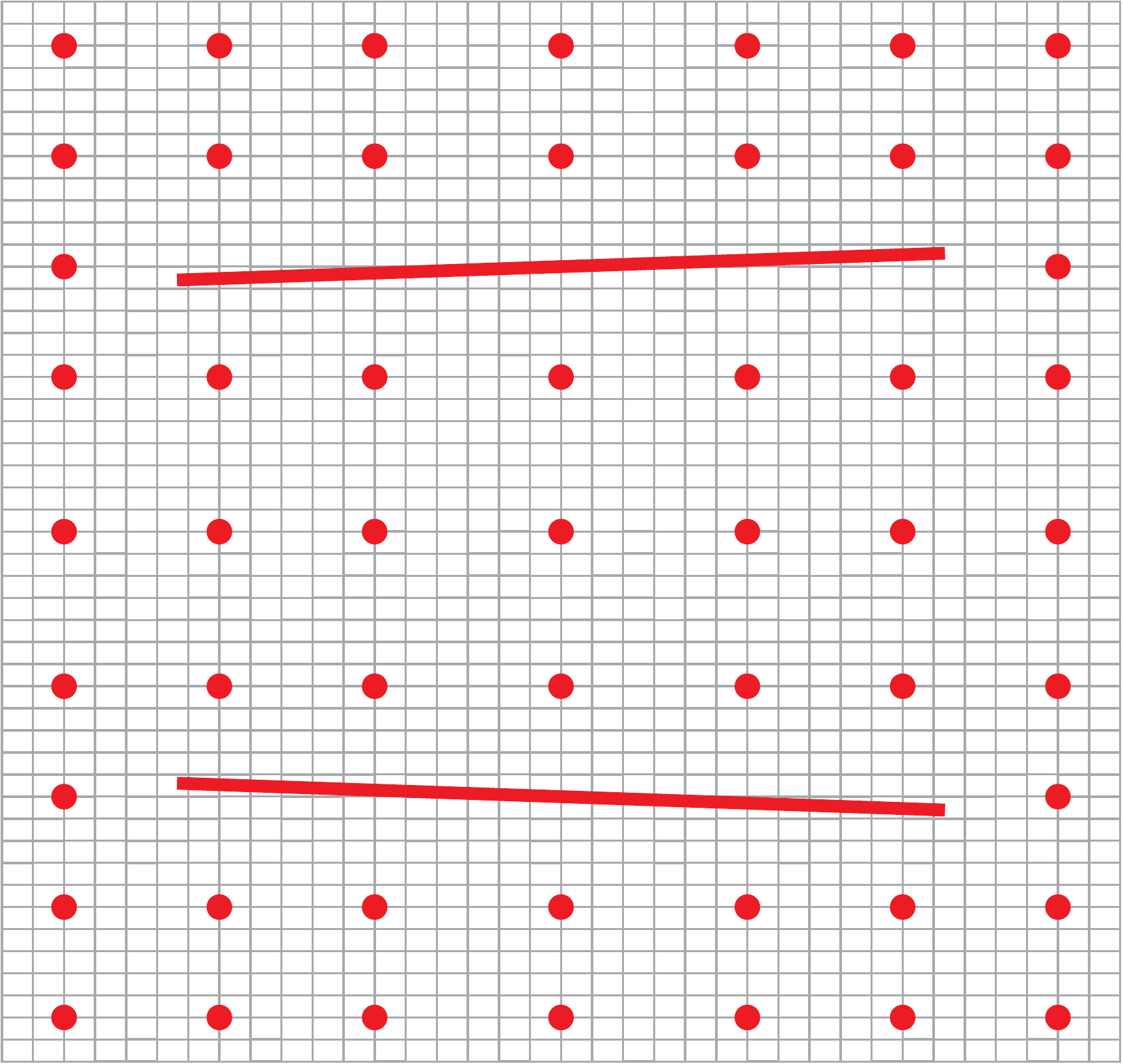}
\caption{The pattern of aluminum targets placed on the cathode
for each micromegas module is shown. The locations of the pads
are projected onto the pattern, showing that the discs are
located at the corners of the pads.}\label{fig:calib-design}
\end{figure}

In order to meet the requirements stated above, thin aluminum discs, 
8~mm in diameter, are glued to the copper surface of the cathode.
Flashing a diffuse pulse of 266 nm light on the cathode will cause
photoelectrons to be emitted from the aluminum but not the copper.
As shown in Fig.~\ref{fig:calib-design}, a 
total of 53~dots are placed in a regular pattern for each micromegas
module, such that each dot is nearly aligned 
to the corner boundaries of 4~interior
pads to optimize the spatial resolution.
With about 100~photoelectrons per dot, the expected spatial resolution is
better than 0.5~mm, dominated by the transverse diffusion.
In order to measure the transverse size of the ionization, two
strips, 4~mm wide, are also included in the pattern at an angle
with respect to the pad boundaries.

The 266~nm light, produced by a frequency quadrupled Nd:YAG 
laser\footnote{Quantel Model ULTRA CFR FHG}, is focused into
quartz fibres\footnote{CeramOptec model UV800/880/980/1200N}
that transport the light to small optical packages
embedded in the inner box module frames that defocus the light
onto the cathode. 
The light from each fibre illuminates a region of the cathode
sampled by 4~micromegas modules, and so a total of 18~fibres
are used for the 3~TPCs.
An electro-mechanical multiplexer was built
to direct the UV light pulses from the laser into any one
of the 18~fibres by moving a mirror.
Therefore a single laser calibration event includes data
for just 4~readout modules.
During beam operations, laser calibration data is collected
in the 3.5~s inter-spill periods.

A motorized variable attenuator is adjusted on the laser so that
typically about 100~photoelectrons are produced from each disc.
The focusing is matched to the 0.8~mm diameter of the fibre ends
in order to efficiently inject the light
while avoiding damage to the fibre surface.

\subsection{Performance}
\label{ss-calib-perf}

Generally good image contrast is seen for the aluminum targets,
provided the surfaces are carefully cleaned.
Typically the aluminum emits 2~photoelectrons/mm$^2$, whereas 
for copper it is less than about 0.03~photoelectrons/mm$^2$.
For regions of one TPC, regions of copper emit a higher density of
photoelectrons for unknown reasons, perhaps due to
surface contamination.

The photoelectrons arrive simultaneously on the micromegas modules which
results in a large current spike between the mesh and the readout pads.
The electric potential of the mesh fluctuates, due to its finite
capacitance, and all pads on the module sense the change in potential,
resulting in a small opposite polarity pulse in time with the
photoelectrons.
By sampling pads which do not collect photoelectrons from the targets,
the appropriate correction can be derived for all pad signals.

The time difference between the laser trigger time and the arrival time
of the signals on the mesh gives a very precise determination of the
drift velocity.
The standard deviation of repeated measurements for a single channel is
roughly 10~ns.
By combining all channels in a module, the system could provide sub-ns 
resolution of relative changes to the drift time with a single event.
The jitter in the delay between the laser trigger and laser pulse, 
however, limits the resolution for drift time to a few~ns for single events. 

\begin{figure}[htp]
\centering
\includegraphics[width=0.45\textwidth]{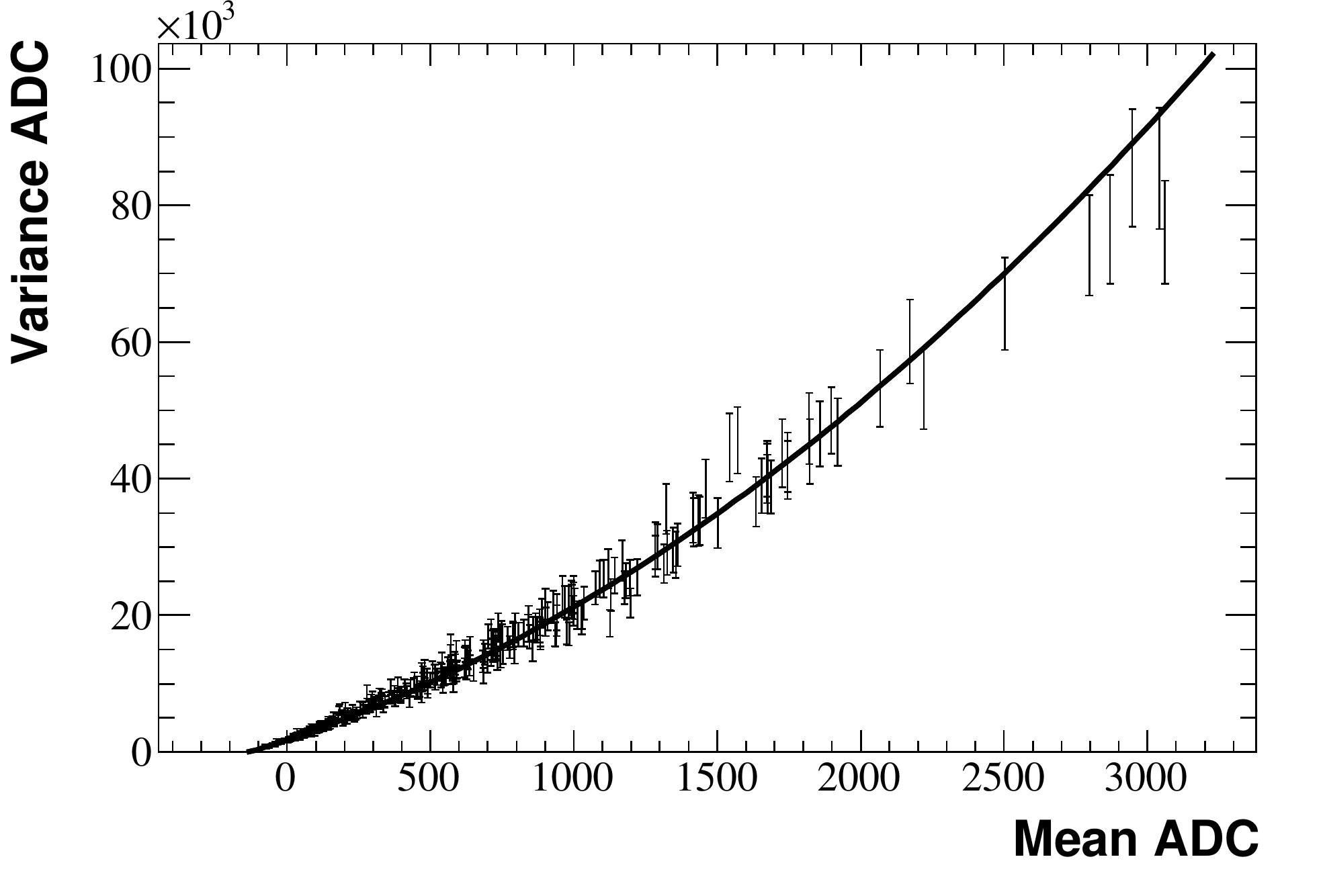}
\caption{The variance in amplitude and
the mean amplitude is shown for repeated measurements by
pads that sample significant photoelectron signals.
The curve shows the result of a fit to estimate the
gain, magnitude of opposite polarity pulses, and laser variation.
}\label{fig:calib-gain}
\end{figure}

The variation in the amplitude of signals from repeated measurements
arises primarily from the photoelectron statistics and gain fluctuations.
A simplified model can describe the relation between the amplitude variance
and mean:
\begin{enumerate}
\item the number of photoelectrons arriving above a particular readout pad 
is given by a Poisson random number $N$, with mean $\nu$
\item for any electron arriving at the micromegas mesh, the number of 
electrons produced in the avalanche is described by an exponential
random number, $G$, with mean $\gamma$.
\item the total energy emitted from the laser is given by a Gaussian
random number with relative standard deviation $\beta$.
\end{enumerate}
In this model the amplitude of a signal for a pad is described by
a random number, $A$.
If the multiplicative conversion factor from collected electrons to ADC
channels is $\alpha$, the expectation value is $E[A]=\alpha\gamma\nu$.
The variance of $A$ has contributions in order
from the three points above:
\begin{eqnarray*}
V[A] &=& (\alpha\gamma)^2V[N]+\alpha^2\nu V[G]+(\beta\alpha\gamma\nu)^2\\
&=& 2\alpha\gamma E[A] + \beta^2 E[A]^2
\end{eqnarray*}
In this model, the variance from gain fluctuations is equal to the
variance from photoelectron production.
To allow for the effect of opposite polarity pulses, the relation is
modified to include an additional offset:
\begin{displaymath}
V[A] =2\alpha\gamma (E[A]-a_0) + \beta^2 (E[A]-a_0)^2
\end{displaymath}
Fig.~\ref{fig:calib-gain} shows a typical result from 
a fit of repeated measurements
to pads on a single module.
The resulting estimates for the system gain, 
$\alpha\gamma\approx 8$~ADC/pe are similar to
the gain measurements made on the module with an Fe$^{55}$ source,
described in section~\ref{ss-mm-testbench}.

The observed distribution of charge from the
strip targets is used to form a maximum likelihood
estimate of the transverse diffusion of the gas.
The standard deviation from repeated measurements
from the same strip target is roughly 15~$\mu$m/$\sqrt{\rm cm}$,
in accord with a Monte Carlo simulation.
Some strips give a much larger diffusion estimate than typically found.
For these strips, neighbouring copper may emit enough
photoelectrons to cause the diffusion constant to be
overestimated.
For that reason, it is expected that the sample of strips
with the smallest diffusion measurements are the most
reliable.
For these, the typical diffusion constants are found to be
0.34~mm/$\sqrt{\rm cm}$ with B=0 and
0.29~mm/$\sqrt{\rm cm}$ with B=0.18 T,
almost 20\% larger than measurements 
from particle tracks, described
in section~\ref{ss-perf-spatial}.
The uncertainty in the strip width leads to a 4\%
systematic uncertainty in these estimates.

\begin{figure}[htp]
\centering
\includegraphics[width=0.45\textwidth]{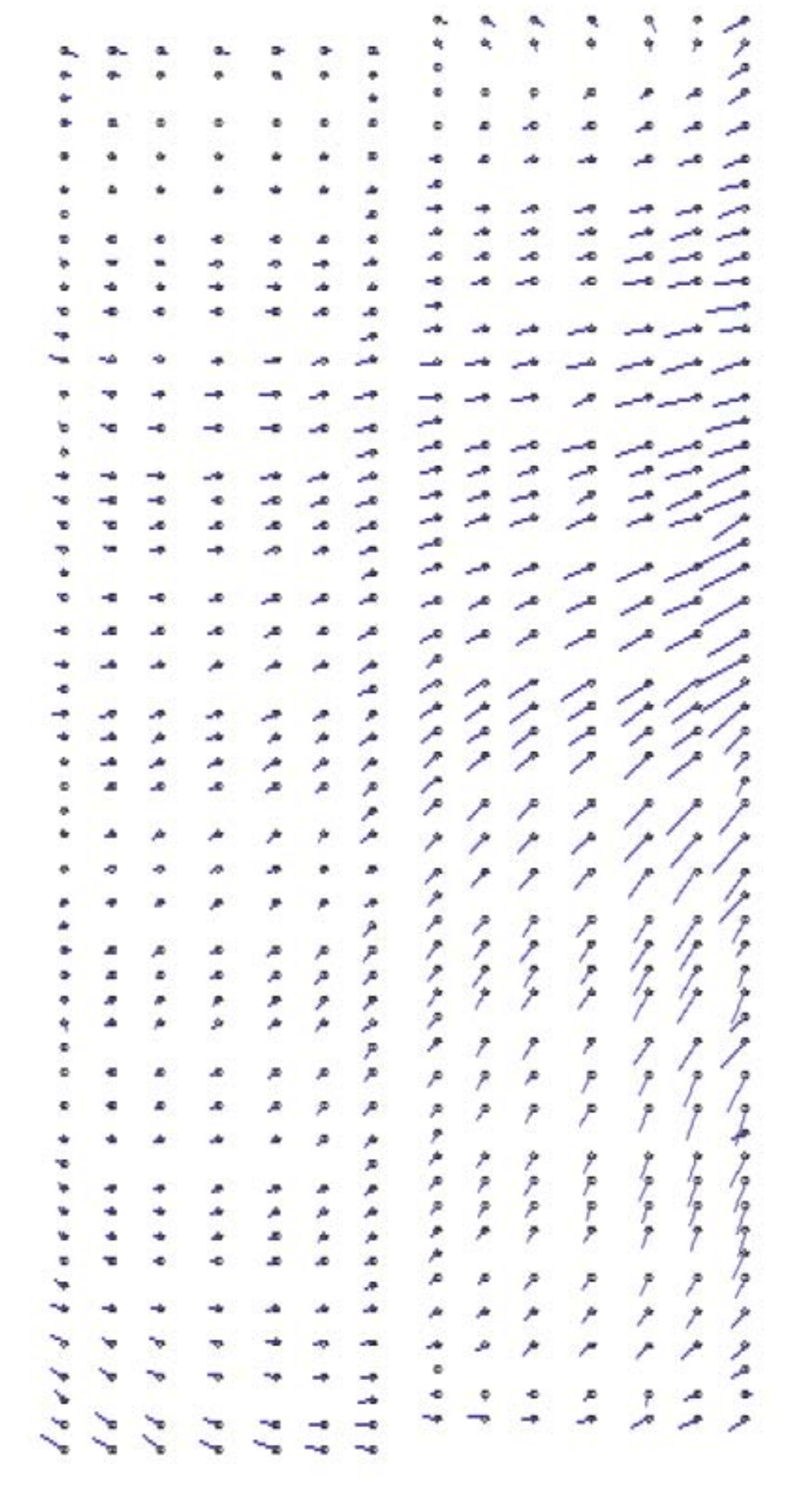}
\caption{The effect of the magnetic field on the
drift of electrons from the cathode is illustrated for the
most downstream TPC, which suffers from the largest
magnetic field distortions.
The dots indicate the nominal locations of the target images
for an entire endplate.
The lines represent the displacement of the target images
when the magnetic field is turned on. The magnitude of the
displacements are magnified by a factor of 10 to make them
visible on this scale.
Normally the displacements are less than 1~mm, but for
this TPC, the displacements are as large as 5~mm.
Note that the aspect ratio of the figure has been adjusted
for better illustration of the effect.
}\label{fig:calib-pin}
\end{figure}

The sharing of charge across four neighbouring pads is
used to form a maximum likelihood
estimate for the centre of the image of each disc target.
The standard deviation from repeated measurements
from the same target is roughly 0.5~mm,
in accord with a Monte Carlo simulation.
The offsets of the dot centres from their surveyed
locations provides information about electric
field distortions, using data taken with magnetic field off.
By comparing field on with field off data, the
magnet field distortions can be measured, as
illustrated in Fig.~\ref{fig:calib-pin}.
The study of field distortions and implementation of
field corrections are underway at the time this paper
was being prepared.

%% file: chap07.tex
\section{TPC performance}
\label{sec-perf}

\begin{figure}[htp]
\centering
\ifx\figstyle\bw
 \includegraphics[width=0.45\textwidth]{fig20_bw.png}
\else
 \includegraphics[width=0.45\textwidth]{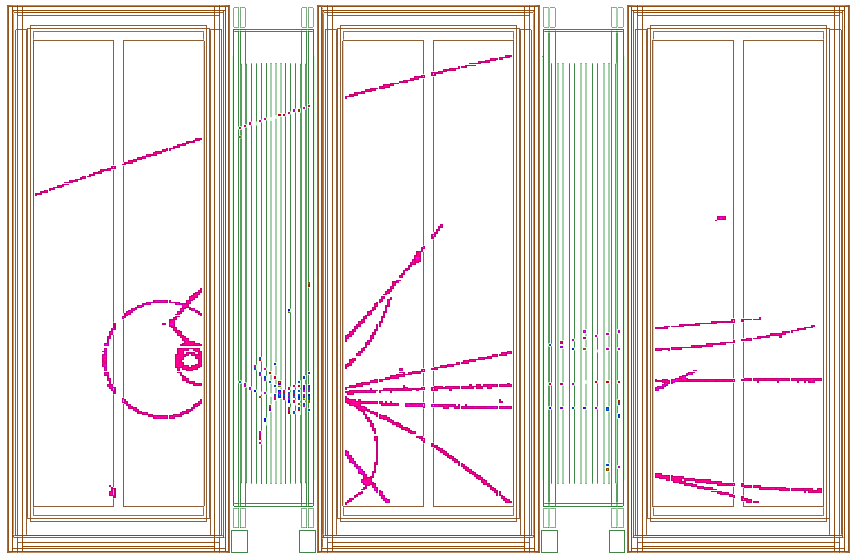}
\fi
\caption{An event recorded in the tracker section of the
off-axis near detector. One neutrino interacted in front of the
first TPC, and a second neutrino undergoes a deep inelastic scatter
in the first fine grained scintillator detector.
More typical neutrino interactions involve only a few particle tracks.
}\label{fig:perf-event}
\end{figure}

Since late 2009, the 3 TPCs have been in place within the
off-axis near detector for the T2K experiment.
Neutrino, cosmic ray, and calibration events have been
recorded and processed, such as the event shown in Fig.~\ref{fig:perf-event}.
This section shows the initial performance achieved by
the TPCs, after basic corrections are applied to account
for gain variation resulting from gas density changes and
to account for module misalignment as determined by survey measurements.

\subsection{Basic performance}
\label{ss-perf-basic}

\subsubsection{Electronics}

Current runs use the 120~fC charge range and the 200~ns shaping time of the AFTER chip. Pad sampling rate is typically set to 25~MHz, i.e. the 511 time-bins of the SCA provide a 20~$\mu$s time window. In these conditions, the typical noise on each pad is 4-5~ADC counts rms, roughly 800 electrons equivalent.

Beam events and cosmic ray events are readout in zero-suppressed mode while the uncompressed readout mode is used to acquire pedestal and laser calibration events. With thresholds set to 4.5 standard deviations above pedestal noise level, the typical average event size for cosmic and beam events is about 60~kB corresponding to a reduction factor of 1900 compared to raw data. Very low channel occupancy and low noise readout electronics greatly simplify the data reduction task, and the simple algorithm implemented in the front-end is sufficient to bring event data to a manageable size. Pedestal events are exploited off-line to compute the average baseline and noise level of each channel. A per-channel pedestal equalization constant and threshold are computed and loaded in the front-end electronics. For laser calibration events, only the four FEMs corresponding to the area illuminated by the laser are readout and only 50 time-bins out of 511 are acquired leading to fixed-size events of 700~kB. At the design acquisition rate of 20~Hz (including 0.5~Hz of laser events), the global throughput of the TPCs to the global DAQ is below 2~MB/s. The system is far from its bandwidth limits. The typical average latency for TPC data acquisition is 33~ms and 52~ms for beam/cosmic and laser calibration events respectively. The power consumption of the 6 FECs and the FEM used to readout a detector module is about 7~A at 4.5~V. In total, the TPC front-end electronics dissipates 2.3~kW, i.e. 18~mW per channel. In addition, about 500~W are dissipated in the 16~m long cables that bring power from the power supply crates located at the service stage level under the magnet to the TPC front-end electronics. 

The development of TPC readout electronics took five years from the specifications of the new ASIC until full board production, test and complete installation in-situ. The complete readout system has been in operation for over 5000~hours and no failure of the electronics has been observed so far.

\subsubsection{Temperature stability}

\begin{figure}[htp]
\centering
\ifx\figstyle\bw
 \includegraphics[width=0.45\textwidth]{fig21_bw.png}
\else
 \includegraphics[width=0.45\textwidth]{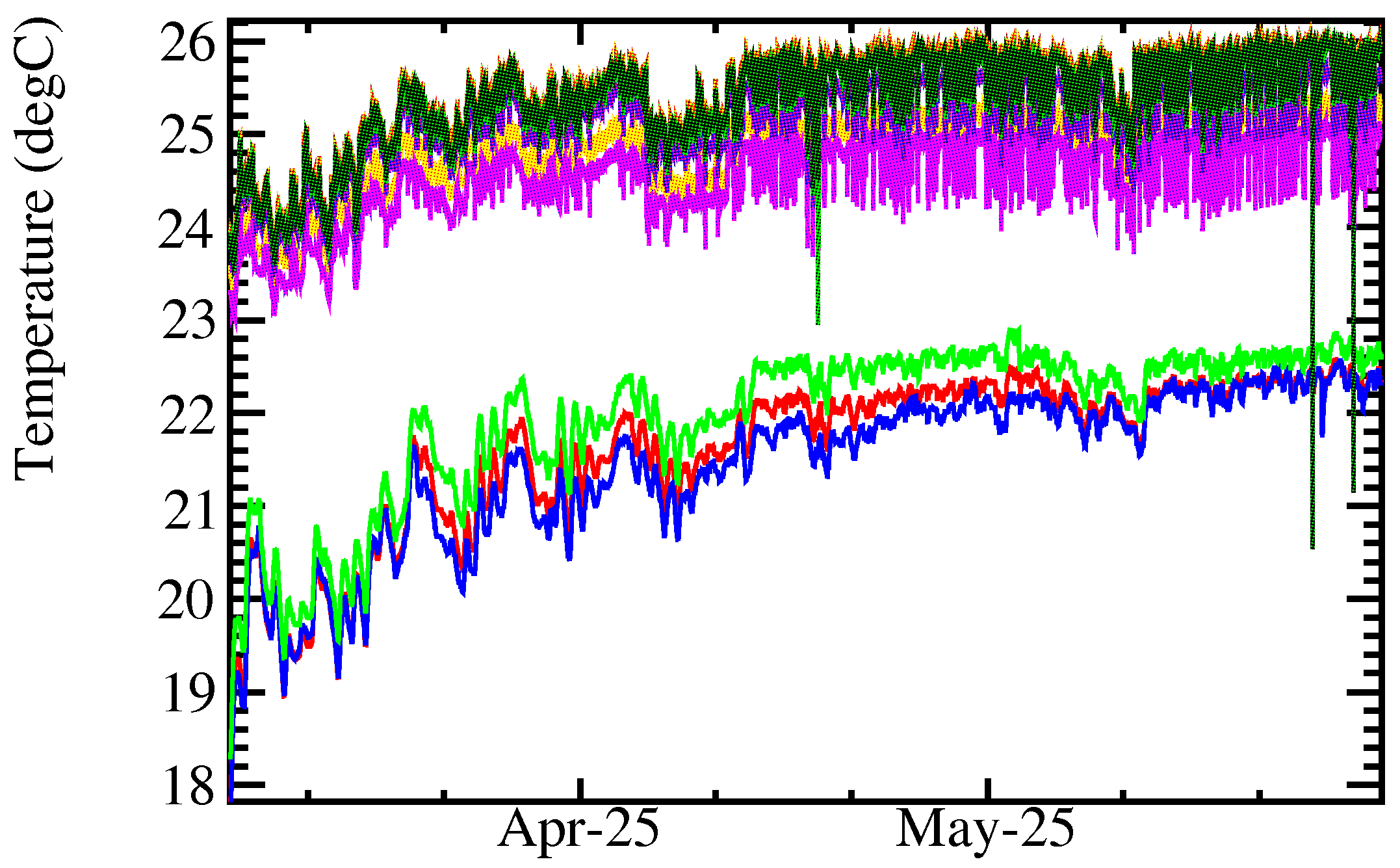}
\fi
\caption{
Long term temperature histories are shown for the temperatures of the
3~TPCs as measured near the gas input to the inner volumes (lower lines) and
the front end electronics (upper lines).
}\label{fig:perf-temp}
\end{figure}

The typical temperature measured on the FECs and the FEM are $26 ^{\circ}$C and $24 ^{\circ}$C respectively and remain stable over time within a couple of degrees.
This is illustrated in Fig.~\ref{fig:perf-temp}

\subsubsection{Gain stability}
\label{ss-perf-stability}

\begin{figure}[htp]
\centering
\ifx\figstyle\bw
 \includegraphics[width=0.45\textwidth]{fig22_bw.png}
\else
 \includegraphics[width=0.45\textwidth]{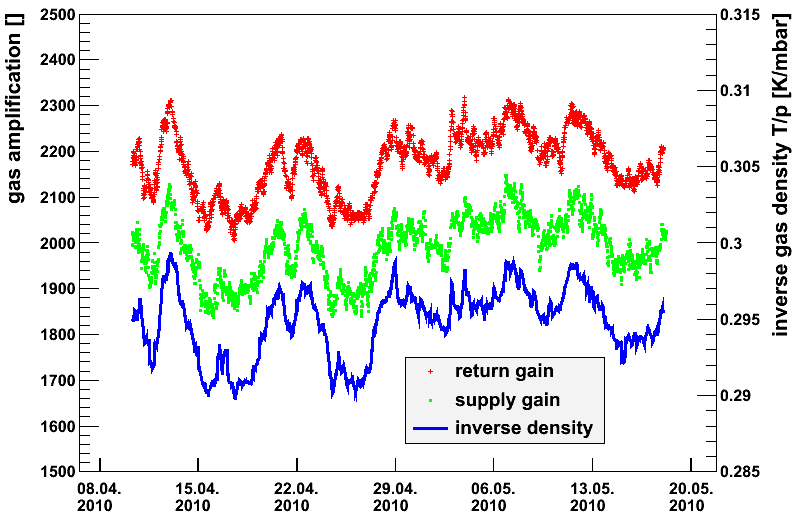}
\fi
\caption{
Gain measured by the monitor chambers over a period of 6 weeks is shown
by the upper two sets of points, for the return and supply gas to the TPCs.
The two monitor chambers have not been cross calibrated, resulting in a
constant offset between the two measurements.
The lower curve shows the variation in the inverse gas density over the
same period (using the scale on the right).
The variation in gas gain is primarily due to
atmospheric pressure changes.
}
\label{fig:perf-gainstability}
\end{figure}

Fig.\ref{fig:perf-gainstability} shows a
six-week history for gain as measured by the
monitor chambers for the gas supplied to the TPCs and returned from the
TPCs.
As an overlay the inverse gas density $T/p$ is plotted.
It is seen that the gain variation over this period is less than $\pm$10\%,
and is mostly due to the gas density variation, primarily due to
atmospheric pressure changes.
This inverse gas density is used to correct the measured gain
value of the gas monitor chambers and the TPCs.
A correction for density changes is given by
\begin{displaymath}
    g_{corr} = \frac{g_{meas}}{1 + (\frac{T/p}{T_0/p_0} - 1)\cdot s}
\end{displaymath}
with $T_0 = 298.15$~K and $p_0 = 1013$~mbar.
The slope $s$ describes the relative change of the gain per
relative change of $T/p$.
This correction is necessary because the temperature and pressure
of the TPCs and the monitor chambers
differ due to different barometric altitudes
and climate conditions.
After applying the correction to this data,
the remaining variation, due to other factors such as gas composition,
is below 1\%.
No evidence is seen for electron attachment in the TPC drift volume
when comparing the mean signal amplitudes for cosmic ray
tracks at short and long drift distances.

\subsection{Tracking performance}
\label{ss-perf-spatial}

\begin{figure}[htp]
\centering
\ifx\figstyle\bw
 \includegraphics[width=0.45\textwidth]{fig23_bw.png}
\else
 \includegraphics[width=0.45\textwidth]{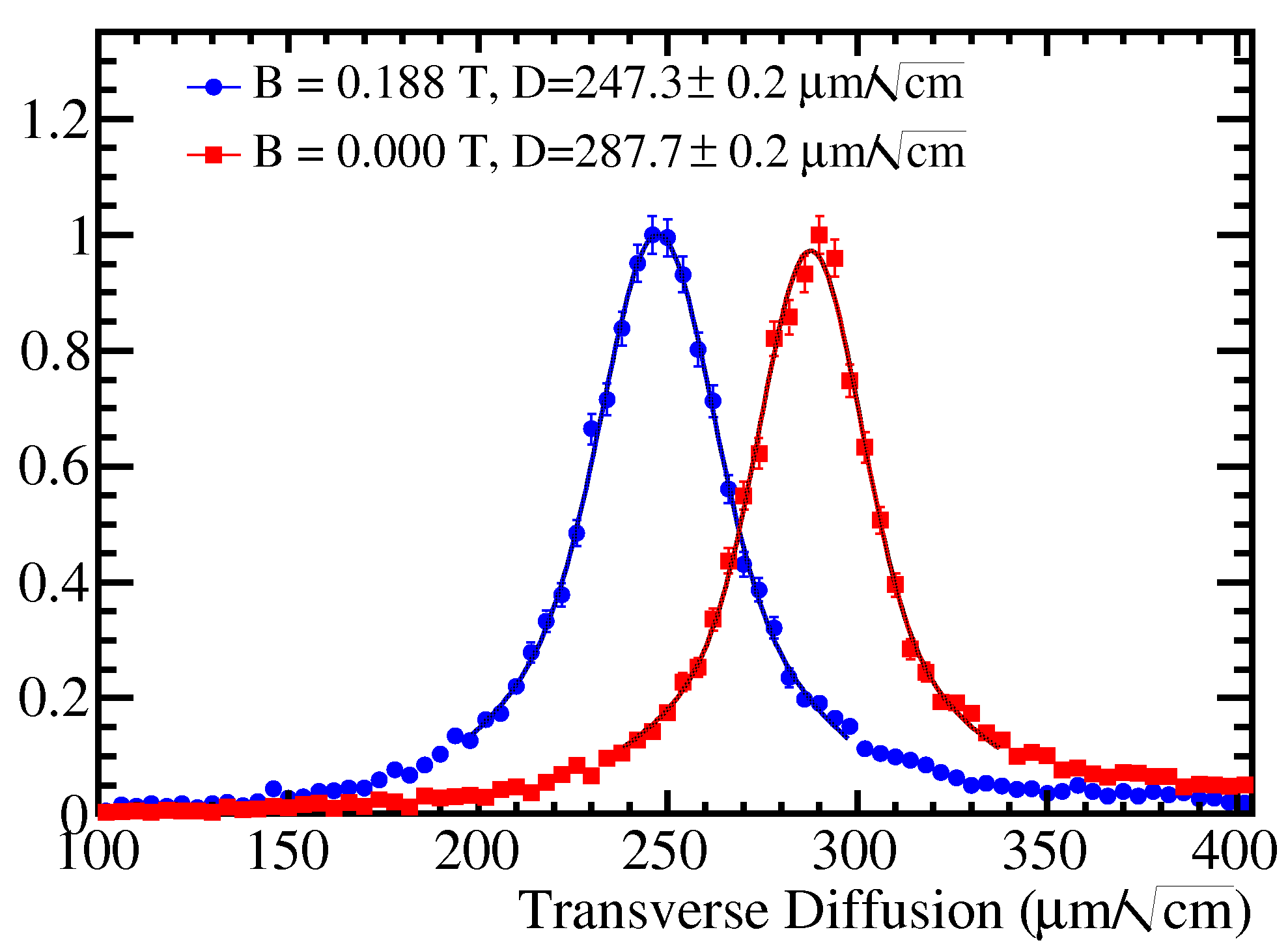}
\fi
\caption{
As part of the maximum likelihood track fit,
the diffusion constant for the mean drift distance of the
track is allowed to vary.
The change in diffusion along the length of the track, due to differences
in drift distance, is fixed in the fit.
This plot shows the distribution of diffusion constant estimates from
samples of cosmic rays with mean drift
distance of more than 30~cm with magnetic field on and off.
The quoted uncertainties are statistical only.
}\label{fig:perf-diffusion}
\end{figure}

Track reconstruction is performed by separate methods for track finding
and track fitting.
Signals in neighbouring pads consistent with arising from the same
particle are grouped to form a track of ionization.
Clusters are formed consisting of neighbouring pads within a
column (row) for roughly horizontal (vertical) tracks.
The likelihood of the observed charge sharing between the pads
within the clusters is maximized to estimate the track parameters
and the width of the ionization track.\cite{Karlen:2005iw}
This allows the diffusion properties of the gas to be measured
from a set of tracks as illustrated in Fig.~\ref{fig:perf-diffusion}.

\subsubsection{Spatial resolution}

The spatial resolution is estimated by comparing the transverse coordinate resulting 
from the global track fit to the one obtained with a single cluster
fit when the other track parameters (angles and curvature) are fixed to the result of 
the global track fit. The residual distribution is fit to a normal distribution providing 
the values of the spatial resolution and bias.

\begin{figure}[ht]
\centering
\includegraphics[width=0.45\textwidth]{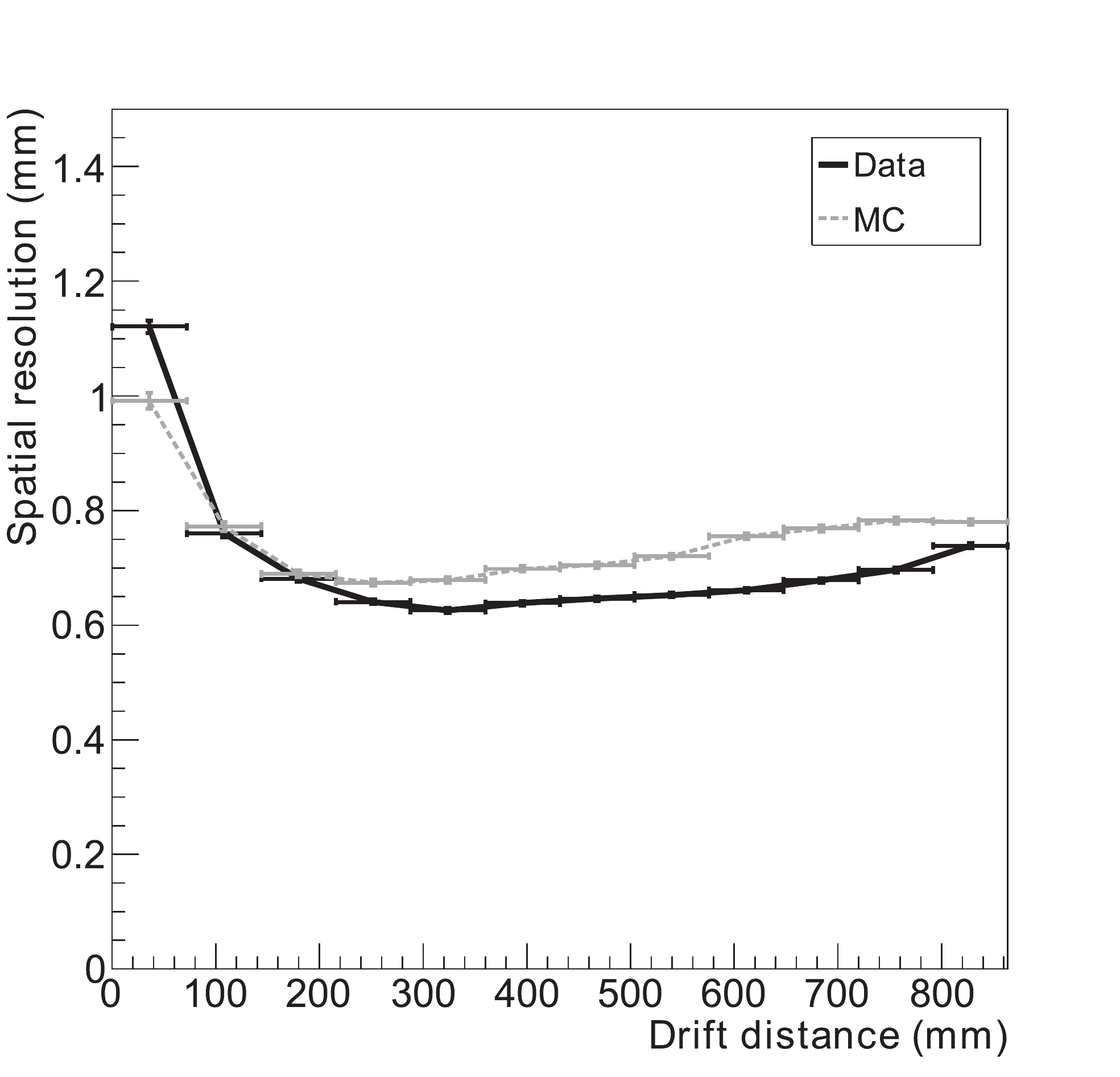}
\caption{Spatial resolution per cluster as function of the drift distance.
Black points (continuous line) show the results computed from data
and grey points (dashed line) show the results from simulations.
}\label{fig:PointResolutionDriftAll}
\end{figure}

\begin{figure}[ht]
\centering
\includegraphics[width=0.45\textwidth]{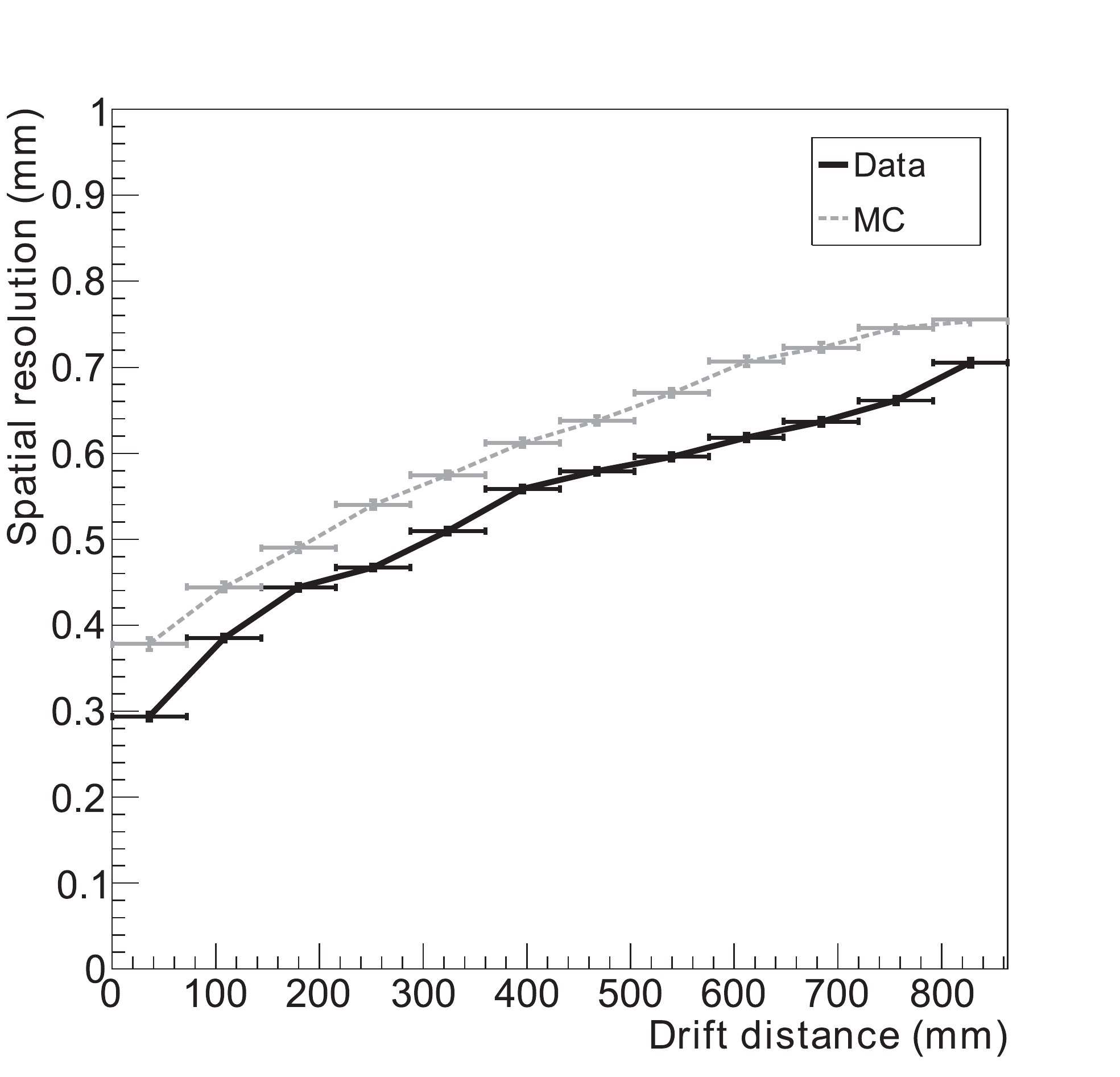}
\caption{Spatial resolution as function of the drift distance for
clusters with 2 detector pads. Black points (continuous line)
show the results computed from data and grey points (dashed line)
show the results from simulations.
}\label{fig:PointResolutionDrift2Pad}
\end{figure}

\begin{figure}[ht]
\centering
\includegraphics[width=0.45\textwidth]{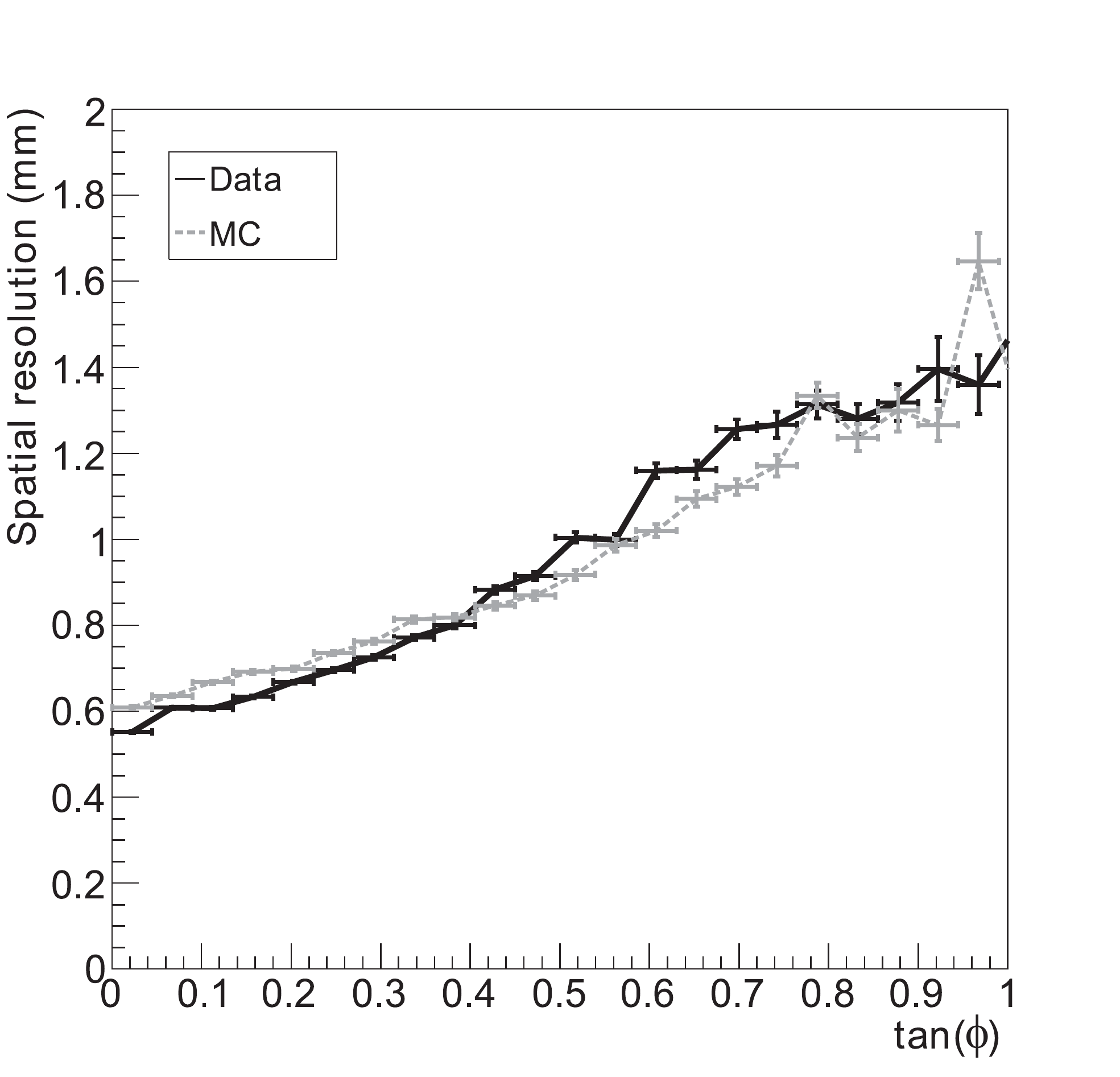}
\caption{Spatial resolution as function of the tangent of the angle
away from the horizontal plane for all drift
distances and number of pads per cluster. Black points (continuous
line) show the results computed from data and grey points (dashed
line) show the results from simulations.
}\label{fig:PointResolutionAngle}
\end{figure}

The spatial resolution for tracks is shown
as a function of drift distance in Fig.~\ref{fig:PointResolutionDriftAll}. 
The degraded resolution at short drift distances is due to the larger fraction
of single pad clusters that occur for this case
where tracks are well aligned with the pad boundaries and the diffusion
is insufficient to cause the signals to be distributed to two pads in
a column.
The spatial resolution for track clusters consisting of
two pads is shown as a function of drift distance
in Fig.~\ref{fig:PointResolutionDrift2Pad} and shows a clear
dependence on diffusion.
The resolution as function of the angle away from the horizontal plane
is shown in  Fig.~\ref{fig:PointResolutionAngle}.
The strong dependence on angle is due to the ionization fluctuations
along the track, which increase the variance of charge sharing in a cluster
for tracks at an angle to pad boundaries.
The simulation incorporates most of the important
detector effects, including transverse and longitudinal diffusion 
and a parametrization of the electronics response.
There is generally good agreement between the simulated and
measured spatial resolution.

\begin{figure}[ht]
\centering
\includegraphics[width=0.45\textwidth]{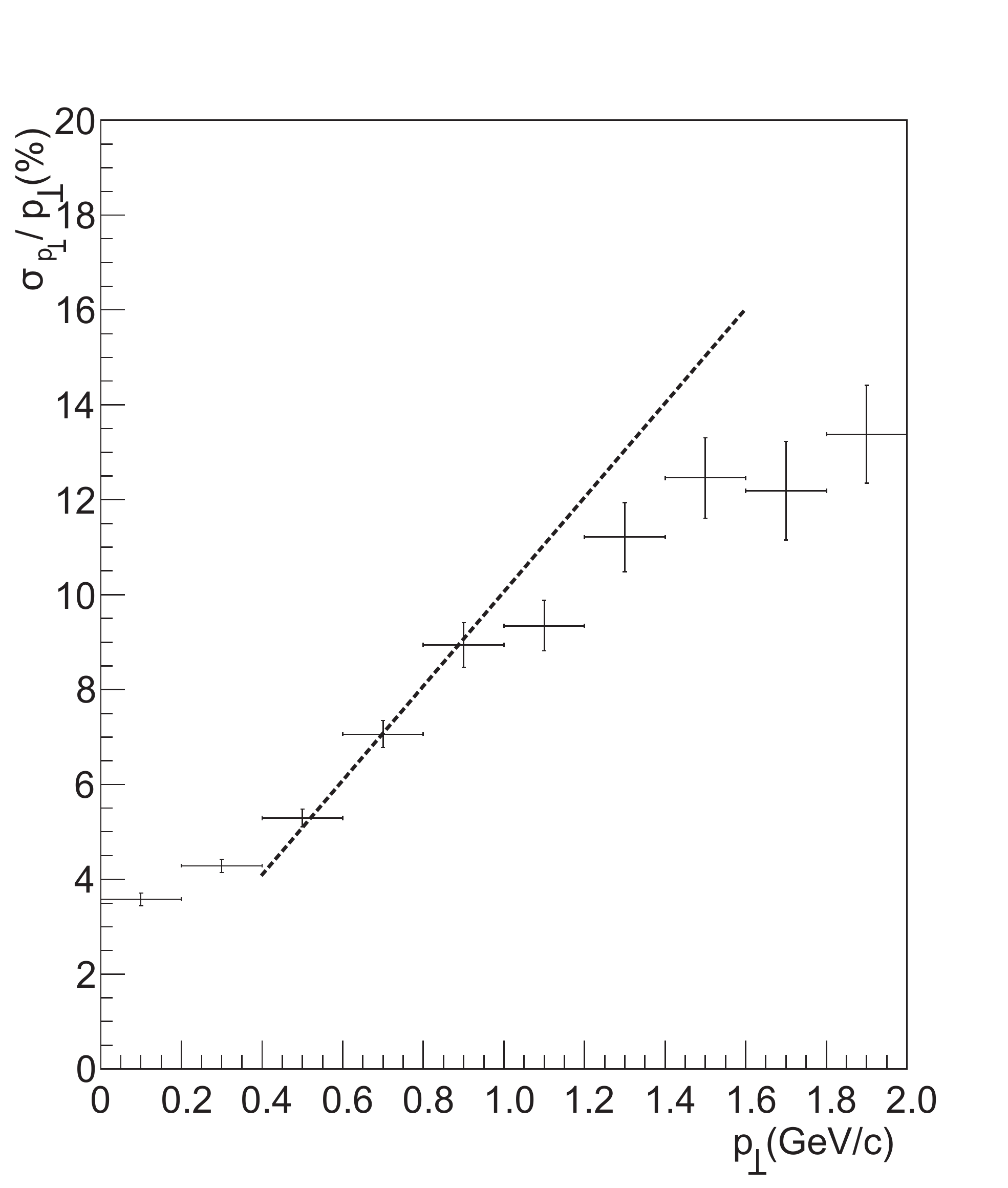}
\caption{Momentum resolution for a single TPC is shown as a function
of momentum perpendicular to the magnetic field as predicted by 
the Monte Carlo simulation of muons generated with the 
standard neutrino event generator of T2K. The tracks are selected 
to cross at least 50 out of the 72 pad columns of the TPC volume.
The dashed lines represents the momentum resolution goal.
}\label{fig:MomResolution}
\end{figure}

The momentum resolution for a single TPC,
computed with a Monte Carlo sample of 
simulated neutrino events, is shown in
Fig.~\ref{fig:MomResolution}.  
This sample includes all muons
which leave tracks that are sampled by at least 
50 of the 72 pad columns in a single TPC.
The TPC design goal was to achieve a relative resolution of
about 0.1~$p_\perp$/(GeV/c).
The simulation indicates that the measured spatial resolution is 
sufficient to attain that goal.

Using data samples taken with the magnet off,
the mean residual per TPC pad column is used to study the spatial point 
distortions produced by misalignments and nonuniform electric
fields.
After including corrections derived from an optical survey of
the micromegas module locations,
the means are below 0.1\,mm, apart from one readout plane
where some means are in the range 0.1 to 0.2\,mm.

\subsection{Particle identification}
\label{ss-perf-pid}

Particle identification in the TPC uses a truncated mean of measurements of energy loss of charged particles in the gas.
For each cluster a measured charge is defined as the sum of the detected charge
on all the pads in the cluster.
This charge is corrected for variation of the gas temperature and pressure as explained in section~\ref{ss-perf-stability}. Clusters at the edge of the Micromegas or  close to the central cathode are
rejected as an unknown fraction of the charge in these clusters has not been collected on the sensitive area.
The linear charge density of the track is estimated for each cluster
by taking into account the length of the track segment corresponding
to a pad column.
The lowest 70\% of the values are used to compute the truncated mean,
an optimized approach found through Monte Carlo simulation and
test beam studies.
A correction is applied to take into account the number of
clusters
used in the determination of the truncated mean.
The measured energy loss per unit length is used to calculate the
``pull'', $\delta_E(i)$,
the number of standard deviations the measurement is away from the
the expected value for particle type $i$ at the observed momentum.

\begin{figure}[ht]
\centering
\includegraphics[width=0.45\textwidth]{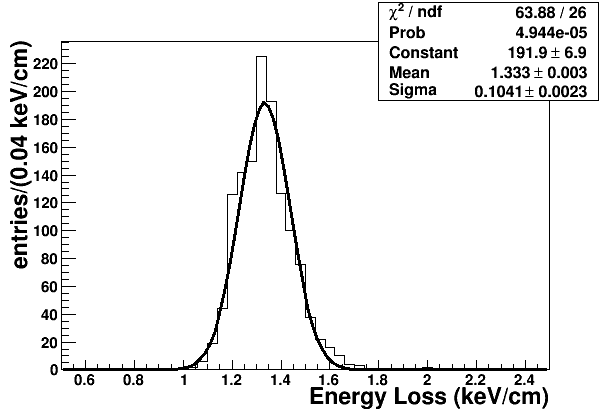}
\caption{Distribution of the energy loss for negatively charged particles with momenta between 400 and 500 MeV/c.
}\label{fig:perf-pidres}
\end{figure}

\begin{figure}[ht]
\centering
\ifx\figstyle\bw
 \includegraphics[width=0.45\textwidth]{fig29_bw.pdf}
\else
 \includegraphics[width=0.45\textwidth]{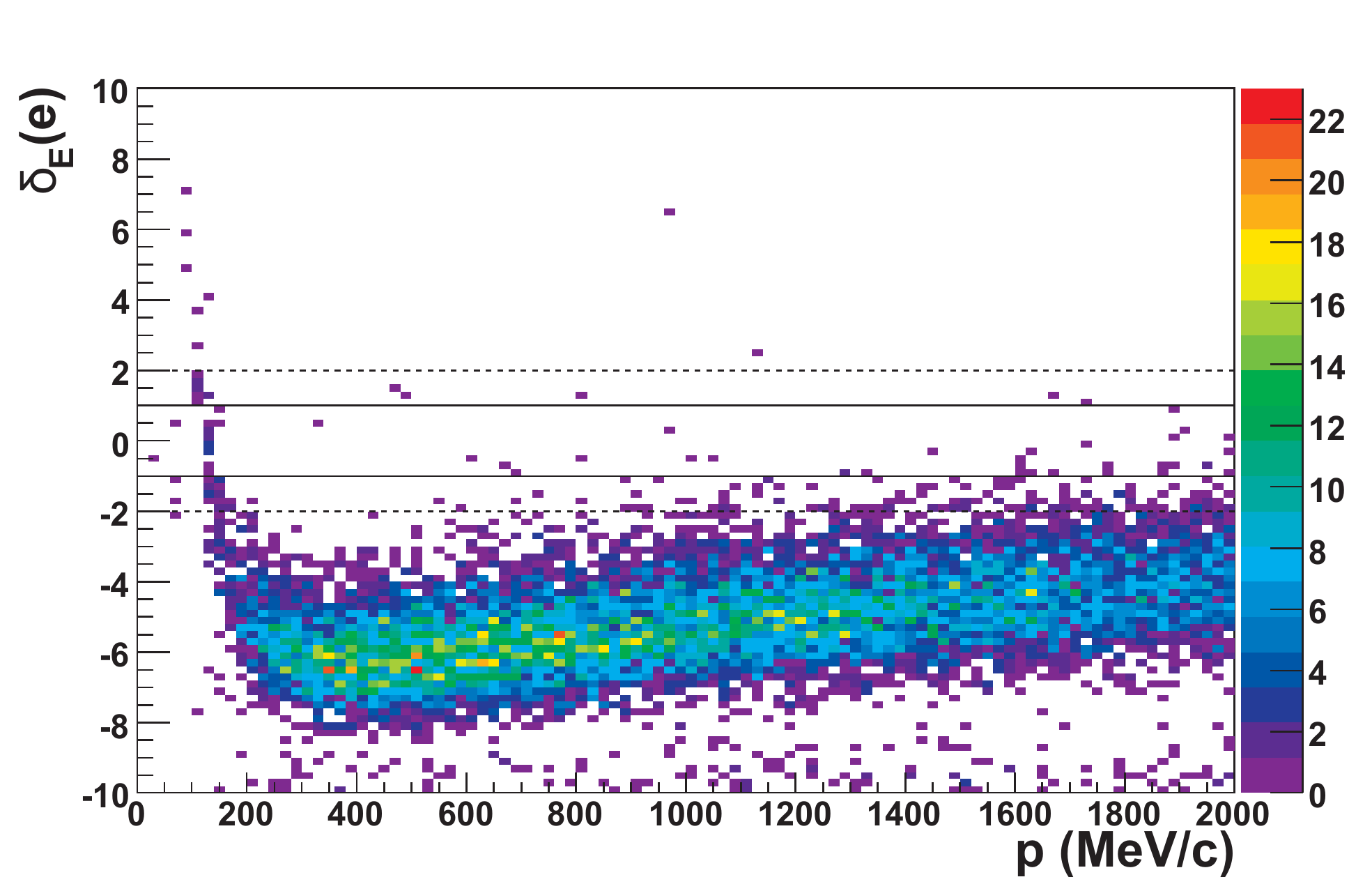}
\fi
\caption{Distribution of the energy loss pull in the electron hypothesis for a sample of through going muons. The solid and dashed lines indicate $|\delta_E(e)|<1$ and $|\delta_E(e)|<2$ respectively.
}\label{fig:perf-misidmu}
\end{figure}

The distribution of the deposited energy obtained using this method is shown in Fig.~\ref{fig:perf-pidres}. The resolution is of $7.8\pm0.2\%$ for minimum ionizing particles,
better than the 10\% requirement for the T2K TPCs.
This resolution allows muons to be distinguished 
from electrons in the TPCs: the probability of identifying a
muon as an electron is $0.2\%$ for $-1<\delta_E(e)<2$ and 
tracks below 1~GeV/c, as shown in Fig.~\ref{fig:perf-misidmu}.

\begin{figure}[ht]
\centering
\ifx\figstyle\bw
 \includegraphics[width=0.45\textwidth]{fig30_bw.pdf}
\else
 \includegraphics[width=0.45\textwidth]{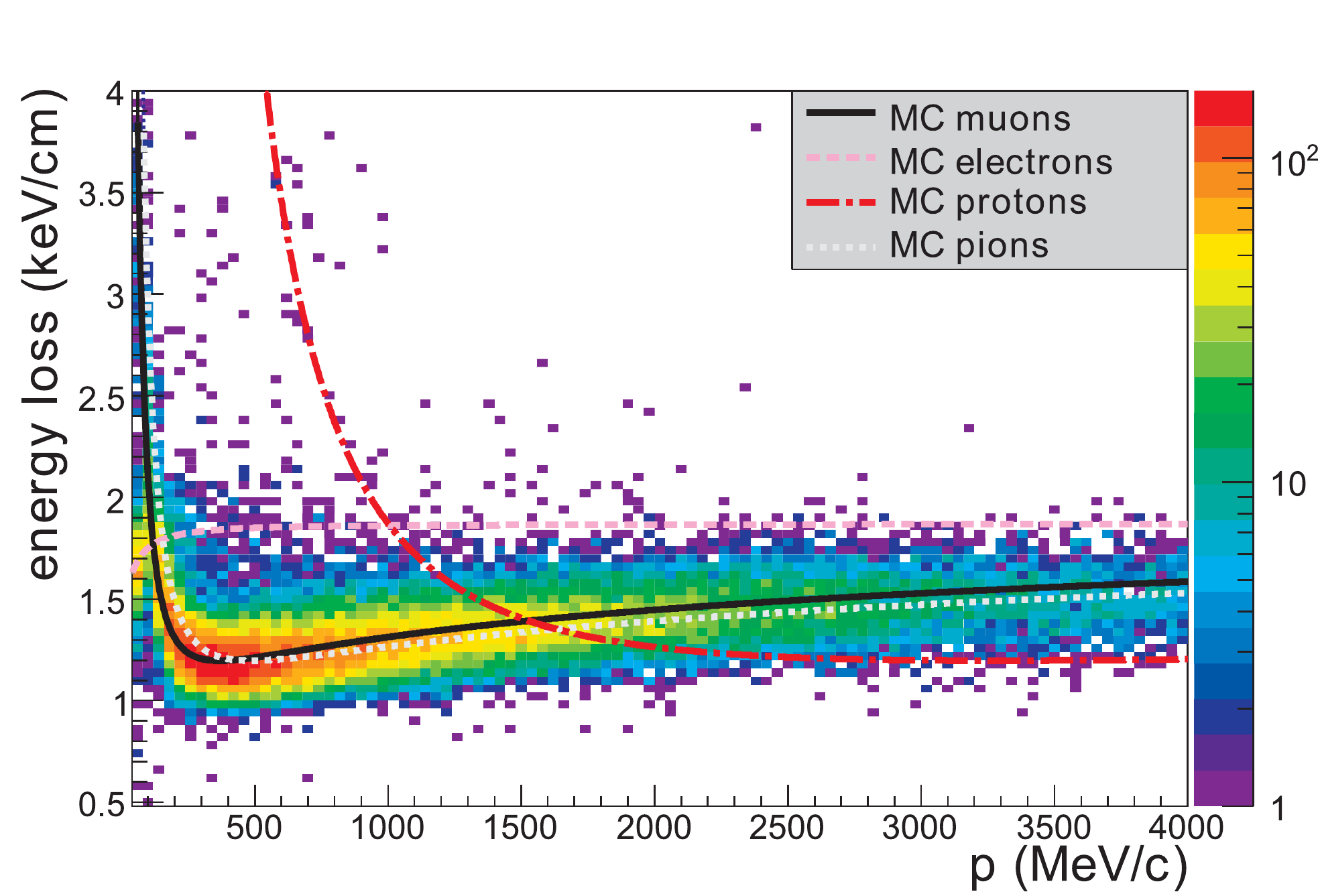}
\fi
\caption{Distribution of the energy loss as a function of the momentum for negatively charged particles produced in neutrino interactions, compared to the expected curves for muons, electrons, protons, and pions.
}\label{fig:perf-CTvspneg}
\end{figure}

\begin{figure}[ht]
\centering
\ifx\figstyle\bw
 \includegraphics[width=0.45\textwidth]{fig31_bw.pdf}
\else
 \includegraphics[width=0.45\textwidth]{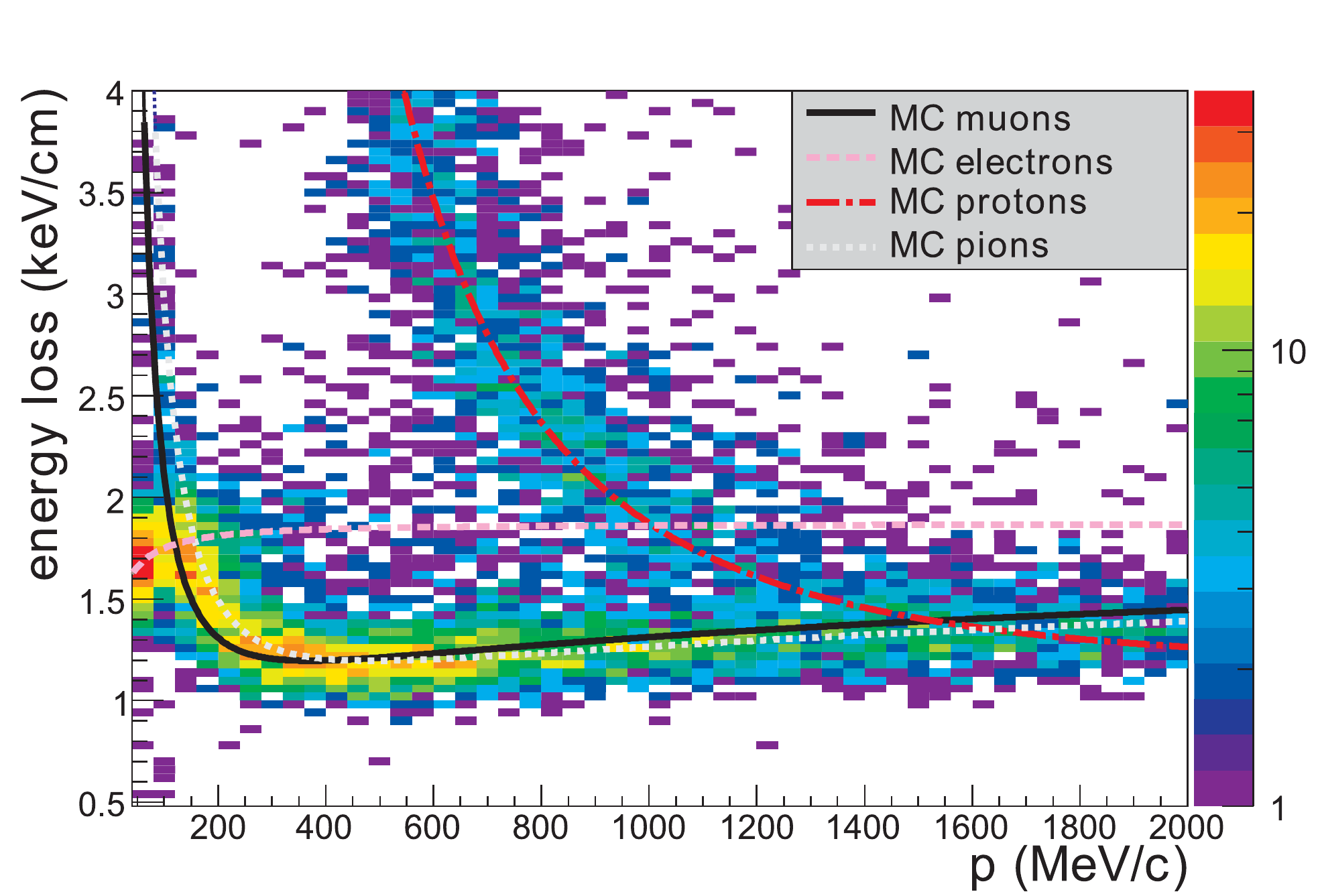}
\fi
\caption{Distribution of the energy loss as a function of the momentum for positively charged particles produced in neutrino interactions, compared to the expected curves for muons, electrons, protons, and pions.
}\label{fig:perf-CTvsppos}
\end{figure}

The distributions of the energy loss as a function of the momentum for data taken during the first T2K physics run are shown in Fig.~\ref{fig:perf-CTvspneg} and Fig.~\ref{fig:perf-CTvsppos} for negatively and positively charged particles respectively. These events mainly contain through-going muons and neutrino interactions in ND280. The data are compared to the expected curves for muons, electrons, pions and protons: the different particle species are clearly visible in the TPC. For negatively charged particles, mainly muons with few low momentum electrons are observed while in the positively charged sample protons, pions and positrons are seen.

%% file: chap08.tex
\section{Conclusion}

Over the period between 2005-2009, the T2K near detector TPCs 
and its subsystems were designed, constructed, 
tested in beam at TRIUMF, transported to JPARC,
installed and brought into operation.
Prior to the construction, prototypes of the TPCs and subsystems had 
been built for verification of design and performance.
The TPCs were ready for the first physics data taking of the
T2K experiment in 2010, and the spatial and energy loss
resolution goals have been achieved.
In the years to come,
the TPCs and the near detector tracker will make important
contributions to the understanding of neutrino oscillations.

%% file: chap09.tex
\section{Acknowledgements}

We are indebted to the
CERN-EN-ICE-DEM department which was responsible for 
the anode PCB production and mesh integration,
and in particular to R. de Oliveira and O. Pizzirusso who played
a major role in the quality and reliability of the T2K TPC bulk micromegas.

We would like to thank the support of the following agencies that
made the T2K TPC project possible:
National Research Council (NRC) Canada,
Natural Sciences and Engineering Research Council (NSERC) of Canada,
Commissariat \`a l'\'Energie Atomique (CEA) France,
Deutsche Forschungsgemeinschaft (DFG) Germany, and
Ministerio de Educaci\'on y Ciencia (MEC) Spain.